\documentclass[aps,pre,reprint, amsmath, amssymb,superscriptaddress]{revtex4-1}
 
\usepackage{bm}
\usepackage[retainorgcmds]{IEEEtrantools}
\usepackage{graphicx}
\usepackage{mathrsfs}
\usepackage{amsmath}
\usepackage{amsfonts}
\usepackage{amssymb}
\usepackage{color}
\usepackage{times,txfonts}
\usepackage{nicefrac}
\usepackage{ragged2e}
\usepackage{tikz}
\usepackage{braket}

\usepackage[colorlinks=true,linkcolor=blue,urlcolor=blue,citecolor=blue,pdfusetitle]{hyperref}

\usepackage{blindtext}

\usepackage[framemethod=TikZ]{mdframed}
\usepackage[T1]{fontenc}
\mdfdefinestyle{MyFrame}{%
    linecolor=black,
    outerlinewidth=1pt,
    roundcorner=0pt,
    innertopmargin=0.77\baselineskip,
    innerbottommargin=0.75\baselineskip,
    innerrightmargin=14pt,
    innerleftmargin=14pt,
    backgroundcolor=gray!20!white}
\mdfdefinestyle{MyFrameTitled}{%
    linecolor=black,
    outerlinewidth=1pt,
    roundcorner=0pt,
    innertopmargin=0.77\baselineskip,
    innerbottommargin=0.75\baselineskip,
    innerrightmargin=14pt,
    innerleftmargin=14pt,
    backgroundcolor=gray!20!white,
     frametitlebackgroundcolor  =black!15,
    frametitlerule =true,
}

\usepackage{amssymb}
\usepackage{pifont}

\begin{document}

\title{Thermodynamics of a continuously monitored double quantum dot heat engine \\in the repeated interactions framework}

\date{\today}

\author{Laetitia P. Bettmann}
\email{bettmanl@tcd.ie}
\affiliation{School of Physics, Trinity College Dublin, College Green, Dublin 2, Ireland}
\author{Michael J. Kewming}
\email{kewmingm@tcd.ie}
\affiliation{School of Physics, Trinity College Dublin, College Green, Dublin 2, Ireland}
\author{John Goold}
\email{gooldj@tcd.ie}
\affiliation{School of Physics, Trinity College Dublin, College Green, Dublin 2, Ireland}
\begin{abstract}
Understanding the thermodynamic role of measurement in quantum mechanical systems is a burgeoning field of study.
In this article, we study a double quantum dot (DQD) connected to two macroscopic fermionic thermal reservoirs. We assume that the DQD is continuously monitored by a quantum point contact (QPC), which serves as a charge detector. 
Starting from a minimalist microscopic model for the QPC and reservoirs, we show that the local master equation of the DQD can alternatively be derived in the framework of repeated interactions and that this framework guarantees a thermodynamically consistent description of the DQD and its environment (including the QPC).
We analyze the effect of the measurement strength and identify a regime in which particle transport through the DQD is both assisted and stabilized by dephasing. 
We also find that in this regime the entropic cost of driving the particle current with fixed relative fluctuations through the DQD is reduced. We thus conclude that under continuous measurement a more constant particle current may be achieved at a fixed entropic cost.
\end{abstract}
\maketitle{}
\section{Introduction}
\label{sec:int}
In recent decades, technological developments have led to the creation of artificial mesoscopic and nanoscopic steady-state heat engines \cite{benenti_fundamental_2017, myers_quantum_2022, pekola_colloquium_2021,pekola_towards_2015,arrachea_energy_2022}.
Without the need for moving parts in such engines, stochastic energy obtained in the form of a heat transfer is converted into useful work by currents of microscopic particles. 
Applications are diverse, ranging from atomic and molecular junctions \cite{dubi_colloquium_2011} to quantum dots \cite{josefsson_quantum-dot_2018}. 
At the scale at which these devices operate, not only do average values of thermodynamic quantities matter, but thermal and quantum fluctuations become just as relevant and must be taken into account for a complete thermodynamic understanding \cite{esposito_nonequilibrium_2009, campisi_fluctuation_2014}.
To study the devices experimentally, one often relies on continuous quantum measurement to obtain information about the temporal evolution of the internal state.
At the same time, even with the most non-invasive detectors such as quantum point contacts (QPCs) acting as charge sensors \cite{levinson_dephasing_1997, aleiner_dephasing_1997, stodolsky_measurement_1999, buttiker_charge_2000}, performing weak measurements inevitably leads to measurement backaction. 
Moreover, by monitoring an observable incompatible with energy, the energy transport in quantum devices can be altered too \cite{bhandari_continuous_2022}.
This points to the fundamental role that measurement and the subsequent quantum trajectory description arising from measurement play in the interpretation of stochastic quantum thermodynamics quantities \cite{Horowitz_2012, Manzano_Nonequilibrium_2015, Campisi_2015, elouard_role_2017,Elouard_2017, Manzano_quantum_2018,  di_stefano_nonequilibrium_2018, strasberg_operational_2019,Alonso_thermodynamics_2016, belenchia_entropy_2020, Rossi_experimental_2020, Menczel_quantum_2020, miller_quantum_2020, Liu_stochastic_2020, kewming_entropy_2022, Manzano_2022}.

While dephasing occurs in coherent quantum systems in many scenarios, such as electron-phonon interaction \cite{fedichkin_error_2004-1, fedichkin_study_2005} and  background charge noise \cite{itakura_dephasing_2003}, it also arises from capacitive coupling to QPCs  \cite{fujisawa_bidirectional_2006, ihn_quantum_2009}. In the latter scenario, dephasing is a direct consequence of the measurement backaction of the QPC on the monitored quantum system \cite{gurvitz_measurements_1997, levinson_dephasing_1997, aleiner_dephasing_1997, stodolsky_measurement_1999, buttiker_charge_2000}.
Interestingly, there are situations where, counter-intuitively, quantum transport is aided by dephasing, e.g. \cite{contreras-pulido_dephasing-assisted_2014, plenio_dephasing-assisted_2008, rebentrost_environment-assisted_2009,
scholes_perspective_2014,zerah-harush_effects_2020,kilgour_charge_2015,kilgour_inelastic_2016,sowa_environment-assisted_2017,
lacerda_dephasing_2021,mendoza-arenas_dephasing_2013, engel_evidence_2007,  collini_coherently_2010, znidaric_dephasing_2017, chiaracane_dephasing-enhanced_2022, caruso_highly_2009, caruso_fast_2016, chin_coherence_2012, collini_spectroscopic_2013,biggerstaff_enhancing_2016}, by suppressing coherent single-particle interference effects.
Exploiting this mechanism to improve transport efficiency is therefore interesting for applications in quantum technology, including controlled quantum systems, but also because it challenges the conception that disturbances due to couplings to the environment, under all circumstances, hinder performance. Interestingly, the presence of a dephasing noisy environment has been shown to be also beneficial for performance in some biological systems, such as photosynthetic systems \cite{faisal_noise_2008, adolphs_how_2006, gaab_effects_2004, mohseni_environment-assisted_2008, rebentrost_role_2009, rebentrost_environment-assisted_2009}.

A widely used tool to study the temporal evolution as well as steady-states of quantum systems are Gorini - Kossakowski - Sudarshan - Lindblad (GKSL) master equations (MEs) \cite{lindblad_generators_1976,gorini_completely_1976}.
Often the effect of an environment on the quantum system of interest can be approximated as local.
Then, one assumes that the dissipators, which account for the interaction of the system with the individual components of the environment, can be derived independently of each other.
A discussion of local versus global MEs can be found in Refs. \cite{cattaneo_local_2019, gonzalez_testing_2017,hofer_markovian_2017}. 
Importantly, as argued in Ref. \cite{levy_local_2014}, a local GKSL ME --- for which global detailed balance does not generally hold \cite{chiara_reconciliation_2018} --- does not guarantee a consistent thermodynamic description of non-equilibrium steady state (NESS), if the mechanism for its emergence is not fully taken into account. 
This issue was addressed by Barra \cite{barra_thermodynamic_2015} who---building off previous works \cite{esposito_entropy_2010, reeb_improved_2014}---developed a consistent thermodynamic framework for calculating the average flows of work, heat and entropy in the repeated interactions framework (collisional models) \cite{attal_repeated_2006, karevski_quantum_2009, ciccarello_quantum_2022, campbell_collision_2021, scarani_thermalizing_2002, ziman_diluting_2002} for systems driven locally by interactions with the environment. This framework is an intuitive and convenient methodology which requires assumptions on the environment similar to those of the local GKSL ME. However, it is important to point out that the compatibility of local MEs with thermodynamics is an active area of research and there are a variety of alternative approaches \cite{trushechkin_perturbative_2016, strasberg_operational_2019, soret_thermodynamic_2022, hewgill_quantum_2021,potts_thermodynamically_2021, cattaneo_local_2019}.
In the present work, we investigate the influence of continuous measurement via a QPC on the NESS transport and thermodynamic properties of a double quantum dot (DQD) \cite{van_der_wiel_electron_2002} coupled to two independent macroscopic thermal fermionic reservoirs.
Importantly, DQDs exhibit quantum coherence which makes them susceptible to dephasing due to measurement-induced backaction. 
A similar set up was studied also within the stochastic thermodynamics approach in Ref.~\cite{cuetara_double_2015}.
Here, we use a thermodynamically consistent formalism \cite{barra_thermodynamic_2015} in the framework of repeated interactions, starting from a minimalist microscopic model of the QPC and the reservoirs.
We show that the commonly used additive local GKSL ME for the DQD, accounting for both dissipation to the reservoirs as well as dephasing due to the coupling to the QPC, can be alternatively derived in the repeated interactions framework. 
Thus, the characteristics of the environment relevant to the time evolution of the DQD state and its NESS --- within the assumptions made in deriving the local GKSL ME --- are well captured in our minimalist model, suggesting that it can be used as a starting point for more sophisticated models in the future. We find that the net output power of the DQD operated as a heat engine is, in certain regimes, assisted by dephasing and can be tuned via the QPC's measurement strength.
Further, we study the entropic cost of precision in the particle current through the DQD within the context of a thermodynamic uncertainty relation (TUR) \cite{barato_thermodynamic_2015, pietzonka_universal_2018, pietzonka_finite-time_2017, pietzonka_universal_2016, horowitz_proof_2017}. 
TURs set a fundamental lower bound on the trade-off between the relative fluctuations of a thermodynamic current in classical Markovian NESS heat engines and the entropy production rate.
Moreover, TURs valid for quantum systems were developed in Ref. \cite{timpanaro_thermodynamic_2019} and subsequently studied for continuous measurements, and open quantum systems \cite{carollo_unraveling_2019, hasegawa_quantum_2020, hasegawa_thermodynamic_2021, van_vu_thermodynamics_2022}.
Importantly, recent works on TURs have been using the DQD as a working model \cite{agarwalla_assessing_2018, liu_thermodynamic_2019, Prech_2022}, albeit without the presence of a QPC.
We find that in the parameter regime in which the output power is increased due to the dephasing measurement-induced backaction, the precision of the particle current through the DQD, quantified by its relative fluctuations, is enhanced relative to the entropic cost of driving it. 
The fundamental lower bound set by the TUR, however, remains intact.


The paper is organised as follows: First, we study the NESS of the DQD in the absence of continuous measurement (\ref{sec: DQD}). We introduce the microscopic Hamiltonian describing the DQD, the two reservoirs, and their local interaction with the DQD, and then give the well-known GKSL ME for the DQD (\ref{sec: DQD Hamiltonian}). Next, we discuss the issue of the apparent thermodynamic inconsistency of the local GKSL equation for this set-up (\ref{sec: thermodynamic inconsistency}). We then show that the GKSL ME for the DQD can be alternatively derived in the framework of repeated interactions (\ref{sec: GKSL DQD FRI}).  Finally, we compute the NESS flows of work, heat and entropy production following Ref. \cite{barra_thermodynamic_2015} and show that our result is consistent with the first and second law of thermodynamics (\ref{sec: thermo DQD}).
Next, we introduce continuous measurement of the DQD via a QPC. We first give the microscopic Hamiltonian (\ref{sec: Hamiltonian DQD+QPC}) and then state the unconditional GKSL ME for the reduced density matrix for the DQD, now also accounting for dephasing due to measurement backaction. We then propose a minimalist composite unit for the QPC in the framework of repeated interactions  and show that the GKSL ME derived in this framework coincides with the GKSL ME derived starting from the microscopic Hamiltonian (\ref{sec: GKSL DQD+QPC FRI}). Next, we compute the NESS flows of work, heat and entropy production in the presence of the QPC and show that our result is consistent with the first and second law of thermodynamics (\ref{sec: thermo DQD+QPC}). We address the parameter regimes allowing for enhanced power output via dephasing assisted particle transport in Sec. (\ref{sec: DAT}). We show the consequences of the latter to reduce relative fluctuations of the particle current at a fixed entropic cost (\ref{sec: TUR DQD}). Finally, we conclude by summarizing and giving the outlook (\ref{sec: conclusion}).
%
\section{DQD operated as heat engine}
\label{sec: DQD}

Before investigating the effect of measurement-induced backaction due to the QPC on the boundary-driven DQD NESS, we first provide a pedagogical overview on computing the particle and energy fluxes of the NESS using the framework of repeated interactions, in the absence of a QPC. 
We show that although the time evolution of the reduced DQD density matrix in this approach is governed by a local GKSL ME that does not obey global detailed balance, the average NESS energy fluxes between the DQD and the reservoirs are compatible with the first and second laws of thermodynamics.
\begin{figure}
\begin{center}
\includegraphics[width=\linewidth]{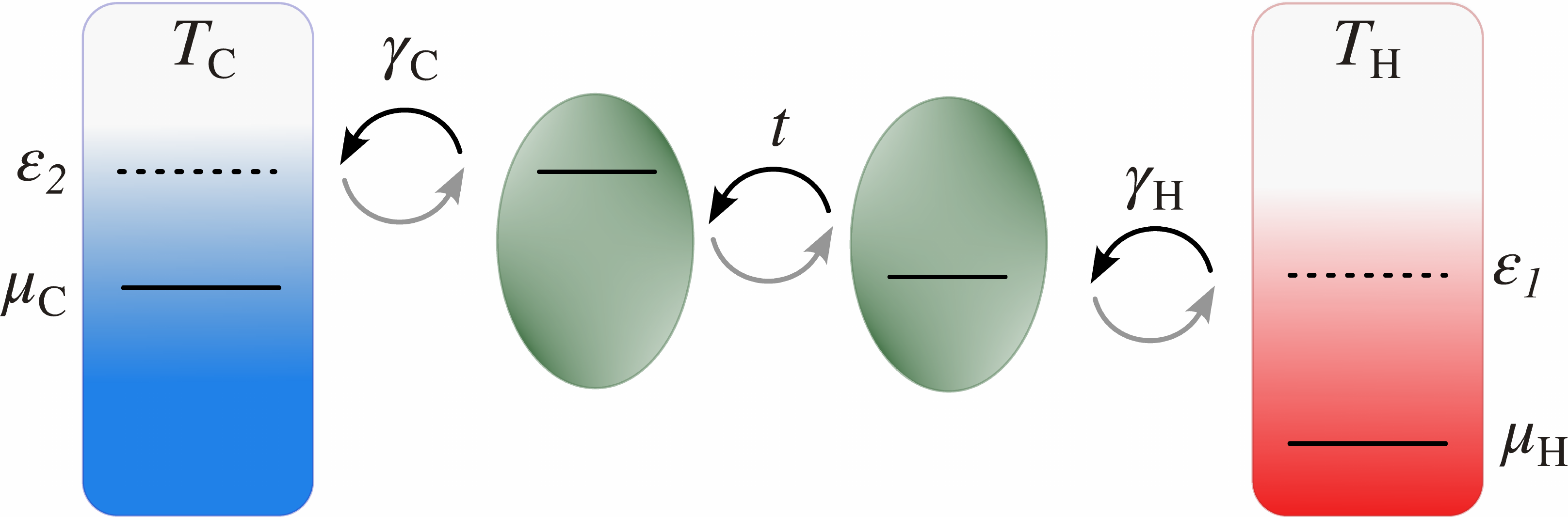}
  \caption{The DQD consists of two quantum dots coupled via a tunneling interaction of amplitude $t$. Each quantum dot is locally exchanging energy and particles with an independent, macroscopic thermal fermionic reservoir, at rate $\gamma_\mathrm{H(C)}$. The reservoirs are fully characterised by their respective chemical potential $\mu_\mathrm{H(C)}$ and temperature $T_\mathrm{H(C)}$. 
  }
 \label{fig:DQD_full}
\end{center}
\end{figure}

\subsection{DQD and the GKSL ME}
\label{sec: DQD Hamiltonian}
The global microscopic Hamiltonian of the DQD in contact with the two reservoirs, a resonant-level transport model, is of the form
\begin{equation}
    \hat{H}=\hat{H}_\mathrm{DQD}+\sum_{\alpha\in\lbrace \mathrm{H,C}\rbrace} \hat{H}_\mathrm{\alpha}+\hat{H}_\mathrm{DQD,R},
\end{equation}
where $\hat{H}_\mathrm{DQD}$ denotes the DQD Hamiltonian,
\begin{equation}
    \label{eq: H_reservoir_full}
    \hat{H}_\alpha = \sum_{k} \epsilon_{\alpha,k}\hat{c}_{\alpha,k}^\dagger \hat{c}_{\alpha,k}
\end{equation}
is the reservoir $\alpha$ Hamiltonian expressed in terms of the creation (annihilation) operators $\hat{c}_{\alpha,k}^\dagger$  ($\hat{c}_{\alpha,k}$) of the reservoir single-particle states with energy $\epsilon_{\alpha,k}$ and $\hat{H}_\mathrm{DQD, R}$ is the tunneling Hamiltonian between the DQD and the reservoirs. The DQD Hamiltonian is given by
\begin{align}
\label{eq: H_DQD}
 \hat{H}_\mathrm{DQD} & = \epsilon_1 \hat{c}_1^\dagger \hat{c}_1 + \epsilon_2 \hat{c}_2^\dagger \hat{c}_2 + t \left(\hat{c}_1^\dagger \hat{c}_2+\hat{c}_2^\dagger \hat{c}_1\right) .
\end{align}
It is expressed in terms of the creation (annihilation) operators $\hat{c}_{1(2)}^\dagger$ ($\hat{c}_{1(2)}$) of single quantum dot states with energies $\epsilon_{1(2)}$, and $t$ denotes the interdot tunneling amplitude. The tunneling Hamiltonian between the DQD and the reservoirs is given by
\begin{align}
\label{eq: tunneling_reservoirs}
\begin{split}
    \hat{H}_\mathrm{DQD,R} =& \sum_{k} g_{\mathrm{H},k} \hat{c}_1^\dagger \hat{c}_{\mathrm{H},k} + g^*_{\mathrm{H},k}\hat{c}^\dagger_{\mathrm{H},k}\hat{c}_1 \\
&+ \sum_{k} g_{\mathrm{C},k}\hat{c}_2^\dagger \hat{c}_{\mathrm{C},k} + g^*_{\mathrm{C},k}\hat{c}^\dagger_{\mathrm{C},k}\hat{c}_2,
\end{split}
\end{align}
where $g_{\alpha,k}$ denotes the  tunneling rate between mode $k$ in reservoir $\alpha$ and the respective quantum dot. 
The set-up is such that the DQD is driven into a NESS in the long-time limit by its coupling to the two reservoirs with chemical potentials $\mu_\mathrm{H(C)}$ and temperatures $T_\mathrm{H(C)}$ at its boundaries. 

Usually, the NESS is derived from a GKSL master equation, which is only valid under the following assumptions \cite{breuer_quantum_2003}: weak coupling of the reservoirs to the DQD (Born approximation), the reservoirs are memory-less (Markov approximation), and the bandwidths of the reservoirs are much larger than those of the DQD. The last assumption yields frequency-independent interactions between the reservoirs and the DQD (wide-band limit).
Finally, we suppose that the reservoirs are kept in local thermal equilibrium states and are static and uncorrelated with the DQD and among each other at all times.
Tracing out the reservoir degrees of freedom one obtains the additive local GKSL ME governing the time evolution of the DQD, $\hat{\rho}_\mathrm{DQD}=\mathrm{Tr_R}\left[\hat{\rho}_\mathrm{tot}\right]$, ($\hbar=1$)
\begin{align}
\label{eq: me_DQD}
\begin{split}
     \dot{\hat{\rho}}_\mathrm{DQD}=&-i[\hat{H}_\mathrm{DQD},\hat{\rho}_\mathrm{DQD}]  \\ 
    &+\underbrace{\gamma_\mathrm{H} f_\mathrm{H}\left(\epsilon_1\right)\mathcal{D}[\hat{c}_1^\dagger] \hat{\rho}_\mathrm{DQD}
    +\gamma_\mathrm{H} \left(1-f_\mathrm{H}\left(\epsilon_1\right)\right)\mathcal{D}[\hat{c}_1]\hat{\rho}_\mathrm{DQD}}_{\mathcal{L}_\mathrm{H}\left(\hat{\rho}_\mathrm{DQD}\right)}\\
    &+ \underbrace{\gamma_\mathrm{C} f_\mathrm{C}\left(\epsilon_2\right)\mathcal{D}[\hat{c}_2^\dagger] \hat{\rho}_\mathrm{DQD}
    +\gamma_\mathrm{C} \left(1-f_\mathrm{C}\left(\epsilon_2\right)\right)\mathcal{D}[\hat{c}_2]\hat{\rho}_\mathrm{DQD}}_{\mathcal{L}_\mathrm{C}\left(\hat{\rho}_\mathrm{DQD}\right)},\\
    &= \mathcal{L}(\hat{\rho}_\mathrm{DQD}),
\end{split}
\end{align}
where the reservoir Fermi functions $f_\mathrm{H(C)}\left(\epsilon_{1(2)}\right)=\left[ 1+\exp\left(\left(\epsilon_{1(2)}-\mu_\mathrm{H(C)}\right)/k_\mathrm{B}T_\mathrm{H(C)}\right)\right]^{-1}$, where $k_\mathrm{B}$ is the Boltzmann constant, are evaluated at the quantum dot energies $\epsilon_{1(2)}$ (resonant tunneling), and the tunneling rates between the DQD and the two reservoirs are denoted by $\gamma_\mathrm{H(C)}$.
The set-up as well as the interactions are schematically depicted in Fig.~(\ref{fig:DQD_full}).
The dissipators are defined as $\mathcal{D}[\hat{L}]\hat{\rho}_\mathrm{DQD}=\hat{L}\hat{\rho}_\mathrm{DQD} \hat{L}^\dagger - \frac{1}{2}\lbrace L^\dagger \hat{L},\hat{\rho}_\mathrm{DQD}\rbrace$. Here, we use the common notations $\left[\cdot,\cdot\right]$ for the commutator and $\lbrace\cdot,\cdot\rbrace$ for the anti-commutator. The jump operators in Eq. (\ref{eq: me_DQD}) are local and account for the boundary-driving of the DQD by the reservoirs since they drive transitions between single quantum dot eigenstates at the boundary rather than between delocalized energy eigenstates of $\hat{H}_\mathrm{DQD}$. In this approach it is assumed that the dissipators for the two reservoirs can be derived independently of each other. While this approximation is commonly made, it is important to point out that in general the exact evolution cannot be expressed as a sum of dissipators describing the action of each reservoir alone \cite{mitchison_non-additive_2018}. Moreover, the interaction $\hat{H}_\mathrm{DQD,R}$ does not commute with the DQD Hamiltonian $\hat{H}_\mathrm{DQD}$, i.e. $\left[\hat{H}_\mathrm{DQD,R},\hat{H}_\mathrm{DQD}\right] \neq 0$. This is because the jump operators $\hat{L}$ here are not eigenoperators of $\hat{H}_\mathrm{DQD}$, due to the inter-system coupling \cite{chiara_reconciliation_2018}. Thus, the local GKSL ME (\ref{eq: me_DQD}) derived from the interaction $\hat{H}_\mathrm{DQD,R}$ violates global detailed balance, although locally detailed balance holds. 
%
\subsection{Thermodynamic inconsistency of the local GKSL ME}
\label{sec: thermodynamic inconsistency}
In this section, we follow the discussion in the work by Levy and Kosloff \cite{levy_local_2014} on the the apparent violation of the second law of thermodynamics starting from the local GKSL ME (\ref{eq: me_DQD}) for the DQD coupled to two reservoirs.
Assume for this example the absence of a chemical potential gradient. For simplicity we set $\mu_\mathrm{H},\mu_\mathrm{C}=0$.
Commonly, the heat current from the hot reservoir into the DQD is defined as 
\begin{align}
\begin{split}
    \dot{Q}_\mathrm{H}=& \mathrm{Tr}\left[\hat{H}_\mathrm{DQD}\mathcal{L}_\mathrm{H}\left(\hat{\rho}_\mathrm{DQD}\right)\right].
\end{split}
\end{align}
Evaluating the trace, it may be rewritten as
\begin{equation}
    \dot{Q}_\mathrm{H}= \left(f_\mathrm{H}(\epsilon_1)-f_\mathrm{C}(\epsilon_2)\right)  \mathcal{F},
\end{equation}
where $\mathcal{F}$ is a function of the DQD parameters and is always positive. Clearly, for $\beta_\mathrm{C}\epsilon_2<\beta_\mathrm{H}\epsilon_1$, $\left(f_\mathrm{H}(\epsilon_1)-f_\mathrm{C}(\epsilon_2)\right) <0$, since $\mu_\mathrm{H},\mu_\mathrm{C}=0$, so that the heat current from the hot reservoir is negative. Since in this NESS $\dot{Q}_\mathrm{H}=-\dot{Q}_\mathrm{C}$, the second law of thermodynamics appears to be violated, as the entropy production rate, for $\dot{Q}_\mathrm{H}<0$,
\begin{align}
    \sigma = \left(\beta_\mathrm{C}-\beta_\mathrm{H}\right)\dot{Q}_\mathrm{H}<0.
\end{align}
This apparent violation arises because the above definition of the heat current from the hot reservoir is valid only for global MEs which satisfy the condition of global detailed balance. 
The above example illustrates that the local GKSL ME alone may not be suitable for describing the thermodynamics of the DQD and other approaches are needed.
%
\subsection{GKSL ME in the framework of repeated interactions}
\label{sec: GKSL DQD FRI}
A description of boundary-driven systems such as the above reconciled with the second law of thermodynamics, without having to compromise on using a local GKSL ME, was put forward by Barra in Ref.~\cite{barra_thermodynamic_2015} for the framework of repeated interactions. We will briefly review it here. Note, however, that a variety of alternative approaches exist \cite{trushechkin_perturbative_2016, strasberg_operational_2019, soret_thermodynamic_2022, hewgill_quantum_2021,potts_thermodynamically_2021, cattaneo_local_2019}. 

\begin{figure}
\begin{center}
\includegraphics[width=\linewidth]{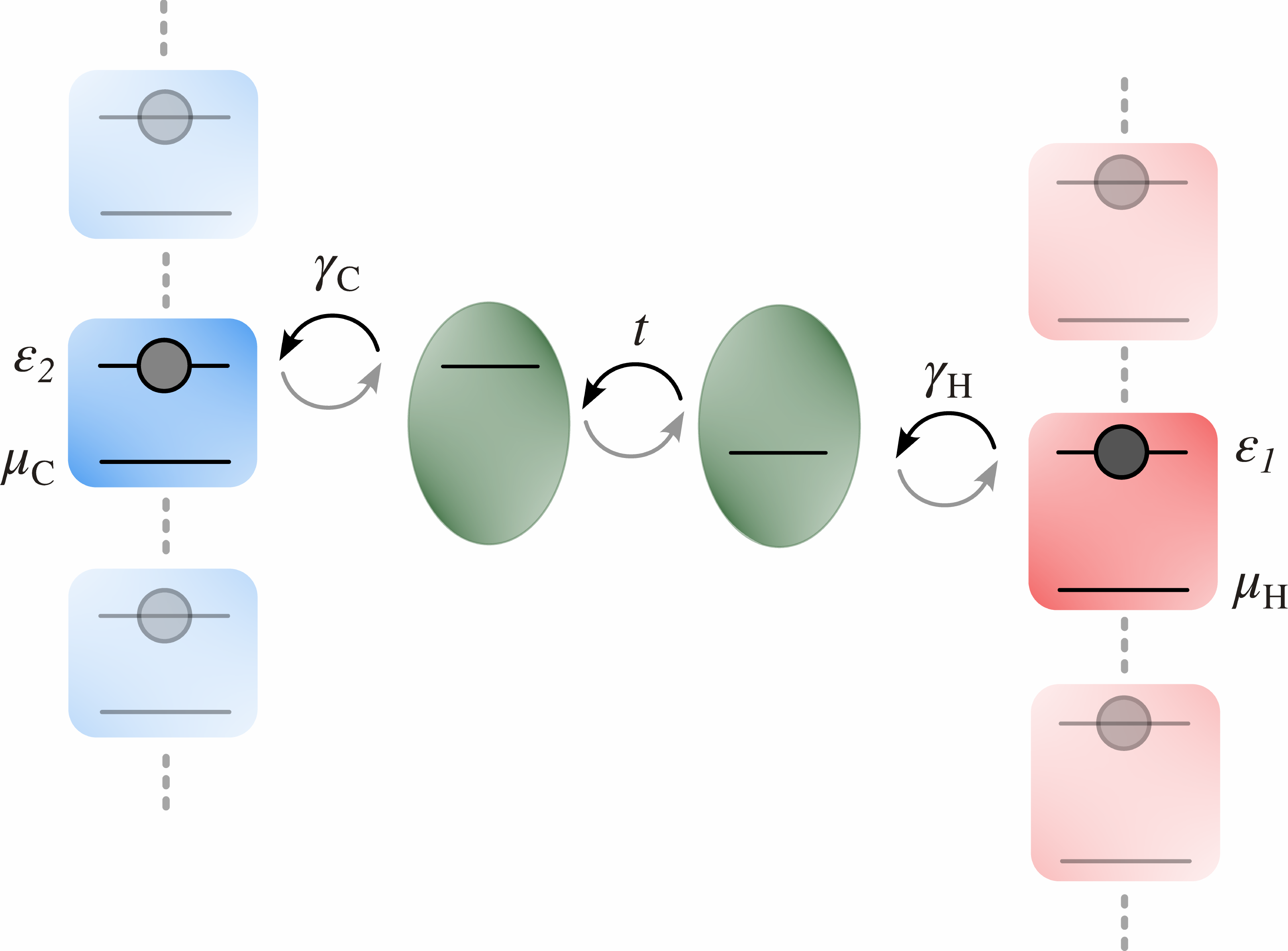}
  \caption{In the framework of repeated interactions, the hot and cold reservoirs are replaced by periodically refreshed units consisting of single qubits with energy splitting $\epsilon_1-\mu_H$ and $\epsilon_2-\mu_C$, respectively. For the interaction time $\tau$, a single unit per reservoir interacts with the DQD. Subsequently the two units are replaced by new units with the same initial state. The inital population of the qubit levels are set by the chemical potentials and temperatures of the reservoirs we aim to model.}
 \label{fig:DQD_minimal}
\end{center}
\end{figure}

The framework of repeated interactions provides a prescription to model a system's interaction with an environment in terms of replacing it by a collection of identical units interacting with the system one after the other for a fixed interaction time $\tau$. Initially the system is uncorrelated with all interaction units and, importantly, the interaction units remain uncorrelated among each other for all times. The system then evolves for an interaction time $\tau$ in the presence of the interaction with a single unit, subsequently the unit is replaced, and the system then interacts with the fresh unit, again for an interval of length $\tau$, and so on.
In Fig. (\ref{fig:DQD_minimal}) we schematically show how we model the two thermal reservoirs by units, each consisting of a thermal qubit with an energy splitting of $\epsilon_1-\mu_\mathrm{H}$ and $\epsilon_2-\mu_\mathrm{C}$, respectively, during each interaction interval. The populations of the qubit levels are set by the chemical potentials and temperatures, $\mu_\mathrm{H(C)}$ and $T_\mathrm{H(C)}$, of the original reservoirs.
This microscopic set-up is described by the bare Hamiltonians for the DQD $\hat{H}_\mathrm{DQD}$ defined in Eq.~(\ref{eq: H_DQD}) and the qubit-units for the two reservoirs
\begin{align}
\label{eq: H_unit_reservoirs}
    \hat{H}_\alpha & = (\epsilon_\alpha-\mu_\alpha) \hat{c}_\alpha^\dagger \hat{c}_\alpha,
\end{align}
where $\alpha \in \lbrace\mathrm{H,C}\rbrace$, and the (single) interactions between the DQD and reservoir qubits, via
\begin{align}
\label{eq: single_ints_DQD_H}
    \hat{v}_\mathrm{H} &= \sqrt{\gamma_\mathrm{H}} \left(\hat{c}_\mathrm{H}^\dagger \hat{c}_1 + \hat{c}_1^\dagger \hat{c}_\mathrm{H} \right), \\
\label{eq: single_ints_DQD_C}
    \hat{v}_\mathrm{C} &= \sqrt{\gamma_\mathrm{C}} \left(\hat{c}_\mathrm{C}^\dagger \hat{c}_2 + \hat{c}_2^\dagger \hat{c}_\mathrm{C} \right).
\end{align}
The total Hamiltonian is given by
\begin{equation}
    \hat{H}=\hat{H}_\mathrm{DQD} + \sum_\alpha \hat{H}_\alpha + \sum_\alpha \hat{V}_\alpha
\end{equation}
where $\hat{V}_\alpha=\hat{v}_\alpha/\sqrt{\tau}$. Importantly, the periodic refreshing of the units introduces time-dependence into the total Hamiltonian. Barra shows that because of this, an inherent external work is required to produce the dissipative evolution described by the local GKSL equation \cite{barra_thermodynamic_2015}.
Since the units representing the two reservoirs consist of a single thermal qubit each, their respective density matrix is initialised to 
\begin{align}
\label{eq: dens_mat_H}
    \hat{\omega}_{\beta_\mathrm{H}} &= \left(\begin{array}{cc}
         f_\mathrm{H}(\epsilon_1)&  0\\
         0 & 1-f_\mathrm{H}(\epsilon_1)
    \end{array}\right),\\
    \label{eq: dens_mat_C}
    \hat{\omega}_{\beta_\mathrm{C}} &= \left(\begin{array}{cc}
         f_\mathrm{C}(\epsilon_2)&  0\\
         0 & 1-f_\mathrm{C}(\epsilon_2)
    \end{array}\right).
\end{align}
The total density matrix $\hat{\rho}$ at the start of each subsequent interaction is given by
\begin{equation}
   \hat{\rho} = \hat{\rho}_\mathrm{DQD} \otimes \underbrace{\hat{\omega}_{\beta_\mathrm{H}} \otimes \hat{\omega}_{\beta_\mathrm{C}}}_{\hat{\omega}_\mathrm{R}}. 
\end{equation}
After initialisation $\hat{\rho}$ evolves unitarily for the duration of the interaction time $\tau$ with
\begin{equation}
\label{eq: unitary}
    \hat{U}=\exp\left(-i\tau \hat{H} \right).
\end{equation}
After the $n^\mathrm{th}$ interaction, the difference $\Delta \hat{\rho}_\mathrm{DQD}\left(n\tau\right)= \hat{\rho}_\mathrm{DQD}(n\tau+\tau)-\hat{\rho}_\mathrm{DQD}(n\tau)$ is given by
\begin{equation}
    \Delta \hat{\rho}_\mathrm{DQD}\left(n\tau\right) = \mathrm{Tr}_\mathrm{E}\left[\hat{U}\hat{\rho}_\mathrm{DQD}\left(n\tau\right)\otimes\hat{\omega}_\mathrm{R} \hat{U}^\dagger - \hat{\rho}_\mathrm{DQD}\left(n\tau\right)\otimes\hat{\omega}_\mathrm{R}\right].
\end{equation} 
\\
The GKSL ME is recovered by expanding the unitary $\hat{U}$ to second order in $\tau$ and taking the limit of infinitely many interactions $n\rightarrow \infty$ and infinitesimal interaction time $\tau \rightarrow 0$. One then finds
\begin{equation}
\label{eq: ME_RI}
    \dot{\hat{\rho}}_\mathrm{DQD}=-i\left[\hat{H}_\mathrm{DQD},\hat{\rho}_\mathrm{DQD}\right]+\sum_{\alpha \in \lbrace\mathrm{H}, \mathrm{C} \rbrace} \mathcal{D}_\alpha\left(\hat{\rho}_\mathrm{DQD}\right).
\end{equation}
Importantly, the action of the reservoirs on the system emerges as a sum of dissipators due to the interaction with individual reservoirs defined by
\begin{align}
\label{eq: Dalpha}
\begin{split}
    \mathcal{D}_\alpha\left(\hat{\rho}_\mathrm{DQD}\right) =& \mathrm{Tr}_\alpha\left[\hat{v}_\alpha\left(\hat{\rho}_\mathrm{DQD}\otimes\hat{\omega}_{\beta_\alpha}\right)\hat{v}_\alpha\right]\\
    &-\frac{1}{2}\mathrm{Tr}_\alpha\lbrace \hat{v}_\alpha^2,\left(\hat{\rho}_\mathrm{DQD}\otimes\hat{\omega}_{\beta_\alpha}\right)\rbrace.
\end{split}
\end{align}
By plugging Eq.~(\ref{eq: single_ints_DQD_H}) and Eq.~(\ref{eq: single_ints_DQD_C}) into Eq.~(\ref{eq: Dalpha}), we recover the local GKSL ME stated in Eq.~(\ref{eq: me_DQD}).
Therefore, we know that at the level of the DQD, the dynamics dictated by the local GKSL equation and in the framework of repeated interactions are equivalent. However, the utility of the repeated interactions framework allows us to recover a thermodynamically consistent picture, as we now show.
\subsection{Thermodynamic consistency}
\label{sec: thermo DQD}
We are now in a position to compute the flows of heat and work between the DQD and the reservoirs, as well as the entropy production rate in the NESS~\cite{barra_thermodynamic_2015}.

The NESS of the DQD is uniquely defined by
\begin{equation}
\dot{\hat{\rho}}_\mathrm{DQD}^\mathrm{NESS}=\mathcal{L}\left({\hat{\rho}}_\mathrm{DQD}^\mathrm{NESS}\right)=0,
\end{equation}
where the Liouvillian $\mathcal{L}$ is defined in Eq.~(\ref{eq: me_DQD}).
Firstly, we are interested in showing agreement with the first law of thermodynamics.
To this aim, we begin by defining the work and heat currents from the reservoir $\alpha$ as
\begin{align}
\label{eq: heat_curr}
    \dot{Q}_\alpha &= -D_\alpha\left(\hat{H}_\alpha\right), \\
    \label{eq: work_curr}
    \dot{W}_\alpha &= D_\alpha\left(\hat{H}_\mathrm{DQD}+\hat{H}_\alpha\right),
\end{align}
where
\begin{align}
    D_\alpha\left(\hat{A}\right) &= \mathrm{Tr}\left[ \left(\hat{v}_\alpha \hat{A} \hat{v}_\alpha -\frac{1}{2} \lbrace \hat{v}_\alpha^2,\hat{A}\rbrace\right) \hat{\rho}^\mathrm{NESS}_\mathrm{DQD}\otimes \hat{\omega}_{\beta_\alpha}\right]. 
\end{align}
In the NESS, the average particle current from the hot reservoir into the DQD is given by
\begin{align}
\label{eq:jH}
     J^N_\mathrm{H} &= \gamma_\mathrm{H}\left(f_\mathrm{H}(\epsilon_1)-\langle \hat{c}_1^\dagger \hat{c}_1\rangle\right)\,,\\
    &= \frac{4|t|^2 \Delta f\gamma_\mathrm{H}\gamma_\mathrm{C}\tilde{\gamma}}{\gamma_\mathrm{H}\gamma_\mathrm{C}\left(4\Delta \epsilon^2+\tilde{\gamma}^2\right)+4|t|^2\tilde{\gamma}^2},
\end{align}
where $\Delta \epsilon =\epsilon_2-\epsilon_1$ denotes the detuning between the two quantum dots in the DQD, $\Delta f = f_\mathrm{H}(\epsilon_1)-f_\mathrm{C}(\epsilon_2)$ is the difference in the Fermi distributions of the reservoirs evaluated at the quantum dot energies, and $\tilde{\gamma} = (\gamma_{\rm H} + \gamma_{\rm C})$. 
Using Eq.~(\ref{eq: work_curr}), we find that the work currents into the DQD in the NESS are given by
\begin{align}
    \dot{W}_\mathrm{H} &= \left(\frac{\gamma_\mathrm{H}\Delta \epsilon }{ \tilde{\gamma}}+\mu_\mathrm{H}\right) J^N_\mathrm{H}, \\
    \dot{W}_\mathrm{C} &= \left(\frac{\gamma_\mathrm{C}\Delta \epsilon }{ \tilde{\gamma}}-\mu_\mathrm{C}\right) J^N_\mathrm{H}, \\
    \dot{W}_{\mathrm{tot}} &= -\left[\left(\epsilon_1-\mu_\mathrm{H}\right)-\left(\epsilon_2-\mu_\mathrm{C}\right)\right] J^N_\mathrm{H}.
\end{align}
Similarly, the heat currents into the DQD in the NESS, according to Eq.~(\ref{eq: heat_curr}), are found to be
\begin{align}
    \dot{Q}_\mathrm{H} &= \left(\epsilon_1-\mu_\mathrm{H}\right)J^N_\mathrm{H}, \\
    \dot{Q}_\mathrm{C} &= -\left(\epsilon_2-\mu_\mathrm{C}\right)J^N_\mathrm{H}, \\
    \dot{Q}_{\mathrm{tot}} &= \left[\left(\epsilon_1-\mu_\mathrm{H}\right)-\left(\epsilon_2-\mu_\mathrm{C}\right)\right] J^N_\mathrm{H}.
\end{align}
Note the linear dependence of the energy current on the particle current. This so-called tight-coupling limit, where the determinant of the Onsager response matrix vanishes, is a consequence of the weak-coupling approximation made in deriving the GKSL equation and does not hold in general. 
Nevertheless, the tight-coupling limit is of high interest in nanoelectronics since in this limit the system may operate at Carnot efficiency \cite{humphrey_reversible_2002}.

The first law states that $\dot{W}_\mathrm{tot}+\dot{Q}_\mathrm{tot}=\langle \dot{\hat{H}}_\mathrm{DQD} \rangle$. Within this framework, the expectation value of the first derivative with respect to time of the DQD Hamiltonian is $\langle \dot{\hat{H}}_\mathrm{DQD} \rangle=\sum_\alpha D_\alpha\left(\hat{H}_\mathrm{DQD}\right)$. In the NESS, this quantity vanishes. Clearly, $\dot{W}_{\mathrm{tot}}=-\dot{Q}_{\mathrm{tot}}$, so that overall our calculation of heat and work currents into the DQD in this framework is consistent with the first law of thermodynamics.

Moreover, if this device is operated as a heat engine, then its efficiency $\eta$ is defined as the ratio between the time derivative of total work performed by the system DQD and the time derivative of the heat flowing into the DQD from the hot reservoir,
\begin{align}
\label{eq: efficiency_no_QPC}
\begin{split}
    \eta &= \frac{-\dot{W}_{\mathrm{tot}}}{\dot{Q}_{\mathrm{H}}}\,,\\
     &= 1-\frac{\epsilon_2-\mu_\mathrm{C}}{\epsilon_1-\mu_\mathrm{H}}.\\
\end{split}
\end{align}
Note that the DQD operates as a heat engine, if the work current leaving the DQD $-\dot{W}_\mathrm{tot}>0$. This is ensured if $\left(f_\mathrm{H}(\epsilon_1)-f_\mathrm{C}(\epsilon_2)\right)>0$, so that $J^N_\mathrm{H}>0$, and $\left(\epsilon_1-\mu_H\right)-\left(\epsilon_2-\mu_C\right)>0$.
By the first condition,
\begin{equation}
    \frac{\left(\epsilon_2-\mu_\mathrm{C}\right)/T_\mathrm{C}}{\left(\epsilon_1-\mu_\mathrm{H}\right)/T_\mathrm{H}}>1
\end{equation} 
and we find that $\eta$ is bounded from above by the Carnot efficiency $\eta_C$, since
\begin{align}
    \eta  < 1- \frac{T_\mathrm{C}}{T_\mathrm{H}}= \eta_C.
\end{align}

We can finally show that this model satisfies the second law of thermodynamics. 
This can be readily achieved by recognising that in the NESS, the entropy production rate can be expressed in standard thermodynamic form 
\begin{align}
\label{eq: ent_prod}
    \sigma &= -\sum_{\alpha \in \lbrace \mathrm{H,C} \rbrace} \beta_\alpha \dot{Q}_\alpha\\
    &=\left(\frac{\epsilon_2-\mu_\mathrm{C}}{k_\mathrm{B}T_\mathrm{C}}-\frac{\epsilon_1-\mu_\mathrm{H}}{k_\mathrm{B}T_\mathrm{H}}\right)J^{N}_{\rm H}\\
    & \geq 0,
\end{align}
where $\beta_\alpha$ is the inverse temperature of reservoir $\alpha$. One can easily show that it is strictly positive, thus we find that this result is indeed consistent with the second law of thermodynamics.

\section{DQD continuously monitored by QPC}
\label{sec: DQD+QPC}
\begin{figure}
\begin{center}
\includegraphics[width=\linewidth]{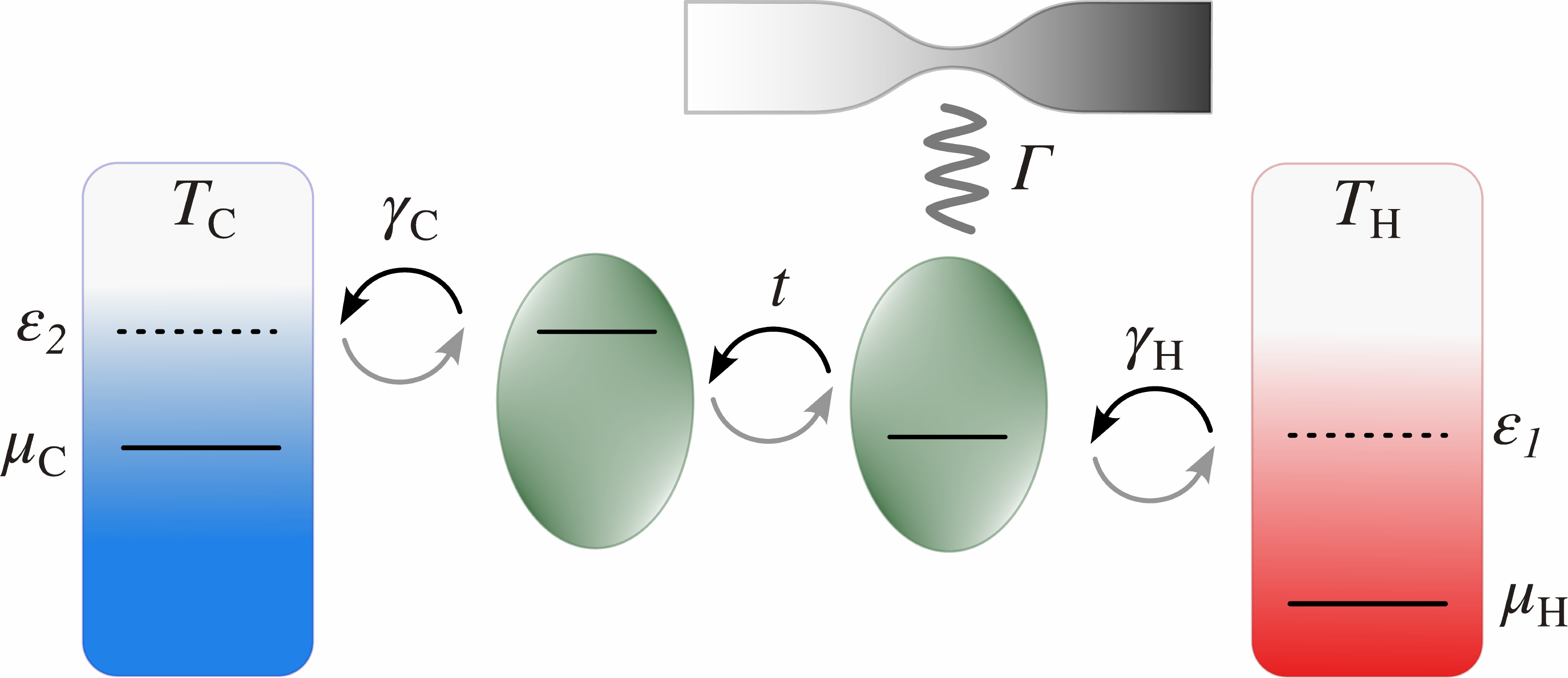}
  \caption{The DQD in the two-terminal set-up is continuously monitored via an interaction between one quantum dot in the DQD and a QPC with measurement strength $\Gamma$. The particle current through the QPC depends on the occupation of the monitored quantum dot.}
 \label{fig:DQD_QPC_full}
\end{center}
\end{figure}

We now illustrate a setup where the occupation of one of the quantum dots in the DQD is continuously monitored by a QPC, see Fig.~(\ref{fig:DQD_QPC_full}). 
The goal of this section is to recover a thermodynamically consistent picture of such a system using the framework of repeated interactions. 
In this way, we consider the QPC as performing boundary-driving on the DQD.

\subsection{Monitored DQD and the GKSL ME}
\label{sec: Hamiltonian DQD+QPC}
The global Hamiltonian of this configuration is given by
\begin{equation}
    \hat{H}=\hat{H}_\mathrm{DQD}+\sum_{\alpha \in \lbrace H,C\rbrace} \hat{H}_\alpha+\hat{H}_\mathrm{QPC}+\hat{H}_\mathrm{DQD,R}+\hat{H}_\mathrm{DQD,QPC},
\end{equation}
where the DQD Hamiltonian $\hat{H}_\mathrm{DQD}$ is defined in Eq.~(\ref{eq: H_DQD}), the Hamiltonian $\hat{H}_\alpha$ of the $\alpha$ reservoir is defined in Eq.~(\ref{eq: H_reservoir_full}), and the tunneling Hamiltonians describing the coupling between the DQD and the thermal reservoirs is defined in Eq.~(\ref{eq: tunneling_reservoirs}). 
Here, the QPC is modeled as a tunnel junction between a left and right reservoir comprised of non-interacting fermions
\begin{align}
\label{eq: H_QPC_full}
\hat{H}_\mathrm{QPC} &= \sum_{\alpha \in \{\rm L,R\},k} \hat{\omega}_{\alpha,k}\hat{a}_k^\dagger \hat{a}_k + \sum_{k, k^\prime} \mathcal{T}_{k,k^\prime} \left(\hat{a}_{\mathrm{R},k}^\dagger \hat{a}_{\mathrm{L},k^\prime} + \hat{a}_{\mathrm{L},k^\prime}^\dagger \hat{a}_{\mathrm{R},k}\right)\,.
\end{align}
In the first term, $\hat{\omega}_{\alpha,k}$ corresponds to the energies for the
left and right reservoir at mode $k$ and $\hat{a}_{\alpha,k}^\dagger$  ($\hat{a}_{\alpha,k}$) are the respective QPC fermionic creation (annihilation) operators.
In the second term $\mathcal{T}_{k,k^\prime}$ is the tunneling matrix element between states $k$ and $k'$ in left and right reservoir.
When the DQD is occupied, then the tunneling rates in the QPC are modulated.
The interaction Hamiltonian between the quantum dot in the DQD and the QPC is given by
\begin{align}
\hat{H}_\mathrm{DQD,QPC} =& \sum_{k, k^\prime} \chi_{k,k^\prime}\hat{c}_1^\dagger \hat{c}_1 \left(\hat{a}_{\mathrm{R},k}^\dagger \hat{a}_{\mathrm{L},k^\prime} + \hat{a}_{\mathrm{L},k^\prime}^\dagger \hat{a}_{\mathrm{R},k}\right).
\end{align}
where $\chi_{k,k'}$ is the change in the tunneling amplitude between states $k$ and $k'$ when the first quantum dot is occupied. 
Now we follow Ref. \cite{goan_continuous_2001} in assuming that the chemical potentials $\mu_\mathrm{L}$ and $\mu_\mathrm{R}$ of the left and right QPC leads, respectively, the resulting bias $eV=\mu_\mathrm{R}-\mu_\mathrm{L}$ and the temperature of the QPC are chosen such that $|eV|,k_\mathrm{B}T_\mathrm{QPC} \ll \mu_\mathrm{L},\mu_\mathrm{R}$. 
Moreover, we assume that both reservoirs in the QPC are at the same, albeit low temperature $T_\mathrm{QPC}$.
Thus, the Fermi functions in the right and left lead of the QPC, $f_\mathrm{R}$ and $f_\mathrm{L}$, respectively, will behave similar to step functions, and the narrow transport window, in which  $f_\mathrm{R} \approx 1$ and  $f_\mathrm{L}\approx 0$, becomes sharp, resulting in unidirectional current in the QPC.
Supposing the wide band limit for both reservoirs and the leads of the QPC and taking the Born-Markov secular approximation \cite{breuer_quantum_2003}, the GKSL ME is then given by \cite{goan_continuous_2001}
\begin{align}
     \dot{\hat{\rho}}_\mathrm{DQD,\Gamma}=\mathcal{L}(\hat{\rho}_\mathrm{DQD,\Gamma})
    + \mathcal{D}\left[ \mathcal{T}_{+} +\chi_{+} \hat{c}_1^\dagger \hat{c}_1\right] \hat{\rho}_\mathrm{DQD,\Gamma} \nonumber \\
    + \mathcal{D}\left[ \mathcal{T}_{-}^{*} +\chi_{-}^{*} \hat{c}_1^\dagger \hat{c}_1\right] \hat{\rho}_\mathrm{DQD,\Gamma}\,,
\end{align}
where $\mathcal{L}(\hat{\rho}_\mathrm{DQD,\Gamma})$ is defined in Eq.~(\ref{eq: me_DQD}), and $\mathcal{T}_{\pm}$ and $\chi_{\pm}$ are assumed to be energy independent tunneling rates along/against the chemical potential gradient and real \cite{goan_continuous_2001}. Note that 
\begin{equation}
    \mathcal{D}\left[ \mathcal{T}_{\pm} +\chi_{\pm} \hat{c}_1^\dagger \hat{c}_1\right] \hat{\rho}_\mathrm{DQD,\Gamma} = \vert \chi_{\pm} \vert^{2} \mathcal{D}\left[ \hat{c}_1^\dagger \hat{c}_1\right] \hat{\rho}_\mathrm{DQD,\Gamma},
\end{equation}
which can be intuitively understood as that the evolution of the DQD density matrix does not depend on the baseline current through the QPC, and allows us to rewrite the master equation as
\begin{align}
\begin{split}
    \dot{\hat{\rho}}_\mathrm{DQD,\Gamma}&=\mathcal{L}(\hat{\rho}_\mathrm{DQD,\Gamma})
    + \Gamma \mathcal{D}\left[\hat{c}_1^\dagger \hat{c}_1\right] \hat{\rho}_\mathrm{DQD,\Gamma} \,,\\
    \label{eq: me_DQD_QPC}
&=\mathcal{L}_\Gamma\left({\hat{\rho}}_\mathrm{DQD,\Gamma}\right)
\end{split}
\end{align}
where $\Gamma = \vert \chi_{+} \vert^{2} + \vert \chi_{-} \vert^{2}$ is the dephasing rate due to the measurement.
We can therefore neglect the contribution due to the QPC dynamics given by $\mathcal{T}$ as it does not change the dynamics of the DQD.
Following \cite{goan_continuous_2001}, we can define an explicit expression for $\Gamma$ in terms of the physical QPC parameters given that
\begin{align}
    \vert \mathcal{T}_{\pm}\vert^{2} &= D_{\pm} = 2\pi  \vert \mathcal{T}_{00} \vert^{2}g_\mathrm{L}g_\mathrm{R} e V_{\pm}\\
    \vert \mathcal{T}_{\pm} + \chi_{\pm}\vert^{2} &= D_{\pm}' = 2\pi \vert\mathcal{T}_{00} + \chi_{00} \vert^{2}g_\mathrm{L}g_\mathrm{R} e V_{\pm}\,,
\end{align}
where $\mathcal{T}_{00}$ and $\chi_{00}$ are the tunneling amplitudes for energies near the chemical potentials, $g_\mathrm{L}$ and $g_\mathrm{R}$ are the energy-independent density of states for the left and right reservoirs, and effective finite temperature external potential bias is 
\begin{equation}
    eV_{\pm} = \frac{\pm eV}{1 - \exp \left(\mp e V/k_{B}T_\mathrm{QPC}\right)}\,.
\end{equation}
Here, the average electron currents through the QPC when the quantum dot is unoccupied and occupied are $eD = e (D_{+}-D_{-})$ and $eD' = e (D_{+}'-D_{-}')$ respectively.
Combining these, and assuming the tunneling rates $\mathcal{T}_{00}$ and $\chi_{00}$ are real, then one can show that the dephasing rate $\Gamma = (\sqrt{D} - \sqrt{D'})^{2}$ is given by the expression
\begin{equation}
    \Gamma = 2\pi g_\mathrm{L}g_\mathrm{R}\chi_{00}^{2} e V \coth \left(\frac{e V}{2 k_{B}T_\mathrm{QPC}}\right)\,.
\end{equation}
There are thus many parameters that can be varied to tune the dephasing rate i.e the temperature $T_\mathrm{QPC}$, $\chi_{00}$, the density of states $g_\mathrm{R(L)}$ and the chemical potential bias $eV$. Note, that even in the limit of zero bias, $\Gamma \neq 0$ due to the fact that the tunneling rate varies depending on the occupation of the quantum dot.

The NESS of the DQD in the presence of the QPC is uniquely defined by
\begin{equation}
\dot{\hat{\rho}}_\mathrm{DQD,\Gamma}^\mathrm{NESS}=\mathcal{L}_\Gamma\left({\hat{\rho}}_\mathrm{DQD,\Gamma}^\mathrm{NESS}\right)=0,
\end{equation}
where the Liouvillian $\mathcal{L}_\Gamma$ is defined in Eq.~(\ref{eq: me_DQD_QPC}).
As studied in section (\ref{sec: thermodynamic inconsistency}), the above local GKSL ME is not guaranteed to be thermodynamically consistent. Therefore, we now develop an approach for the QPC analogous to that in section (\ref{sec: GKSL DQD FRI}) for the thermal reservoirs, which were replaced by a string of identical auxiliary units that interact with the DQD for a finite time and are then replaced.
%
\subsection{GKSL ME in the framework of repeated interactions}
\label{sec: GKSL DQD+QPC FRI}
\begin{figure}
\begin{center}
\includegraphics[width=\linewidth]{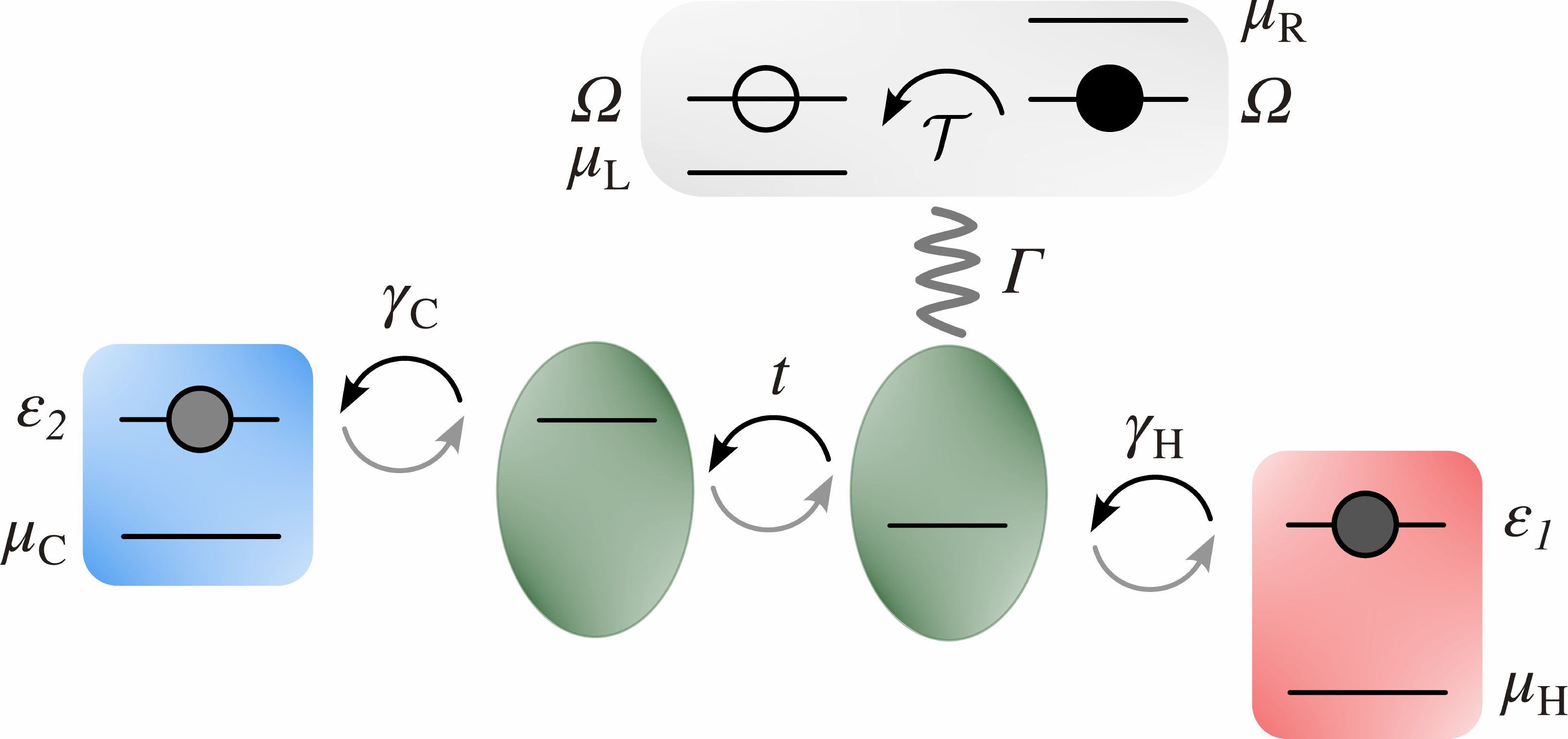}
  \caption{In the framework of repeated interactions, we model the interaction of the DQD with the QPC with periodically refreshed units comprised of two qubits, i.e. one qubit per QPC lead. The energy splitting of the qubits is $(\Omega-\mu_\mathrm{R(L)})$, respectively. They are coupled via a tunneling interaction with amplitude $\mathcal{T}$. The qubits are initialized so that the qubit in the right lead is occupied while the qubit in the left lead is empty, resulting in a unidirectional current from right to left.}
 \label{fig:DQD_QPC_minimal}
\end{center}
\end{figure}

In order to deal with the QPC in the framework of repeated interactions, we recast it as a single unit comprised of two qubits with energy splitting $\Omega-\mu_\mathrm{R(L)}$, respectively, coupled via tunneling with strength $\mathcal{T}$. The bare Hamiltonians of the DQD and reservoir units, Eqs.~(\ref{eq: H_DQD}) and~(\ref{eq: H_unit_reservoirs}), respectively, remain unchanged, while the Hamiltonian of the QPC in Eq.~(\ref{eq: H_QPC_full}) turns into
\begin{align}
    \hat{H}_{\mathrm{QPC}} & = \left(\Omega -\mu_\mathrm{R}\right)\hat{a}_{\mathrm{R}}^\dagger \hat{a}_{\mathrm{R}} + \left(\Omega -\mu_\mathrm{L}\right)\hat{a}_{\mathrm{L}}^\dagger \hat{a}_{\mathrm{L}} 
    + \mathcal{T} \left(\hat{a}_{\mathrm{R}}^\dagger \hat{a}_{\mathrm{L}}+\hat{a}_{\mathrm{L}}^\dagger \hat{a}_{\mathrm{R}} \right),
\end{align}
where we choose $\mu_\mathrm{R}\geq\mu_\mathrm{L}$.
It is assumed that at the beginning of each interaction  the global density matrix is a product state of the individual states of all four unities: DQD, hot and cold reservoir units and the QPC unit. The latter is also assumed to be factorized in terms of two thermal density matrices associated with each L,R qubit (see Fig.~(\ref{fig:DQD_QPC_minimal})), $\hat{\omega}_{\beta_{L}}$ and $\hat{\omega}_{\beta_{R}}$, respectively. 
Thus, at the start of each interaction
\begin{equation}
    \hat{\rho} = \hat{\rho}_\mathrm{DQD,\Gamma} \otimes \underbrace{\hat{\omega}_{\beta_\mathrm{H}} \otimes \hat{\omega}_{\beta_\mathrm{C}}\otimes \hat{\omega}_{\beta_{R}} \otimes \hat{\omega}_{\beta_{L}}}_{\hat{\rho}_\mathrm{E}}.
\end{equation}
where $\hat{\omega}_{\beta_\mathrm{H}}$ and $\hat{\omega}_{\beta_\mathrm{C}}$ are defined in Eq.~(\ref{eq: dens_mat_H}) and (\ref{eq: dens_mat_C}), respectively.
This configuration is representative of the QPC in the limit for which the ME was derived originally, i.e. $|eV|,k_\mathrm{B}T_\mathrm{QPC} \ll \mu_\mathrm{L},\mu_\mathrm{R}$, such that the energy window for transmission is narrow and the current through the QPC is unidirectional. In this limit, and assuming that the temperature of the QPC is low, the Fermi functions of the right and left lead are $f_\mathrm{R}(\Omega)=1$ and $f_\mathrm{L}(\Omega)=0$, respectively, so that we define
\begin{align}
    \hat{\omega}_{\beta_{\mathrm{R}}} &= \left(\begin{array}{cc}
         1 &  0\\
         0 & 0
    \end{array}\right),\\
    \hat{\omega}_{\beta_{\mathrm{L}}} &= \left(\begin{array}{cc}
         0 &  0\\
         0 & 1
    \end{array}\right).
\end{align}
Since we assume that the QPC unit couples only to the first quantum dot with coupling strength $\sqrt{\Gamma}$, we define the (single) interaction between the monitored dot and the QPC via
\begin{align}
\label{eq: int_QPC}
    \hat{v}_{\mathrm{QPC}} &= \sqrt{\Gamma} \hat{c}_1^\dagger \hat{c}_1 \left(\hat{a}_{\mathrm{R}}^\dagger \hat{a}_{\mathrm{L}}+\hat{a}_{\mathrm{L}}^\dagger \hat{a}_{\mathrm{R}}  \right), \\
\end{align}
and the interactions with the hot and cold reservoir, $\hat{v}_\mathrm{H}$ and $\hat{v}_\mathrm{C}$, as defined in Eqs.~(\ref{eq: single_ints_DQD_H}) and  (\ref{eq: single_ints_DQD_C}), respectively.

We are now in a position to study the repeated interactions of the DQD by following the recipe outlined in Sec.~(\ref{sec: GKSL DQD FRI}) but including the QPC.
The formalism developed in Ref. \cite{barra_thermodynamic_2015} assumes that the density matrices of the units are thermal. While the density matrices describing the left and right side of the QPC, $\hat{\omega}_{\beta_{\mathrm{R(L)}}}$, are thermal with respect to their respective bare Hamiltonian $\hat{H}_\mathrm{QPC,R(L)}=\left(\Omega -\mu_{\mathrm{R(L)}}\right)\hat{a}_\mathrm{R(L)}^\dagger\hat{a}_\mathrm{R(L)}$, the total QPC density matrix $\hat{\rho}_\mathrm{QPC}=\hat{\omega}_{\beta_{\mathrm{R}}}\otimes \hat{\omega}_{\beta_{\mathrm{L}}}$ is not thermal with respect to $\hat{H}_\mathrm{QPC}$. This can be easily seen as $\left[\hat{H}_\mathrm{QPC},\hat{\rho}_\mathrm{QPC}\right]\neq 0$, because of the hopping interaction of the two thermal qubits. Nevertheless, Eqs. (\ref{eq: ME_RI},\ref{eq: heat_curr}, \ref{eq: work_curr}) can be analogously derived also for the interaction between the QPC unit proposed here and the DQD.
One can thus easily obtain the GKSL ME for the DQD via
\begin{equation}
    \dot{\hat{\rho}}_\mathrm{DQD,\Gamma}=-i\left[\hat{H}_\mathrm{DQD},\hat{\rho}_\mathrm{DQD,\Gamma}\right]+\sum_{\alpha \in \lbrace\mathrm{H}, \mathrm{C}, \mathrm{QPC} \rbrace} \mathcal{D}_\alpha\left(\hat{\rho}_\mathrm{DQD,\Gamma}\right),
\end{equation}
where $\mathcal{D}_\alpha\left(\hat{\rho}_\mathrm{DQD,\Gamma}\right)$ is defined in Eq.~(\ref{eq: Dalpha}).
In fact, one can readily show that this recovers Eq.~(\ref{eq: me_DQD_QPC}).

\subsection{Thermodynamics of the boundary-driven monitored DQD in the framework of repeated interactions}

\begin{figure}
\begin{center}
\includegraphics[width=\linewidth]{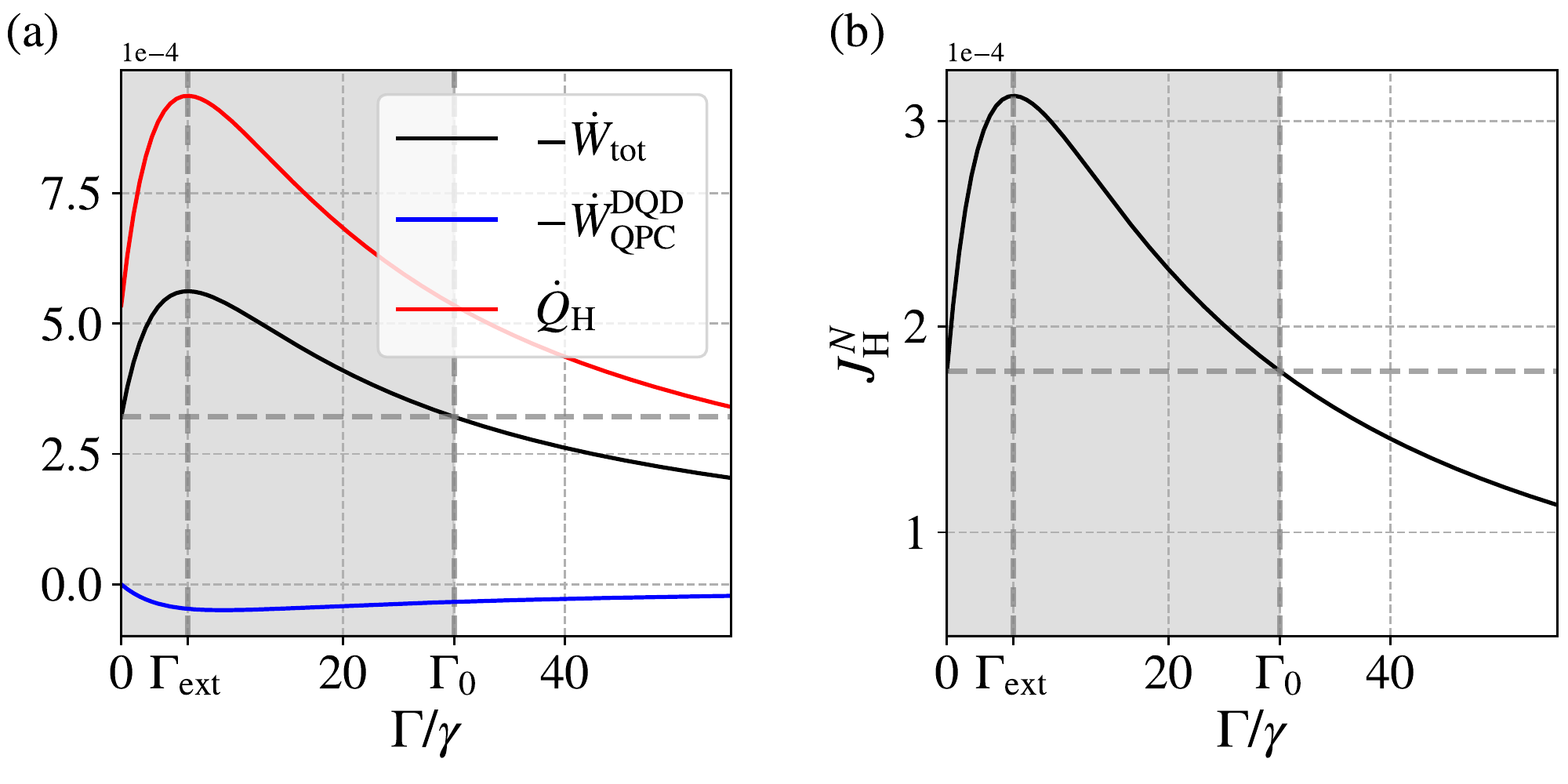}
  \caption{Work, heat and particle currents as a function of the measurement strength. (a) In the NESS, the total work current out of the DQD $-\dot{W}_\mathrm{tot}$ (black solid line) exhibits an extremum at the measurement strength $\Gamma_\mathrm{ext}$ and is larger than its value in the absence of measurement up a measurement strength of $\Gamma_0$, as indicated by the dashed grey lines and the grey box. Here, also the work current between DQD and QPC $-\dot{W}^\mathrm{DQD}_\mathrm{QPC}$ (blue solid line) and the heat current between the hot reservoir and the DQD $\dot{Q}_\mathrm{H}$ (red solid line) are shown.
  (b) The average particle current attains a maximum at the measurement strength $\Gamma_\mathrm{ext}$ and exceeds its magnitude in the absence of measurement up a measurement strength of $\Gamma_0$, as indicated by the dashed grey lines and the grey box.
  Parameters: reservoir temperatures $T_\mathrm{H}$, $T_\mathrm{C}$ = 3, 1 $T=(T_\mathrm{H}+T_\mathrm{C})/2$; reservoir chemical potentials $\mu_H$, $\mu_C$ = 0.5$k_\mathrm{B}T$, 1.5$k_\mathrm{B}T$; quantum dot energies $\epsilon_1$, $\epsilon_2$ = 2$k_\mathrm{B}T$, 2.1$k_\mathrm{B}T$; interdot hopping amplitude $t$ = 0.025$k_\mathrm{B}T$; coupling to reservoirs $\gamma_\mathrm{C}, \gamma_\mathrm{H}$ = $\gamma$ = 0.025$k_\mathrm{B}T$.}
 \label{fig:thermo_fig}
\end{center}
\end{figure}

\label{sec: thermo DQD+QPC}
We now study the influence of continuous measurement via the QPC on the energy currents into the DQD as well as the entropy production rate. 
The first and second law of thermodynamics can be validated, and one can analyse the efficiency of the continuously monitored DQD operated as a heat engine.

Firstly, we confirm that this model is thermodynamically consistent with the first law.
Using the formalism described in section ~(\ref{sec: thermo DQD}) we find the heat and work currents
---Eqs.~(\ref{eq: heat_curr_gamma}) and (\ref{eq: work_curr_gamma}) respectively---
are given by
\begin{align}
\label{eq: heat_curr_gamma}
\begin{split}
   \dot{Q}_\mathrm{QPC} =& - eV \Gamma \langle  \hat{c}_1^\dagger \hat{c}_1\rangle, \\
    \dot{Q}_\mathrm{H} =& \left(\epsilon_1-\mu_H\right)J^N_\mathrm{H,\Gamma}, \\
    \dot{Q}_\mathrm{C} =& -\left(\epsilon_2-\mu_C\right)J^N_\mathrm{H,\Gamma}, \\
\end{split} 
\end{align}
and 
\begin{align}
\label{eq: work_curr_gamma}
\begin{split}
    \dot{W}_\mathrm{QPC} =&\frac{\Gamma\Delta \epsilon }{\Gamma + \tilde{\gamma}} J^N_\mathrm{H,\Gamma}+ eV \Gamma \langle \hat{c}_1^\dagger \hat{c}_1\rangle, \\
    \dot{W}_\mathrm{H} =& \left(\frac{\gamma_\mathrm{H}\Delta \epsilon }{\Gamma \tilde{\gamma}}+\mu_H\right) J^N_\mathrm{H,\Gamma}, \\
    \dot{W}_\mathrm{C} =& \left(\frac{\gamma_\mathrm{C}\Delta \epsilon }{\Gamma + \gamma_\mathrm{C} + \gamma_\mathrm{H}}-\mu_C\right) J^N_\mathrm{H,\Gamma}, \\
\end{split}
\end{align}
where $eV=\mu_\mathrm{R} - \mu_\mathrm{L}$ and the particle current is given by
\begin{align}
\label{eq:jNgamma}
    J^{N}_\mathrm{H,\Gamma} &= \gamma_\mathrm{H}\left(f_\mathrm{H}(\epsilon_1)-\langle \hat{c}_1^\dagger \hat{c}_1\rangle\right),\\
    &=\frac{4 \Delta f\gamma_\mathrm{C} \gamma_\mathrm{H} \left(\Gamma + \tilde{\gamma}\right) |t|^2}{
 \tilde{\gamma} \left(4 \Delta \epsilon ^2 + \left(\Gamma + \tilde{\gamma}\right)^2\right) + 
  4 \tilde{\gamma} \left(\Gamma + \tilde{\gamma}\right) |t|^2}.
\end{align}
Here, we have used the $\Gamma$ subscript to indicate the presence of the QPC.
We can readily show that our results are consistent with the first law of thermodynamics for the NESS since
\begin{align}
\begin{split}
     \langle\dot{\hat{H}}_\mathrm{DQD}\rangle_\mathrm{tot}&= \sum_{\alpha = \mathrm{QPC,H,C}} \dot{W}_\mathrm{\alpha}+ \dot{Q}_\mathrm{\alpha}=0\,.
\end{split}
\end{align}
The QPC's work and heat current can be divided into two contributions: one associated with the particle current through the DQD, $J^N_\mathrm{H,\Gamma}$, and the other associated with the change in particle current through the QPC as a function of the monitored dot occupation, $\Delta J^N_\mathrm{QPC}=\Gamma \langle \hat{c}_1^\dagger \hat{c}_1\rangle$, so that
\begin{align}
    \dot{Q}_\mathrm{QPC} =& - eV \Delta J^N_\mathrm{QPC},\\
    \dot{W}_\mathrm{QPC} =& \frac{\Gamma\Delta \epsilon }{\Gamma + \tilde{\gamma}} J^N_\mathrm{H,\Gamma} + eV \Delta J^N_\mathrm{QPC}\\
    =&\dot{W}^\mathrm{DQD}_\mathrm{QPC}+\dot{W}^\mathrm{QPC}_\mathrm{QPC}.
\end{align}
Note that the energy current associated with $\Delta J^N_\mathrm{QPC}$ follows Watt's law.

Here we are interested in the influence of continuous measurement on the thermodynamics of the DQD operated as a heat engine, in which the particle current through the DQD, $J^N_\mathrm{H,\Gamma}$, is the carrier of energy. Thus, we argue that the heat and work currents due to the change in the QPC particle current, $\Delta J^N_\mathrm{QPC}$, that, importantly, is contained within the QPC unit, must not be taken into account for the efficiency of the DQD heat engine and the entropic cost of driving its particle current. 

We now study the heat and work current related to the particle current throught the DQD as as a function of the measurement strength $\Gamma$, which quantifies the strength of the dephasing backaction.
To this aim, we define
\begin{equation}
    \dot{W}_\mathrm{tot} =\dot{W}_\mathrm{H}+\dot{W}_\mathrm{C}+\dot{W}^\mathrm{DQD}_\mathrm{QPC}.
\end{equation}
We find that all thermodynamic currents are either reduced or assisted by the dephasing due to the measurement-induced backaction, as depicted in Fig.~(\ref{fig:thermo_fig}).
Moreover, we notice that the total output power $-\dot{W}_\mathrm{tot}$ is extremal at $\Gamma_{\rm ext}$, which we will revisit shortly. 
It suffices to say that because we are operating the DQD heat engine in the tight-coupling limit, the non-trivial dependence on $\Gamma$ can be traced back to the the particle current Eq.~(\ref{eq:jNgamma}). 

This raises the question whether there is any advantage in the efficiency of this heat engine due to $\Gamma$. 
As argued above, one can define the efficiency of the engine as in Eq.~(\ref{eq: efficiency_no_QPC}) taking into account energy currents associated with the particle current through the DQD.
In the limit of tight coupling, the efficiency is independent of the dephasing strength---but this may not hold if the weak-coupling assumption is relaxed.

Lastly, the entropy production rate associated with the thermodynamic cost of driving the NESS particle current through the DQD is given by
\begin{align}
\begin{split}
    \sigma_\mathrm{DQD} =& -\left(\beta_H \dot{Q}_\mathrm{H} + \beta_C \dot{Q}_\mathrm{C}\right),\\
    =&  \left(\frac{\epsilon_1-\mu_\mathrm{H}}{k_\mathrm{B}T_\mathrm{H}}-\frac{\epsilon_2-\mu_\mathrm{C}}{k_\mathrm{B}T_\mathrm{C}}\right)J^{N}_{\rm H,\Gamma}\geq  0.
\end{split}
\end{align}
The total entropy production rate $\sigma_\mathrm{tot}=\sigma_\mathrm{DQD}+\sigma_\mathrm{QPC}$ is in agreement with the second law of thermodynamics since 
\begin{align}
\begin{split}
    \sigma_\mathrm{QPC} =& -\beta_\mathrm{QPC}\dot{Q}_\mathrm{QPC},\\
    =&  \beta_\mathrm{QPC} eV J^N_\mathrm{QPC} \geq  0.
\end{split}
\end{align}

%

\subsection{Dephasing enhanced particle transport}
\label{sec: DAT}
\begin{figure}
\begin{center}
\includegraphics[width=\linewidth]{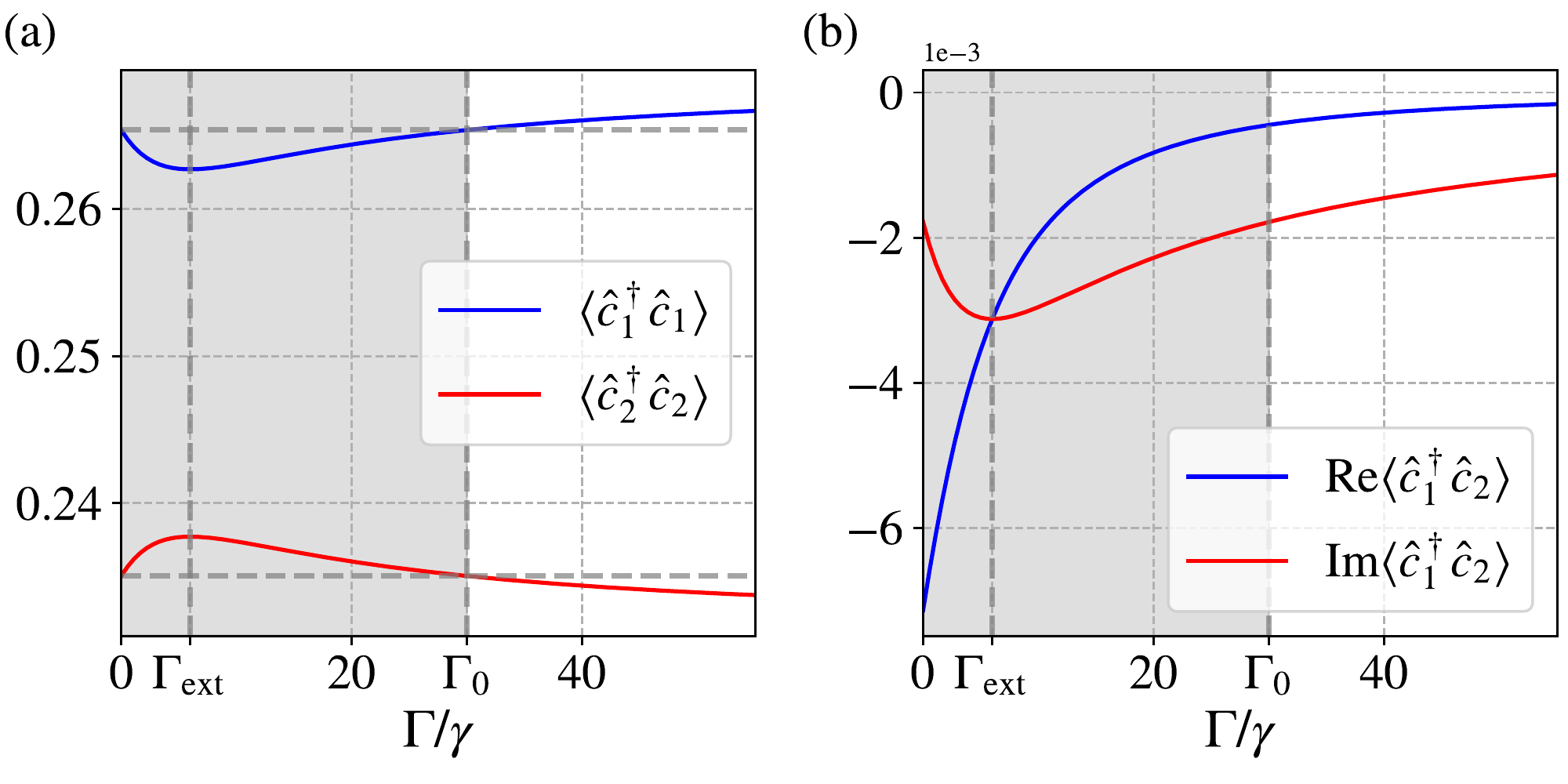}
  \caption{Quantum dot occupations and coherences as a function of the measurement strength. (a) The single quantum dot occupations $n_{1(2)}=\langle \hat{c}_{1(2)}^\dagger \hat{c}_{1(2)}\rangle$ exhibit an extremum at the measurement strength $\Gamma_\mathrm{ext}$ and return to their original value in the absence of measurement at measurement strength $\Gamma_0$, as indicated by the dashed grey lines and the grey box. (b) At $\Gamma_\mathrm{ext}$ the real and imaginary part of the coherence $\langle \hat{c}_1^\dagger \hat{c}_2\rangle$ coincide and $\mathrm{Im}\langle \hat{c}_1^\dagger \hat{c}_2\rangle$, which in the NESS is proportional to the average particle current, exhibits an extremum.
  Parameters: reservoir temperatures $T_\mathrm{H}$, $T_\mathrm{C}$ = 3, 1 $T=(T_\mathrm{H}+T_\mathrm{C})/2$; reservoir chemical potentials $\mu_H$, $\mu_C$ = 0.5$k_\mathrm{B}T$, 1.5$k_\mathrm{B}T$; quantum dot energies $\epsilon_1$, $\epsilon_2$ = 2$k_\mathrm{B}T$, 2.1$k_\mathrm{B}T$; interdot hopping amplitude $t$ = 0.025$k_\mathrm{B}T$; coupling to reservoirs $\gamma_\mathrm{C}, \gamma_\mathrm{H}$ = $\gamma$ = 0.025$k_\mathrm{B}T$.}
 \label{fig:densmat_meas_strength_overview}
\end{center}
\end{figure}
In the previous section we observed a non-trivial behaviour of the particle current $J^{N}_{\rm H,\Gamma}$ as a function of the measurement strength $\Gamma$. 
The measurement introduces dephasing in the DQD which typically suppresses the coherences in the DQD density matrix \cite{levinson_dephasing_1997}. Thus, for large measurement strength we expect the particle current to vanish due to the quantum Zeno effect \cite{Itano_quantum_1990} leading to localization. 
For a significant range in measurement strength that depends on the system's parameters as well as the reservoir couplings we find, however, dephasing enhanced particle transport \cite{contreras-pulido_dephasing-assisted_2014, plenio_dephasing-assisted_2008, rebentrost_environment-assisted_2009, lacerda_dephasing_2021,mendoza-arenas_dephasing_2013, engel_evidence_2007,  collini_coherently_2010}.
This can be understood by studying the particle current in the NESS.
We start from the Heisenberg equations of motion for the single quantum dot occupations, $n_{1(2)}=\langle \hat{c}_{1(2)}^\dagger \hat{c}_{1(2)}\rangle$, and coherence $\langle \hat{c}_{1}^\dagger \hat{c}_{2}\rangle$
\begin{align}
\begin{split}
     \frac{\mathrm{d}\langle \hat{c}_1^\dagger \hat{c}_1\rangle}{\mathrm{d}t}&= \gamma_\mathrm{H}\left(f_\mathrm{H}-\langle \hat{c}_1^\dagger \hat{c}_1\rangle\right)
    +i\left(t \langle \hat{c}_1^\dagger \hat{c}_2\rangle +t^* \langle \hat{c}_2^\dagger \hat{c}_1\rangle \right), \\
    &= J^N_\mathrm{H,\Gamma} + J^N_\mathrm{12}\,,
\end{split}
   \\
    \begin{split}
    \frac{\mathrm{d}\langle \hat{c}_1^\dagger \hat{c}_2\rangle}{\mathrm{d}t}&= \left[ i \Delta \epsilon-\frac{\tilde{\gamma}+\Gamma}{2} \right] \langle \hat{c}_1^\dagger \hat{c}_2\rangle+it^*\left( \langle \hat{c}_2^\dagger \hat{c}_2\rangle+\langle \hat{c}_1^\dagger \hat{c}_1\rangle\right),
    \end{split}
     \\
    \begin{split}
        \frac{\mathrm{d}\langle \hat{c}_2^\dagger \hat{c}_1\rangle}{\mathrm{d}t}&=\left[ -i \Delta \epsilon-\frac{\tilde{\gamma} +\Gamma}{2} \right] \langle \hat{c}_2^\dagger \hat{c}_1\rangle -it\left( \langle \hat{c}_2^\dagger \hat{c}_2\rangle+\langle \hat{c}_1^\dagger \hat{c}_1\rangle\right),
    \end{split}
    \\
    \begin{split}
        \frac{\mathrm{d}\langle \hat{c}_2^\dagger \hat{c}_2\rangle}{\mathrm{d}t}&= \gamma_\mathrm{C}\left(f_\mathrm{C}-\langle \hat{c}_2^\dagger \hat{c}_2\rangle\right) 
    -i\left(t \langle \hat{c}_1^\dagger \hat{c}_2\rangle +t^* \langle \hat{c}_2^\dagger \hat{c}_1\rangle \right)\,,\\
    &= J^N_\mathrm{C,\Gamma} - J^N_\mathrm{12},
    \end{split}
\end{align}
where $J^N_\mathrm{H/C,\Gamma}$ denotes the NESS average particle current flowing into the DQD from the hot/cold side and $J^N_\mathrm{12}$ denotes the NESS average particle current from the first to the second quantum dot in the DQD. In the NESS we set all the time derivatives to zero and clearly see $J^N_\mathrm{H,\Gamma}=-J^N_\mathrm{C,\Gamma}=-J^N_\mathrm{12}$ and therefore in the following we restrict to the discussion of $J^N_\mathrm{H,\Gamma}$.
%
%
The effect of the measurement strength $\Gamma$ on the single quantum dot occupations and the coherence are shown in Fig.~(\ref{fig:densmat_meas_strength_overview}). Note that while the single quantum dot occupations $n_{1(2)}$ depend on $\Gamma$, the total DQD occupation $n=n_1+n_2$ is conserved.  
This can be understood intuitively: the QPC performs work on the DQD as seen in Fig. (\ref{fig:thermo_fig}).
As a result, the populations of eigenstates of the DQD Hamiltonian $H_{\mathrm{DQD}}$ with total occupation $n=1$ are redistributed among the two single quantum dot eigenstates. 
Since the QPC does not exchange particles with the DQD, the interaction conserves the total occupation $n$. 
Solving the above system of equations, we  find that the single quantum dot occupations $n_1$ and $n_2$ are extremal at the measurement strength
\begin{equation}
    \Gamma_{\mathrm{ext}} =  2 |\Delta \epsilon| - \tilde{\gamma}.
\end{equation}
Thus, for any choice of $\Delta \epsilon$ and $\tilde{\gamma}$ for which $\Gamma_{\mathrm{ext}}>0$ the particle current can be enhanced by increasing the measurement strength.
Further, we find that the measurement strength at which the dot occupations match those in the absence
 of measurement, i.e.  $n_{1/2}\left(\Gamma_0\right)=n_{1/2}\left(0\right)$, is given by
\begin{equation}
    \Gamma_0 = \frac{4 \Delta\epsilon^2}{\tilde{\gamma}}-\tilde{\gamma},
\end{equation}
meaning that the particle current is assisted up to $\Gamma_0$.
%
Consequently, the particle current from the hot reservoir into the DQD exhibits a maximum at $\Gamma_\mathrm{ext}$ and is enhanced up to a measurement strength $\Gamma_{0}$, as shown in Fig.~(\ref{fig:thermo_fig}). Hence, a boost in the particle current can obtained via continuous monitoring of the occupation of the first quantum dot.

\subsection{Thermodynamic uncertainty relation}
\label{sec: TUR DQD}
%
\begin{figure}
\begin{center}
\includegraphics[width=\linewidth]{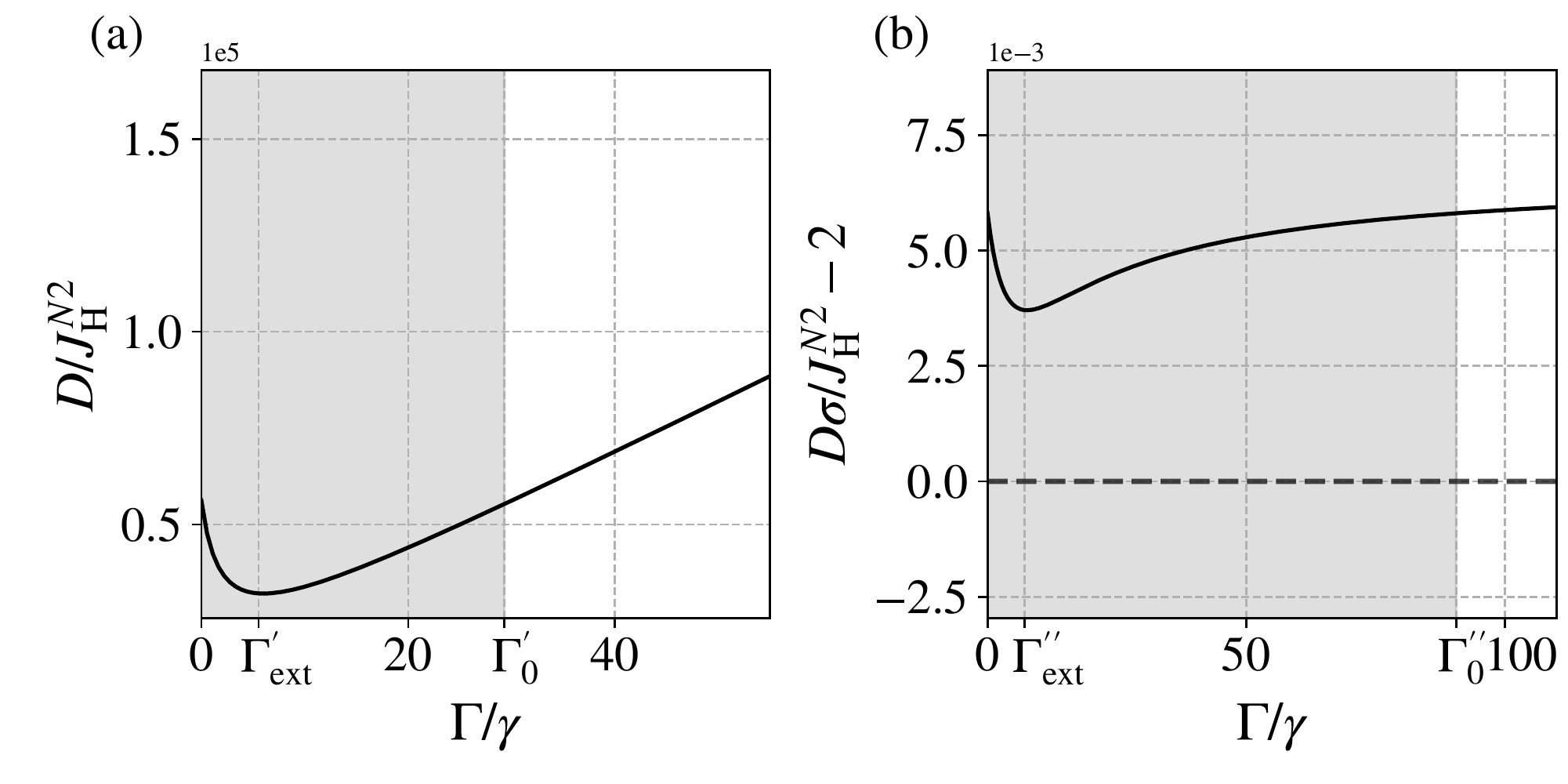}
  \caption{Relative fluctuations and the thermodynamic uncertainty relation ratio (TURR) of the particle current as a function of the measurement strength. (a) Dephasing due to measurement-induced backaction can reduce the relative fluctuations of the particle current up to a measurement stength of $\Gamma^{\prime}_0$. The relative fluctuations are minimal at measurement strength $\Gamma^{\prime}_\mathrm{ext}$.
  (b) Similarly, the TURR expressing the trade-off between the entropic cost and the relative fluctuations of the particle current is minimal at $\Gamma^{\prime\prime}_\mathrm{ext}$ and is reduced up to $\Gamma^{\prime\prime}_0$, which is significantly larger than $\Gamma_0$ and $\Gamma^{\prime}_0$.
  Parameters: reservoir temperatures $T_\mathrm{H}$, $T_\mathrm{C}$ = 3, 1 $T=(T_\mathrm{H}+T_\mathrm{C})/2$; reservoir chemical potentials $\mu_H$, $\mu_C$ = 0.5$k_\mathrm{B}T$, 1.5$k_\mathrm{B}T$; quantum dot energies $\epsilon_1$, $\epsilon_2$ = 2$k_\mathrm{B}T$, 2.1$k_\mathrm{B}T$; interdot hopping amplitude $t$ = 0.025$k_\mathrm{B}T$; coupling to reservoirs $\gamma_\mathrm{C}, \gamma_\mathrm{H}$ = $\gamma$ = 0.025$k_\mathrm{B}T$.}
 \label{fig:current_relfluc2}
\end{center}
\end{figure}
As a final consideration, we study how TURs \cite{barato_thermodynamic_2015, pietzonka_universal_2018, pietzonka_finite-time_2017, pietzonka_universal_2016, horowitz_proof_2017} are impacted by the interplay between the average particle current $J^{N}_{\rm H, \Gamma}$ and the measurement strength $\Gamma$. These relations describe a trade-off between the relative fluctuations of a fluctuating thermodynamic current in classical Markovian NESS heat engines and the entropic cost associated with their driving, and are commonly expressed as
\begin{equation}
\label{eq: TUR}
    \frac{D\sigma}{J^2}\geq 2,
\end{equation}
where $J$ is the average of the fluctuating current, $D$ is its diffusion coefficient and $\sigma$ denotes the entropy production rate. In the following we will refer to $D\sigma/J^2$ as the TUR ratio (TURR). Eq.~(\ref{eq: TUR}) strictly holds only for classical systems, and has been shown to be violated in some quantum systems \cite{kalaee_violating_2021, cangemi_violation_2020, agarwalla_assessing_2018,ptaszynski_coherence-enhanced_2018, agarwalla_assessing_2018, liu_thermodynamic_2019, Prech_2022}. It should be noted, however, that the role of quantum coherence (as is conjectured to be responsible for such violations) can be either constructive or destructive for the fluctuations of thermodynamic currents, depending on the context \cite{rignon-bret_thermodynamics_2021}.


Here, we numerically study the TUR for the particle current $J^{N}_\mathrm{H, \Gamma}$ where the diffusion coefficient $D$ can be calculated using well known methods \cite{kewming_diverging_2022, esposito_nonequilibrium_2009, schaller_open_2014-1,wiseman_quantum_2009} using the power spectrum $S(\omega)$, which often is more accessible experimentally because it describes the amount of power in each frequency component of the measured signal, evaluated at zero frequency $D=S(\omega = 0)$.
It is important to note that we are not computing the power spectrum with respect to the measured signal in the QPC, but rather relating to the counting statistics between the system and the hot reservoir, i.e in the particle current $J^{N}_\mathrm{H, \Gamma}$.
Analytically, it can be expressed in a compact vector notation \cite{kewming_diverging_2022}
\begin{equation}
    D= S(0) = M-2\bra{1}\mathcal{L}_1\mathcal{L}^+\mathcal{L}_1\ket{\hat{\rho}},
\end{equation} 
where $\bra{1}$ is the vectorised identity, $\ket{\hat{\rho}}$ is the vectorised steady-state density matrix and
\begin{equation}
    \mathcal{L}_1 \hat{\rho}= \sum_k \mu_k \hat{L}_k \hat{\rho} \hat{L}_k^\dagger.
\end{equation}
The operator $\hat{L}_k$ is the measurement operator with which the system's density matrix is updated conditioned on a click in the detector with associated measurement outcome $\mu_k$. 
$\mathcal{L}^+$ is the Drazin inverse of the Lindbladian  $\mathcal{L}$ of the GKSL ME and is defined as
\begin{equation}
    \mathcal{L}^+ = \sum_j \frac{1}{\lambda_j}\ket{x_j}\bra{y_j}.
\end{equation}
Here, $\lambda_j$, $\ket{x_j}$ and $\bra{y_j}$ are the non-zero eigenvalues, the right eigenvectors and the left eigenvectors of $\mathcal{L}$ in matrix form, respectively.
Finally, the dynamical activity $M$ is defined as
\begin{equation}
    M = \sum_k \mu_k^2 \mathrm{Tr}\left[\hat{L}_k\hat{\rho} \hat{L}_k^\dagger\right].
\end{equation}
Here, we numerically compute the diffusion coefficient of the fluctuating particle current between the DQD and the hot reservoir. To this end we choose the two measurement operators $\hat{L}_k \in \left( \sqrt{\gamma_\mathrm{H}f_\mathrm{H}(\epsilon_1)} \hat{c}^\dagger_\mathrm{1}, \sqrt{\gamma_\mathrm{H}\left(1-f_\mathrm{H}(\epsilon_1)\right)} \hat{c}_\mathrm{1} \right)$ with their associated measurement outcomes $\mu_k \in  \left(+1,-1\right)$ to represent the detection of single particle hopping into and out of the DQD from the hot reservoir, respectively.
Again, we note this does not correspond to the statistics of the observed current measured by the QPC, but rather the counting statistics between the system and the hot reservoir. 
We find that the relative fluctuations of the particle current, given by $D/{\left(J^{N}_{\mathrm{H,\Gamma}}\right)}^{2}$, can be reduced up to a certain measurement strength $\Gamma^{\prime}_0$, translating to a stabilization of the particle current, as shown in Fig.~(\ref{fig:current_relfluc2}), given that the parameters describing the system and its coupling to the environment are chosen appropriately.
Note, that since we compute $D$ only numerically, we cannot provide analytical expressions for the measurement strength $\Gamma^{\prime}_\mathrm{ext}$ at which the relative fluctuations are minimal and the measurement strength $\Gamma^{\prime}_0$ at which they return to their value in the absence of measurement.

Furthermore, we find that the TURR $D\sigma_\mathrm{DQD}/{J^{N}_\mathrm{H,\Gamma}}^2$ of the particle current can be pushed towards its lower bound for arbitrarily weak measurement strength $\Gamma$, as depicted in Fig. (\ref{fig:current_relfluc2}), in certain parameter regimes. 
Our results suggest that the TURR may be reduced by the dephasing measurement backaction even beyond the regime of dephasing assisted transport, i.e. for $\Gamma>\Gamma_0$.
Further, note that here the TURR changes at constant engine efficiency. This, however, certainly is a pathological effect arising due to tight-coupling between the energy and particle current.

\section{Discussion}
\label{sec: conclusion}
When considering quantum heat engines, fluctuations in thermodynamic currents can be as important as their respective mean values. Increasing precision by reducing fluctuations of a current on small scales is of great interest, also from a fundamental thermodynamics point of view, as the fluctuations of the current are closely related to the entropic cost of driving it, as expressed in TURs. 
Here, we investigated the influence of continuous measurement via a QPC on the relative fluctuations of a thermodynamically driven NESS particle current and its entropic cost in a minimalist model that exhibits quantum coherence: the DQD in a two-terminal setup. 
We show that the local GSKL ME describing the non-unitary evolution of the DQD in the presence of interactions with two thermal reservoirs, and under continuous measurement via a QPC, can be derived in the framework of repeated interactions.
This holds in the limit where the Born-Markov secular approximation is valid and particle transport in the QPC is unidirectional. 
We showed that under this framework the first and second law of thermodynamics are satisfied and that the efficiency of the DQD operating as a heat engine is bounded by the Carnot efficiency.
We argued here that the entropy production rate associated with driving the particle current through the DQD can be computed from the heat currents from the thermal reservoirs and their respective temperatures alone. This is noteworthy because this entropy production rate is also consistent with the second law of thermodynamics, but most importantly is independent of the QPC temperature, which is often assumed to be near zero. In this limit, the entropy production rate in the QPC diverges. The question here is thus how the entropy production in a more realistic model of the measurement device can be adequately taken into account in the thermodynamic description of the continuously monitored system --- and the question remains open.

Finally, using the GKSL ME to compute the NESS particle current, we find dephasing assisted particle transport accompanied by a reduction in relative fluctuations up to a certain measurement strength.
This may mean that the cost of driving a particle current given its signal to noise ratio can be reduced by measurement-induced dephasing. 
In this way, the entropic cost of the precision of the particle flux, which is often more accessible experimentally than any energy flux, may be reduced, while the lower bound set by the TUR remains intact. 
Our model, although minimalist, reproduces the commonly used local GKSL ME for this setup while allowing the calculation of energy fluxes into the DQD in a thermodynamically consistent manner. Thus, it may serve as a starting point for more sophisticated models that will have a higher level of detail in the future, such as realistic spectral densities for the reservoirs, as is possible, for example, within the mesoscopic leads approach \cite{lacerda_quantum_2022, brenes_tensor-network_2020, Brenes_2022} or using more realistic collision based models \cite{purkayastha_periodically_2021, purkayastha_periodically_2022, wojtowicz_accumulative_2022}.

\section{Acknowledgements}
The authors acknowledge Mark Mitchison and Krissia Zawadzki for fruitful discussions and feedback. LPB and JG acknowledge SFI for support through the Frontiers for the Future project. MJK acknowledges the financial support from a Marie Sk\l odwoska-Curie Fellowship (Grant No. 101065974). JG is funded by a Science Foundation Ireland-Royal Society University Research Fellowship. This work was also supported by the European Research Council Starting Grant ODYSSEY (Grant Agreement No. 758403). 
\bibliography{references_1.bib}

\begin{thebibliography}{113}%
\makeatletter
\providecommand \@ifxundefined [1]{%
 \@ifx{#1\undefined}
}%
\providecommand \@ifnum [1]{%
 \ifnum #1\expandafter \@firstoftwo
 \else \expandafter \@secondoftwo
 \fi
}%
\providecommand \@ifx [1]{%
 \ifx #1\expandafter \@firstoftwo
 \else \expandafter \@secondoftwo
 \fi
}%
\providecommand \natexlab [1]{#1}%
\providecommand \enquote  [1]{``#1''}%
\providecommand \bibnamefont  [1]{#1}%
\providecommand \bibfnamefont [1]{#1}%
\providecommand \citenamefont [1]{#1}%
\providecommand \href@noop [0]{\@secondoftwo}%
\providecommand \href [0]{\begingroup \@sanitize@url \@href}%
\providecommand \@href[1]{\@@startlink{#1}\@@href}%
\providecommand \@@href[1]{\endgroup#1\@@endlink}%
\providecommand \@sanitize@url [0]{\catcode `\\12\catcode `\$12\catcode
  `\&12\catcode `\#12\catcode `\^12\catcode `\_12\catcode `\%12\relax}%
\providecommand \@@startlink[1]{}%
\providecommand \@@endlink[0]{}%
\providecommand \url  [0]{\begingroup\@sanitize@url \@url }%
\providecommand \@url [1]{\endgroup\@href {#1}{\urlprefix }}%
\providecommand \urlprefix  [0]{URL }%
\providecommand \Eprint [0]{\href }%
\providecommand \doibase [0]{http://dx.doi.org/}%
\providecommand \selectlanguage [0]{\@gobble}%
\providecommand \bibinfo  [0]{\@secondoftwo}%
\providecommand \bibfield  [0]{\@secondoftwo}%
\providecommand \translation [1]{[#1]}%
\providecommand \BibitemOpen [0]{}%
\providecommand \bibitemStop [0]{}%
\providecommand \bibitemNoStop [0]{.\EOS\space}%
\providecommand \EOS [0]{\spacefactor3000\relax}%
\providecommand \BibitemShut  [1]{\csname bibitem#1\endcsname}%
\let\auto@bib@innerbib\@empty
\bibitem [{\citenamefont {Benenti}\ \emph {et~al.}(2017)\citenamefont
  {Benenti}, \citenamefont {Casati}, \citenamefont {Saito},\ and\ \citenamefont
  {Whitney}}]{benenti_fundamental_2017}%
  \BibitemOpen
  \bibfield  {author} {\bibinfo {author} {\bibfnamefont {G.}~\bibnamefont
  {Benenti}}, \bibinfo {author} {\bibfnamefont {G.}~\bibnamefont {Casati}},
  \bibinfo {author} {\bibfnamefont {K.}~\bibnamefont {Saito}}, \ and\ \bibinfo
  {author} {\bibfnamefont {R.~S.}\ \bibnamefont {Whitney}},\ }\href {\doibase
  10.1016/j.physrep.2017.05.008} {\bibfield  {journal} {\bibinfo  {journal}
  {Physics Reports}\ }\bibinfo {series} {Fundamental aspects of steady-state
  conversion of heat to work at the nanoscale},\ \textbf {\bibinfo {volume}
  {694}},\ \bibinfo {pages} {1} (\bibinfo {year} {2017})}\BibitemShut {NoStop}%
\bibitem [{\citenamefont {Myers}\ \emph {et~al.}(2022)\citenamefont {Myers},
  \citenamefont {Abah},\ and\ \citenamefont {Deffner}}]{myers_quantum_2022}%
  \BibitemOpen
  \bibfield  {author} {\bibinfo {author} {\bibfnamefont {N.~M.}\ \bibnamefont
  {Myers}}, \bibinfo {author} {\bibfnamefont {O.}~\bibnamefont {Abah}}, \ and\
  \bibinfo {author} {\bibfnamefont {S.}~\bibnamefont {Deffner}},\ }\href
  {\doibase 10.1116/5.0083192} {\bibfield  {journal} {\bibinfo  {journal} {AVS
  Quantum Science}\ }\textbf {\bibinfo {volume} {4}},\ \bibinfo {pages}
  {027101} (\bibinfo {year} {2022})}\BibitemShut {NoStop}%
\bibitem [{\citenamefont {Pekola}\ and\ \citenamefont
  {Karimi}(2021)}]{pekola_colloquium_2021}%
  \BibitemOpen
  \bibfield  {author} {\bibinfo {author} {\bibfnamefont {J.~P.}\ \bibnamefont
  {Pekola}}\ and\ \bibinfo {author} {\bibfnamefont {B.}~\bibnamefont
  {Karimi}},\ }\href {\doibase 10.1103/RevModPhys.93.041001} {\bibfield
  {journal} {\bibinfo  {journal} {Reviews of Modern Physics}\ }\textbf
  {\bibinfo {volume} {93}},\ \bibinfo {pages} {041001} (\bibinfo {year}
  {2021})}\BibitemShut {NoStop}%
\bibitem [{\citenamefont {Pekola}(2015)}]{pekola_towards_2015}%
  \BibitemOpen
  \bibfield  {author} {\bibinfo {author} {\bibfnamefont {J.~P.}\ \bibnamefont
  {Pekola}},\ }\href {\doibase 10.1038/nphys3169} {\bibfield  {journal}
  {\bibinfo  {journal} {Nature Physics}\ }\textbf {\bibinfo {volume} {11}},\
  \bibinfo {pages} {118} (\bibinfo {year} {2015})}\BibitemShut {NoStop}%
\bibitem [{\citenamefont {Arrachea}(2022)}]{arrachea_energy_2022}%
  \BibitemOpen
  \bibfield  {author} {\bibinfo {author} {\bibfnamefont {L.}~\bibnamefont
  {Arrachea}},\ }\href {\doibase 10.48550/arXiv.2205.14200} {\enquote {\bibinfo
  {title} {Energy dynamics, heat production and heat-work conversion with
  qubits: towards the development of quantum machines},}\ } (\bibinfo {year}
  {2022})\BibitemShut {NoStop}%
\bibitem [{\citenamefont {Dubi}\ and\ \citenamefont
  {Di Ventra}(2011)}]{dubi_colloquium_2011}%
  \BibitemOpen
  \bibfield  {author} {\bibinfo {author} {\bibfnamefont {Y.}~\bibnamefont
  {Dubi}}\ and\ \bibinfo {author} {\bibfnamefont {M.}~\bibnamefont
  {Di Ventra}},\ }\href {\doibase 10.1103/RevModPhys.83.131} {\bibfield
  {journal} {\bibinfo  {journal} {Reviews of Modern Physics}\ }\textbf
  {\bibinfo {volume} {83}},\ \bibinfo {pages} {131} (\bibinfo {year}
  {2011})}\BibitemShut {NoStop}%
\bibitem [{\citenamefont {Josefsson}\ \emph {et~al.}(2018)\citenamefont
  {Josefsson}, \citenamefont {Svilans}, \citenamefont {Burke}, \citenamefont
  {Hoffmann}, \citenamefont {Fahlvik}, \citenamefont {Thelander}, \citenamefont
  {Leijnse},\ and\ \citenamefont {Linke}}]{josefsson_quantum-dot_2018}%
  \BibitemOpen
  \bibfield  {author} {\bibinfo {author} {\bibfnamefont {M.}~\bibnamefont
  {Josefsson}}, \bibinfo {author} {\bibfnamefont {A.}~\bibnamefont {Svilans}},
  \bibinfo {author} {\bibfnamefont {A.~M.}\ \bibnamefont {Burke}}, \bibinfo
  {author} {\bibfnamefont {E.~A.}\ \bibnamefont {Hoffmann}}, \bibinfo {author}
  {\bibfnamefont {S.}~\bibnamefont {Fahlvik}}, \bibinfo {author} {\bibfnamefont
  {C.}~\bibnamefont {Thelander}}, \bibinfo {author} {\bibfnamefont
  {M.}~\bibnamefont {Leijnse}}, \ and\ \bibinfo {author} {\bibfnamefont
  {H.}~\bibnamefont {Linke}},\ }\href {\doibase 10.1038/s41565-018-0200-5}
  {\bibfield  {journal} {\bibinfo  {journal} {Nature Nanotechnology}\ }\textbf
  {\bibinfo {volume} {13}},\ \bibinfo {pages} {920} (\bibinfo {year}
  {2018})}\BibitemShut {NoStop}%
\bibitem [{\citenamefont {Esposito}\ \emph {et~al.}(2009)\citenamefont
  {Esposito}, \citenamefont {Harbola},\ and\ \citenamefont
  {Mukamel}}]{esposito_nonequilibrium_2009}%
  \BibitemOpen
  \bibfield  {author} {\bibinfo {author} {\bibfnamefont {M.}~\bibnamefont
  {Esposito}}, \bibinfo {author} {\bibfnamefont {U.}~\bibnamefont {Harbola}}, \
  and\ \bibinfo {author} {\bibfnamefont {S.}~\bibnamefont {Mukamel}},\ }\href
  {\doibase 10.1103/RevModPhys.81.1665} {\bibfield  {journal} {\bibinfo
  {journal} {Reviews of Modern Physics}\ }\textbf {\bibinfo {volume} {81}},\
  \bibinfo {pages} {1665} (\bibinfo {year} {2009})}\BibitemShut {NoStop}%
\bibitem [{\citenamefont {Campisi}(2014)}]{campisi_fluctuation_2014}%
  \BibitemOpen
  \bibfield  {author} {\bibinfo {author} {\bibfnamefont {M.}~\bibnamefont
  {Campisi}},\ }\href {\doibase 10.1088/1751-8113/47/24/245001} {\bibfield
  {journal} {\bibinfo  {journal} {Journal of Physics A: Mathematical and
  Theoretical}\ }\textbf {\bibinfo {volume} {47}},\ \bibinfo {pages} {245001}
  (\bibinfo {year} {2014})}\BibitemShut {NoStop}%
\bibitem [{\citenamefont {Levinson}(1997)}]{levinson_dephasing_1997}%
  \BibitemOpen
  \bibfield  {author} {\bibinfo {author} {\bibfnamefont {Y.}~\bibnamefont
  {Levinson}},\ }\href {\doibase 10.1209/epl/i1997-00351-x} {\bibfield
  {journal} {\bibinfo  {journal} {Europhysics Letters (EPL)}\ }\textbf
  {\bibinfo {volume} {39}},\ \bibinfo {pages} {299} (\bibinfo {year}
  {1997})}\BibitemShut {NoStop}%
\bibitem [{\citenamefont {Aleiner}\ \emph {et~al.}(1997)\citenamefont
  {Aleiner}, \citenamefont {Wingreen},\ and\ \citenamefont
  {Meir}}]{aleiner_dephasing_1997}%
  \BibitemOpen
  \bibfield  {author} {\bibinfo {author} {\bibfnamefont {I.~L.}\ \bibnamefont
  {Aleiner}}, \bibinfo {author} {\bibfnamefont {N.~S.}\ \bibnamefont
  {Wingreen}}, \ and\ \bibinfo {author} {\bibfnamefont {Y.}~\bibnamefont
  {Meir}},\ }\href {\doibase 10.1103/PhysRevLett.79.3740} {\bibfield  {journal}
  {\bibinfo  {journal} {Physical Review Letters}\ }\textbf {\bibinfo {volume}
  {79}},\ \bibinfo {pages} {3740} (\bibinfo {year} {1997})}\BibitemShut
  {NoStop}%
\bibitem [{\citenamefont {Stodolsky}(1999)}]{stodolsky_measurement_1999}%
  \BibitemOpen
  \bibfield  {author} {\bibinfo {author} {\bibfnamefont {L.}~\bibnamefont
  {Stodolsky}},\ }\href {\doibase 10.1016/S0370-2693(99)00659-0} {\bibfield
  {journal} {\bibinfo  {journal} {Physics Letters B}\ }\textbf {\bibinfo
  {volume} {459}},\ \bibinfo {pages} {193} (\bibinfo {year}
  {1999})}\BibitemShut {NoStop}%
\bibitem [{\citenamefont {Büttiker}\ and\ \citenamefont
  {Martin}(2000)}]{buttiker_charge_2000}%
  \BibitemOpen
  \bibfield  {author} {\bibinfo {author} {\bibfnamefont {M.}~\bibnamefont
  {Büttiker}}\ and\ \bibinfo {author} {\bibfnamefont {A.~M.}\ \bibnamefont
  {Martin}},\ }\href {\doibase 10.1103/PhysRevB.61.2737} {\bibfield  {journal}
  {\bibinfo  {journal} {Physical Review B}\ }\textbf {\bibinfo {volume} {61}},\
  \bibinfo {pages} {2737} (\bibinfo {year} {2000})}\BibitemShut {NoStop}%
\bibitem [{\citenamefont {Bhandari}\ and\ \citenamefont
  {Jordan}(2022)}]{bhandari_continuous_2022}%
  \BibitemOpen
  \bibfield  {author} {\bibinfo {author} {\bibfnamefont {B.}~\bibnamefont
  {Bhandari}}\ and\ \bibinfo {author} {\bibfnamefont {A.~N.}\ \bibnamefont
  {Jordan}},\ }\href {\doibase 10.1103/PhysRevResearch.4.033103} {\bibfield
  {journal} {\bibinfo  {journal} {Physical Review Research}\ }\textbf {\bibinfo
  {volume} {4}},\ \bibinfo {pages} {033103} (\bibinfo {year}
  {2022})}\BibitemShut {NoStop}%
\bibitem [{\citenamefont {Horowitz}(2012)}]{Horowitz_2012}%
  \BibitemOpen
  \bibfield  {author} {\bibinfo {author} {\bibfnamefont {J.~M.}\ \bibnamefont
  {Horowitz}},\ }\href {\doibase 10.1103/PhysRevE.85.031110} {\bibfield
  {journal} {\bibinfo  {journal} {Phys. Rev. E}\ }\textbf {\bibinfo {volume}
  {85}},\ \bibinfo {pages} {031110} (\bibinfo {year} {2012})}\BibitemShut
  {NoStop}%
\bibitem [{\citenamefont {Manzano}\ \emph {et~al.}(2015)\citenamefont
  {Manzano}, \citenamefont {Horowitz},\ and\ \citenamefont
  {Parrondo}}]{Manzano_Nonequilibrium_2015}%
  \BibitemOpen
  \bibfield  {author} {\bibinfo {author} {\bibfnamefont {G.}~\bibnamefont
  {Manzano}}, \bibinfo {author} {\bibfnamefont {J.~M.}\ \bibnamefont
  {Horowitz}}, \ and\ \bibinfo {author} {\bibfnamefont {J.~M.~R.}\ \bibnamefont
  {Parrondo}},\ }\href {\doibase 10.1103/PhysRevE.92.032129} {\bibfield
  {journal} {\bibinfo  {journal} {Phys. Rev. E}\ }\textbf {\bibinfo {volume}
  {92}},\ \bibinfo {pages} {032129} (\bibinfo {year} {2015})}\BibitemShut
  {NoStop}%
\bibitem [{\citenamefont {Campisi}\ \emph {et~al.}(2015)\citenamefont
  {Campisi}, \citenamefont {Pekola},\ and\ \citenamefont
  {Fazio}}]{Campisi_2015}%
  \BibitemOpen
  \bibfield  {author} {\bibinfo {author} {\bibfnamefont {M.}~\bibnamefont
  {Campisi}}, \bibinfo {author} {\bibfnamefont {J.}~\bibnamefont {Pekola}}, \
  and\ \bibinfo {author} {\bibfnamefont {R.}~\bibnamefont {Fazio}},\ }\href
  {\doibase 10.1088/1367-2630/17/3/035012} {\bibfield  {journal} {\bibinfo
  {journal} {New Journal of Physics}\ }\textbf {\bibinfo {volume} {17}},\
  \bibinfo {pages} {035012} (\bibinfo {year} {2015})}\BibitemShut {NoStop}%
\bibitem [{\citenamefont {Elouard}\ \emph
  {et~al.}(2017{\natexlab{a}})\citenamefont {Elouard}, \citenamefont
  {Herrera-Martí}, \citenamefont {Clusel},\ and\ \citenamefont
  {Auffèves}}]{elouard_role_2017}%
  \BibitemOpen
  \bibfield  {author} {\bibinfo {author} {\bibfnamefont {C.}~\bibnamefont
  {Elouard}}, \bibinfo {author} {\bibfnamefont {D.~A.}\ \bibnamefont
  {Herrera-Martí}}, \bibinfo {author} {\bibfnamefont {M.}~\bibnamefont
  {Clusel}}, \ and\ \bibinfo {author} {\bibfnamefont {A.}~\bibnamefont
  {Auffèves}},\ }\href {\doibase 10.1038/s41534-017-0008-4} {\bibfield
  {journal} {\bibinfo  {journal} {npj Quantum Information}\ }\textbf {\bibinfo
  {volume} {3}},\ \bibinfo {pages} {1} (\bibinfo {year}
  {2017}{\natexlab{a}})}\BibitemShut {NoStop}%
\bibitem [{\citenamefont {Elouard}\ \emph
  {et~al.}(2017{\natexlab{b}})\citenamefont {Elouard}, \citenamefont
  {Bernardes}, \citenamefont {Carvalho}, \citenamefont {Santos},\ and\
  \citenamefont {Auff{\`{e}}ves}}]{Elouard_2017}%
  \BibitemOpen
  \bibfield  {author} {\bibinfo {author} {\bibfnamefont {C.}~\bibnamefont
  {Elouard}}, \bibinfo {author} {\bibfnamefont {N.~K.}\ \bibnamefont
  {Bernardes}}, \bibinfo {author} {\bibfnamefont {A.~R.~R.}\ \bibnamefont
  {Carvalho}}, \bibinfo {author} {\bibfnamefont {M.~F.}\ \bibnamefont
  {Santos}}, \ and\ \bibinfo {author} {\bibfnamefont {A.}~\bibnamefont
  {Auff{\`{e}}ves}},\ }\href {\doibase 10.1088/1367-2630/aa7fa2} {\bibfield
  {journal} {\bibinfo  {journal} {New Journal of Physics}\ }\textbf {\bibinfo
  {volume} {19}},\ \bibinfo {pages} {103011} (\bibinfo {year}
  {2017}{\natexlab{b}})}\BibitemShut {NoStop}%
\bibitem [{\citenamefont {Manzano}\ \emph {et~al.}(2018)\citenamefont
  {Manzano}, \citenamefont {Horowitz},\ and\ \citenamefont
  {Parrondo}}]{Manzano_quantum_2018}%
  \BibitemOpen
  \bibfield  {author} {\bibinfo {author} {\bibfnamefont {G.}~\bibnamefont
  {Manzano}}, \bibinfo {author} {\bibfnamefont {J.~M.}\ \bibnamefont
  {Horowitz}}, \ and\ \bibinfo {author} {\bibfnamefont {J.~M.~R.}\ \bibnamefont
  {Parrondo}},\ }\href {\doibase 10.1103/PhysRevX.8.031037} {\bibfield
  {journal} {\bibinfo  {journal} {Phys. Rev. X}\ }\textbf {\bibinfo {volume}
  {8}},\ \bibinfo {pages} {031037} (\bibinfo {year} {2018})}\BibitemShut
  {NoStop}%
\bibitem [{\citenamefont {Di~Stefano}\ \emph {et~al.}(2018)\citenamefont
  {Di~Stefano}, \citenamefont {Alonso}, \citenamefont {Lutz}, \citenamefont
  {Falci},\ and\ \citenamefont {Paternostro}}]{di_stefano_nonequilibrium_2018}%
  \BibitemOpen
  \bibfield  {author} {\bibinfo {author} {\bibfnamefont {P.~G.}\ \bibnamefont
  {Di~Stefano}}, \bibinfo {author} {\bibfnamefont {J.~J.}\ \bibnamefont
  {Alonso}}, \bibinfo {author} {\bibfnamefont {E.}~\bibnamefont {Lutz}},
  \bibinfo {author} {\bibfnamefont {G.}~\bibnamefont {Falci}}, \ and\ \bibinfo
  {author} {\bibfnamefont {M.}~\bibnamefont {Paternostro}},\ }\href {\doibase
  10.1103/PhysRevB.98.144514} {\bibfield  {journal} {\bibinfo  {journal}
  {Physical Review B}\ }\textbf {\bibinfo {volume} {98}},\ \bibinfo {pages}
  {144514} (\bibinfo {year} {2018})}\BibitemShut {NoStop}%
\bibitem [{\citenamefont {Strasberg}(2019)}]{strasberg_operational_2019}%
  \BibitemOpen
  \bibfield  {author} {\bibinfo {author} {\bibfnamefont {P.}~\bibnamefont
  {Strasberg}},\ }\href {\doibase 10.1103/PhysRevE.100.022127} {\bibfield
  {journal} {\bibinfo  {journal} {Physical Review E}\ }\textbf {\bibinfo
  {volume} {100}},\ \bibinfo {pages} {022127} (\bibinfo {year}
  {2019})}\BibitemShut {NoStop}%
\bibitem [{\citenamefont {Alonso}\ \emph {et~al.}(2016)\citenamefont {Alonso},
  \citenamefont {Lutz},\ and\ \citenamefont
  {Romito}}]{Alonso_thermodynamics_2016}%
  \BibitemOpen
  \bibfield  {author} {\bibinfo {author} {\bibfnamefont {J.~J.}\ \bibnamefont
  {Alonso}}, \bibinfo {author} {\bibfnamefont {E.}~\bibnamefont {Lutz}}, \ and\
  \bibinfo {author} {\bibfnamefont {A.}~\bibnamefont {Romito}},\ }\href
  {\doibase 10.1103/PhysRevLett.116.080403} {\bibfield  {journal} {\bibinfo
  {journal} {Phys. Rev. Lett.}\ }\textbf {\bibinfo {volume} {116}},\ \bibinfo
  {pages} {080403} (\bibinfo {year} {2016})}\BibitemShut {NoStop}%
\bibitem [{\citenamefont {Belenchia}\ \emph {et~al.}(2020)\citenamefont
  {Belenchia}, \citenamefont {Mancino}, \citenamefont {Landi},\ and\
  \citenamefont {Paternostro}}]{belenchia_entropy_2020}%
  \BibitemOpen
  \bibfield  {author} {\bibinfo {author} {\bibfnamefont {A.}~\bibnamefont
  {Belenchia}}, \bibinfo {author} {\bibfnamefont {L.}~\bibnamefont {Mancino}},
  \bibinfo {author} {\bibfnamefont {G.~T.}\ \bibnamefont {Landi}}, \ and\
  \bibinfo {author} {\bibfnamefont {M.}~\bibnamefont {Paternostro}},\ }\href
  {\doibase 10.1038/s41534-020-00334-6} {\bibfield  {journal} {\bibinfo
  {journal} {npj Quantum Information}\ }\textbf {\bibinfo {volume} {6}},\
  \bibinfo {pages} {1} (\bibinfo {year} {2020})}\BibitemShut {NoStop}%
\bibitem [{\citenamefont {Rossi}\ \emph {et~al.}(2020)\citenamefont {Rossi},
  \citenamefont {Mancino}, \citenamefont {Landi}, \citenamefont {Paternostro},
  \citenamefont {Schliesser},\ and\ \citenamefont
  {Belenchia}}]{Rossi_experimental_2020}%
  \BibitemOpen
  \bibfield  {author} {\bibinfo {author} {\bibfnamefont {M.}~\bibnamefont
  {Rossi}}, \bibinfo {author} {\bibfnamefont {L.}~\bibnamefont {Mancino}},
  \bibinfo {author} {\bibfnamefont {G.~T.}\ \bibnamefont {Landi}}, \bibinfo
  {author} {\bibfnamefont {M.}~\bibnamefont {Paternostro}}, \bibinfo {author}
  {\bibfnamefont {A.}~\bibnamefont {Schliesser}}, \ and\ \bibinfo {author}
  {\bibfnamefont {A.}~\bibnamefont {Belenchia}},\ }\href {\doibase
  10.1103/PhysRevLett.125.080601} {\bibfield  {journal} {\bibinfo  {journal}
  {Phys. Rev. Lett.}\ }\textbf {\bibinfo {volume} {125}},\ \bibinfo {pages}
  {080601} (\bibinfo {year} {2020})}\BibitemShut {NoStop}%
\bibitem [{\citenamefont {Menczel}\ \emph {et~al.}(2020)\citenamefont
  {Menczel}, \citenamefont {Flindt},\ and\ \citenamefont
  {Brandner}}]{Menczel_quantum_2020}%
  \BibitemOpen
  \bibfield  {author} {\bibinfo {author} {\bibfnamefont {P.}~\bibnamefont
  {Menczel}}, \bibinfo {author} {\bibfnamefont {C.}~\bibnamefont {Flindt}}, \
  and\ \bibinfo {author} {\bibfnamefont {K.}~\bibnamefont {Brandner}},\ }\href
  {\doibase 10.1103/PhysRevResearch.2.033449} {\bibfield  {journal} {\bibinfo
  {journal} {Phys. Rev. Research}\ }\textbf {\bibinfo {volume} {2}},\ \bibinfo
  {pages} {033449} (\bibinfo {year} {2020})}\BibitemShut {NoStop}%
\bibitem [{\citenamefont {Miller}\ \emph {et~al.}(2020)\citenamefont {Miller},
  \citenamefont {Guarnieri}, \citenamefont {Mitchison},\ and\ \citenamefont
  {Goold}}]{miller_quantum_2020}%
  \BibitemOpen
  \bibfield  {author} {\bibinfo {author} {\bibfnamefont {H.~J.}\ \bibnamefont
  {Miller}}, \bibinfo {author} {\bibfnamefont {G.}~\bibnamefont {Guarnieri}},
  \bibinfo {author} {\bibfnamefont {M.~T.}\ \bibnamefont {Mitchison}}, \ and\
  \bibinfo {author} {\bibfnamefont {J.}~\bibnamefont {Goold}},\ }\href
  {\doibase 10.1103/PhysRevLett.125.160602} {\bibfield  {journal} {\bibinfo
  {journal} {Physical Review Letters}\ }\textbf {\bibinfo {volume} {125}},\
  \bibinfo {pages} {160602} (\bibinfo {year} {2020})}\BibitemShut {NoStop}%
\bibitem [{\citenamefont {Liu}\ and\ \citenamefont
  {Su}(2020)}]{Liu_stochastic_2020}%
  \BibitemOpen
  \bibfield  {author} {\bibinfo {author} {\bibfnamefont {F.}~\bibnamefont
  {Liu}}\ and\ \bibinfo {author} {\bibfnamefont {S.}~\bibnamefont {Su}},\
  }\href {\doibase 10.1103/PhysRevE.101.062144} {\bibfield  {journal} {\bibinfo
   {journal} {Phys. Rev. E}\ }\textbf {\bibinfo {volume} {101}},\ \bibinfo
  {pages} {062144} (\bibinfo {year} {2020})}\BibitemShut {NoStop}%
\bibitem [{\citenamefont {Kewming}\ and\ \citenamefont
  {Shrapnel}(2022)}]{kewming_entropy_2022}%
  \BibitemOpen
  \bibfield  {author} {\bibinfo {author} {\bibfnamefont {M.~J.}\ \bibnamefont
  {Kewming}}\ and\ \bibinfo {author} {\bibfnamefont {S.}~\bibnamefont
  {Shrapnel}},\ }\href {\doibase 10.22331/q-2022-04-13-685} {\bibfield
  {journal} {\bibinfo  {journal} {Quantum}\ }\textbf {\bibinfo {volume} {6}},\
  \bibinfo {pages} {685} (\bibinfo {year} {2022})}\BibitemShut {NoStop}%
\bibitem [{\citenamefont {Manzano}\ and\ \citenamefont
  {Zambrini}(2022)}]{Manzano_2022}%
  \BibitemOpen
  \bibfield  {author} {\bibinfo {author} {\bibfnamefont {G.}~\bibnamefont
  {Manzano}}\ and\ \bibinfo {author} {\bibfnamefont {R.}~\bibnamefont
  {Zambrini}},\ }\href {\doibase 10.1116/5.0079886} {\bibfield  {journal}
  {\bibinfo  {journal} {{AVS} Quantum Science}\ }\textbf {\bibinfo {volume}
  {4}},\ \bibinfo {pages} {025302} (\bibinfo {year} {2022})}\BibitemShut
  {NoStop}%
\bibitem [{\citenamefont {Fedichkin}\ and\ \citenamefont
  {Fedorov}(2004)}]{fedichkin_error_2004-1}%
  \BibitemOpen
  \bibfield  {author} {\bibinfo {author} {\bibfnamefont {L.}~\bibnamefont
  {Fedichkin}}\ and\ \bibinfo {author} {\bibfnamefont {A.}~\bibnamefont
  {Fedorov}},\ }\href {\doibase 10.1103/PhysRevA.69.032311} {\bibfield
  {journal} {\bibinfo  {journal} {Physical Review A}\ }\textbf {\bibinfo
  {volume} {69}},\ \bibinfo {pages} {032311} (\bibinfo {year}
  {2004})}\BibitemShut {NoStop}%
\bibitem [{\citenamefont {Fedichkin}\ and\ \citenamefont
  {Fedorov}(2005)}]{fedichkin_study_2005}%
  \BibitemOpen
  \bibfield  {author} {\bibinfo {author} {\bibfnamefont {L.}~\bibnamefont
  {Fedichkin}}\ and\ \bibinfo {author} {\bibfnamefont {A.}~\bibnamefont
  {Fedorov}},\ }\href {\doibase 10.1109/TNANO.2004.840156} {\bibfield
  {journal} {\bibinfo  {journal} {IEEE Transactions on Nanotechnology}\
  }\textbf {\bibinfo {volume} {4}},\ \bibinfo {pages} {65} (\bibinfo {year}
  {2005})}\BibitemShut {NoStop}%
\bibitem [{\citenamefont {Itakura}\ and\ \citenamefont
  {Tokura}(2003)}]{itakura_dephasing_2003}%
  \BibitemOpen
  \bibfield  {author} {\bibinfo {author} {\bibfnamefont {T.}~\bibnamefont
  {Itakura}}\ and\ \bibinfo {author} {\bibfnamefont {Y.}~\bibnamefont
  {Tokura}},\ }\href {\doibase 10.1103/PhysRevB.67.195320} {\bibfield
  {journal} {\bibinfo  {journal} {Physical Review B}\ }\textbf {\bibinfo
  {volume} {67}},\ \bibinfo {pages} {195320} (\bibinfo {year}
  {2003})}\BibitemShut {NoStop}%
\bibitem [{\citenamefont {Fujisawa}\ \emph {et~al.}(2006)\citenamefont
  {Fujisawa}, \citenamefont {Hayashi}, \citenamefont {Tomita},\ and\
  \citenamefont {Hirayama}}]{fujisawa_bidirectional_2006}%
  \BibitemOpen
  \bibfield  {author} {\bibinfo {author} {\bibfnamefont {T.}~\bibnamefont
  {Fujisawa}}, \bibinfo {author} {\bibfnamefont {T.}~\bibnamefont {Hayashi}},
  \bibinfo {author} {\bibfnamefont {R.}~\bibnamefont {Tomita}}, \ and\ \bibinfo
  {author} {\bibfnamefont {Y.}~\bibnamefont {Hirayama}},\ }\href {\doibase
  10.1126/science.1126788} {\bibfield  {journal} {\bibinfo  {journal}
  {Science}\ }\textbf {\bibinfo {volume} {312}},\ \bibinfo {pages} {1634}
  (\bibinfo {year} {2006})}\BibitemShut {NoStop}%
\bibitem [{\citenamefont {Ihn}\ \emph {et~al.}(2009)\citenamefont {Ihn},
  \citenamefont {Gustavsson}, \citenamefont {Gasser}, \citenamefont {Küng},
  \citenamefont {Müller}, \citenamefont {Schleser}, \citenamefont {Sigrist},
  \citenamefont {Shorubalko}, \citenamefont {Leturcq},\ and\ \citenamefont
  {Ensslin}}]{ihn_quantum_2009}%
  \BibitemOpen
  \bibfield  {author} {\bibinfo {author} {\bibfnamefont {T.}~\bibnamefont
  {Ihn}}, \bibinfo {author} {\bibfnamefont {S.}~\bibnamefont {Gustavsson}},
  \bibinfo {author} {\bibfnamefont {U.}~\bibnamefont {Gasser}}, \bibinfo
  {author} {\bibfnamefont {B.}~\bibnamefont {Küng}}, \bibinfo {author}
  {\bibfnamefont {T.}~\bibnamefont {Müller}}, \bibinfo {author} {\bibfnamefont
  {R.}~\bibnamefont {Schleser}}, \bibinfo {author} {\bibfnamefont
  {M.}~\bibnamefont {Sigrist}}, \bibinfo {author} {\bibfnamefont
  {I.}~\bibnamefont {Shorubalko}}, \bibinfo {author} {\bibfnamefont
  {R.}~\bibnamefont {Leturcq}}, \ and\ \bibinfo {author} {\bibfnamefont
  {K.}~\bibnamefont {Ensslin}},\ }\href {\doibase 10.1016/j.ssc.2009.04.040}
  {\bibfield  {journal} {\bibinfo  {journal} {Solid State Communications}\
  }\bibinfo {series} {Fundamental {Phenomena} and {Applications} of {Quantum}
  {Dots}},\ \textbf {\bibinfo {volume} {149}},\ \bibinfo {pages} {1419}
  (\bibinfo {year} {2009})}\BibitemShut {NoStop}%
\bibitem [{\citenamefont {Gurvitz}(1997)}]{gurvitz_measurements_1997}%
  \BibitemOpen
  \bibfield  {author} {\bibinfo {author} {\bibfnamefont {S.~A.}\ \bibnamefont
  {Gurvitz}},\ }\href {\doibase 10.1103/PhysRevB.56.15215} {\bibfield
  {journal} {\bibinfo  {journal} {Physical Review B}\ }\textbf {\bibinfo
  {volume} {56}},\ \bibinfo {pages} {15215} (\bibinfo {year}
  {1997})}\BibitemShut {NoStop}%
\bibitem [{\citenamefont {Contreras-Pulido}\ \emph {et~al.}(2014)\citenamefont
  {Contreras-Pulido}, \citenamefont {Bruderer}, \citenamefont {Huelga},\ and\
  \citenamefont {Plenio}}]{contreras-pulido_dephasing-assisted_2014}%
  \BibitemOpen
  \bibfield  {author} {\bibinfo {author} {\bibfnamefont {L.~D.}\ \bibnamefont
  {Contreras-Pulido}}, \bibinfo {author} {\bibfnamefont {M.}~\bibnamefont
  {Bruderer}}, \bibinfo {author} {\bibfnamefont {S.~F.}\ \bibnamefont
  {Huelga}}, \ and\ \bibinfo {author} {\bibfnamefont {M.~B.}\ \bibnamefont
  {Plenio}},\ }\href {\doibase 10.1088/1367-2630/16/11/113061} {\bibfield
  {journal} {\bibinfo  {journal} {New Journal of Physics}\ }\textbf {\bibinfo
  {volume} {16}},\ \bibinfo {pages} {113061} (\bibinfo {year}
  {2014})}\BibitemShut {NoStop}%
\bibitem [{\citenamefont {Plenio}\ and\ \citenamefont
  {Huelga}(2008)}]{plenio_dephasing-assisted_2008}%
  \BibitemOpen
  \bibfield  {author} {\bibinfo {author} {\bibfnamefont {M.~B.}\ \bibnamefont
  {Plenio}}\ and\ \bibinfo {author} {\bibfnamefont {S.~F.}\ \bibnamefont
  {Huelga}},\ }\href {\doibase 10.1088/1367-2630/10/11/113019} {\bibfield
  {journal} {\bibinfo  {journal} {New Journal of Physics}\ }\textbf {\bibinfo
  {volume} {10}},\ \bibinfo {pages} {113019} (\bibinfo {year}
  {2008})}\BibitemShut {NoStop}%
\bibitem [{\citenamefont {Rebentrost}\ \emph
  {et~al.}(2009{\natexlab{a}})\citenamefont {Rebentrost}, \citenamefont
  {Mohseni}, \citenamefont {Kassal}, \citenamefont {Lloyd},\ and\ \citenamefont
  {Aspuru-Guzik}}]{rebentrost_environment-assisted_2009}%
  \BibitemOpen
  \bibfield  {author} {\bibinfo {author} {\bibfnamefont {P.}~\bibnamefont
  {Rebentrost}}, \bibinfo {author} {\bibfnamefont {M.}~\bibnamefont {Mohseni}},
  \bibinfo {author} {\bibfnamefont {I.}~\bibnamefont {Kassal}}, \bibinfo
  {author} {\bibfnamefont {S.}~\bibnamefont {Lloyd}}, \ and\ \bibinfo {author}
  {\bibfnamefont {A.}~\bibnamefont {Aspuru-Guzik}},\ }\href {\doibase
  10.1088/1367-2630/11/3/033003} {\bibfield  {journal} {\bibinfo  {journal}
  {New Journal of Physics}\ }\textbf {\bibinfo {volume} {11}},\ \bibinfo
  {pages} {033003} (\bibinfo {year} {2009}{\natexlab{a}})}\BibitemShut
  {NoStop}%
\bibitem [{\citenamefont {Scholes}\ and\ \citenamefont
  {Smyth}(2014)}]{scholes_perspective_2014}%
  \BibitemOpen
  \bibfield  {author} {\bibinfo {author} {\bibfnamefont {G.~D.}\ \bibnamefont
  {Scholes}}\ and\ \bibinfo {author} {\bibfnamefont {C.}~\bibnamefont
  {Smyth}},\ }\href {\doibase 10.1063/1.4869329} {\bibfield  {journal}
  {\bibinfo  {journal} {The Journal of Chemical Physics}\ }\textbf {\bibinfo
  {volume} {140}},\ \bibinfo {pages} {110901} (\bibinfo {year}
  {2014})}\BibitemShut {NoStop}%
\bibitem [{\citenamefont {Zerah-Harush}\ and\ \citenamefont
  {Dubi}(2020)}]{zerah-harush_effects_2020}%
  \BibitemOpen
  \bibfield  {author} {\bibinfo {author} {\bibfnamefont {E.}~\bibnamefont
  {Zerah-Harush}}\ and\ \bibinfo {author} {\bibfnamefont {Y.}~\bibnamefont
  {Dubi}},\ }\href {\doibase 10.1103/PhysRevResearch.2.023294} {\bibfield
  {journal} {\bibinfo  {journal} {Physical Review Research}\ }\textbf {\bibinfo
  {volume} {2}},\ \bibinfo {pages} {023294} (\bibinfo {year}
  {2020})}\BibitemShut {NoStop}%
\bibitem [{\citenamefont {Kilgour}\ and\ \citenamefont
  {Segal}(2015)}]{kilgour_charge_2015}%
  \BibitemOpen
  \bibfield  {author} {\bibinfo {author} {\bibfnamefont {M.}~\bibnamefont
  {Kilgour}}\ and\ \bibinfo {author} {\bibfnamefont {D.}~\bibnamefont
  {Segal}},\ }\href {\doibase 10.1063/1.4926395} {\bibfield  {journal}
  {\bibinfo  {journal} {The Journal of Chemical Physics}\ }\textbf {\bibinfo
  {volume} {143}},\ \bibinfo {pages} {024111} (\bibinfo {year}
  {2015})}\BibitemShut {NoStop}%
\bibitem [{\citenamefont {Kilgour}\ and\ \citenamefont
  {Segal}(2016)}]{kilgour_inelastic_2016}%
  \BibitemOpen
  \bibfield  {author} {\bibinfo {author} {\bibfnamefont {M.}~\bibnamefont
  {Kilgour}}\ and\ \bibinfo {author} {\bibfnamefont {D.}~\bibnamefont
  {Segal}},\ }\href {\doibase 10.1063/1.4944470} {\bibfield  {journal}
  {\bibinfo  {journal} {The Journal of Chemical Physics}\ }\textbf {\bibinfo
  {volume} {144}},\ \bibinfo {pages} {124107} (\bibinfo {year}
  {2016})}\BibitemShut {NoStop}%
\bibitem [{\citenamefont {Sowa}\ \emph {et~al.}(2017)\citenamefont {Sowa},
  \citenamefont {Mol}, \citenamefont {Briggs},\ and\ \citenamefont
  {Gauger}}]{sowa_environment-assisted_2017}%
  \BibitemOpen
  \bibfield  {author} {\bibinfo {author} {\bibfnamefont {J.~K.}\ \bibnamefont
  {Sowa}}, \bibinfo {author} {\bibfnamefont {J.~A.}\ \bibnamefont {Mol}},
  \bibinfo {author} {\bibfnamefont {G.~A.~D.}\ \bibnamefont {Briggs}}, \ and\
  \bibinfo {author} {\bibfnamefont {E.~M.}\ \bibnamefont {Gauger}},\ }\href
  {\doibase 10.1039/C7CP06237K} {\bibfield  {journal} {\bibinfo  {journal}
  {Physical Chemistry Chemical Physics}\ }\textbf {\bibinfo {volume} {19}},\
  \bibinfo {pages} {29534} (\bibinfo {year} {2017})}\BibitemShut {NoStop}%
\bibitem [{\citenamefont {Lacerda}\ \emph {et~al.}(2021)\citenamefont
  {Lacerda}, \citenamefont {Goold},\ and\ \citenamefont
  {Landi}}]{lacerda_dephasing_2021}%
  \BibitemOpen
  \bibfield  {author} {\bibinfo {author} {\bibfnamefont {A.~M.}\ \bibnamefont
  {Lacerda}}, \bibinfo {author} {\bibfnamefont {J.}~\bibnamefont {Goold}}, \
  and\ \bibinfo {author} {\bibfnamefont {G.~T.}\ \bibnamefont {Landi}},\ }\href
  {\doibase 10.1103/PhysRevB.104.174203} {\bibfield  {journal} {\bibinfo
  {journal} {Physical Review B}\ }\textbf {\bibinfo {volume} {104}},\ \bibinfo
  {pages} {174203} (\bibinfo {year} {2021})}\BibitemShut {NoStop}%
\bibitem [{\citenamefont {Mendoza-Arenas}\ \emph {et~al.}(2013)\citenamefont
  {Mendoza-Arenas}, \citenamefont {Grujic}, \citenamefont {Jaksch},\ and\
  \citenamefont {Clark}}]{mendoza-arenas_dephasing_2013}%
  \BibitemOpen
  \bibfield  {author} {\bibinfo {author} {\bibfnamefont {J.~J.}\ \bibnamefont
  {Mendoza-Arenas}}, \bibinfo {author} {\bibfnamefont {T.}~\bibnamefont
  {Grujic}}, \bibinfo {author} {\bibfnamefont {D.}~\bibnamefont {Jaksch}}, \
  and\ \bibinfo {author} {\bibfnamefont {S.~R.}\ \bibnamefont {Clark}},\ }\href
  {\doibase 10.1103/PhysRevB.87.235130} {\bibfield  {journal} {\bibinfo
  {journal} {Physical Review B}\ }\textbf {\bibinfo {volume} {87}},\ \bibinfo
  {pages} {235130} (\bibinfo {year} {2013})}\BibitemShut {NoStop}%
\bibitem [{\citenamefont {Engel}\ \emph {et~al.}(2007)\citenamefont {Engel},
  \citenamefont {Calhoun}, \citenamefont {Read}, \citenamefont {Ahn},
  \citenamefont {Mančal}, \citenamefont {Cheng}, \citenamefont {Blankenship},\
  and\ \citenamefont {Fleming}}]{engel_evidence_2007}%
  \BibitemOpen
  \bibfield  {author} {\bibinfo {author} {\bibfnamefont {G.~S.}\ \bibnamefont
  {Engel}}, \bibinfo {author} {\bibfnamefont {T.~R.}\ \bibnamefont {Calhoun}},
  \bibinfo {author} {\bibfnamefont {E.~L.}\ \bibnamefont {Read}}, \bibinfo
  {author} {\bibfnamefont {T.-K.}\ \bibnamefont {Ahn}}, \bibinfo {author}
  {\bibfnamefont {T.}~\bibnamefont {Mančal}}, \bibinfo {author} {\bibfnamefont
  {Y.-C.}\ \bibnamefont {Cheng}}, \bibinfo {author} {\bibfnamefont {R.~E.}\
  \bibnamefont {Blankenship}}, \ and\ \bibinfo {author} {\bibfnamefont {G.~R.}\
  \bibnamefont {Fleming}},\ }\href {\doibase 10.1038/nature05678} {\bibfield
  {journal} {\bibinfo  {journal} {Nature}\ }\textbf {\bibinfo {volume} {446}},\
  \bibinfo {pages} {782} (\bibinfo {year} {2007})}\BibitemShut {NoStop}%
\bibitem [{\citenamefont {Collini}\ \emph {et~al.}(2010)\citenamefont
  {Collini}, \citenamefont {Wong}, \citenamefont {Wilk}, \citenamefont {Curmi},
  \citenamefont {Brumer},\ and\ \citenamefont
  {Scholes}}]{collini_coherently_2010}%
  \BibitemOpen
  \bibfield  {author} {\bibinfo {author} {\bibfnamefont {E.}~\bibnamefont
  {Collini}}, \bibinfo {author} {\bibfnamefont {C.~Y.}\ \bibnamefont {Wong}},
  \bibinfo {author} {\bibfnamefont {K.~E.}\ \bibnamefont {Wilk}}, \bibinfo
  {author} {\bibfnamefont {P.~M.~G.}\ \bibnamefont {Curmi}}, \bibinfo {author}
  {\bibfnamefont {P.}~\bibnamefont {Brumer}}, \ and\ \bibinfo {author}
  {\bibfnamefont {G.~D.}\ \bibnamefont {Scholes}},\ }\href {\doibase
  10.1038/nature08811} {\bibfield  {journal} {\bibinfo  {journal} {Nature}\
  }\textbf {\bibinfo {volume} {463}},\ \bibinfo {pages} {644} (\bibinfo {year}
  {2010})}\BibitemShut {NoStop}%
\bibitem [{\citenamefont {Žnidarič}\ \emph {et~al.}(2017)\citenamefont
  {Žnidarič}, \citenamefont {Mendoza-Arenas}, \citenamefont {Clark},\ and\
  \citenamefont {Goold}}]{znidaric_dephasing_2017}%
  \BibitemOpen
  \bibfield  {author} {\bibinfo {author} {\bibfnamefont {M.}~\bibnamefont
  {Žnidarič}}, \bibinfo {author} {\bibfnamefont {J.~J.}\ \bibnamefont
  {Mendoza-Arenas}}, \bibinfo {author} {\bibfnamefont {S.~R.}\ \bibnamefont
  {Clark}}, \ and\ \bibinfo {author} {\bibfnamefont {J.}~\bibnamefont
  {Goold}},\ }\href
  {https://onlinelibrary.wiley.com/doi/abs/10.1002/andp.201600298} {\bibfield
  {journal} {\bibinfo  {journal} {Annalen der Physik}\ }\textbf {\bibinfo
  {volume} {529}},\ \bibinfo {pages} {1600298} (\bibinfo {year}
  {2017})}\BibitemShut {NoStop}%
\bibitem [{\citenamefont {Chiaracane}\ \emph {et~al.}(2022)\citenamefont
  {Chiaracane}, \citenamefont {Purkayastha}, \citenamefont {Mitchison},\ and\
  \citenamefont {Goold}}]{chiaracane_dephasing-enhanced_2022}%
  \BibitemOpen
  \bibfield  {author} {\bibinfo {author} {\bibfnamefont {C.}~\bibnamefont
  {Chiaracane}}, \bibinfo {author} {\bibfnamefont {A.}~\bibnamefont
  {Purkayastha}}, \bibinfo {author} {\bibfnamefont {M.~T.}\ \bibnamefont
  {Mitchison}}, \ and\ \bibinfo {author} {\bibfnamefont {J.}~\bibnamefont
  {Goold}},\ }\href {\doibase 10.1103/PhysRevB.105.134203} {\bibfield
  {journal} {\bibinfo  {journal} {Physical Review B}\ }\textbf {\bibinfo
  {volume} {105}},\ \bibinfo {pages} {134203} (\bibinfo {year}
  {2022})}\BibitemShut {NoStop}%
\bibitem [{\citenamefont {Caruso}\ \emph {et~al.}(2009)\citenamefont {Caruso},
  \citenamefont {Chin}, \citenamefont {Datta}, \citenamefont {Huelga},\ and\
  \citenamefont {Plenio}}]{caruso_highly_2009}%
  \BibitemOpen
  \bibfield  {author} {\bibinfo {author} {\bibfnamefont {F.}~\bibnamefont
  {Caruso}}, \bibinfo {author} {\bibfnamefont {A.~W.}\ \bibnamefont {Chin}},
  \bibinfo {author} {\bibfnamefont {A.}~\bibnamefont {Datta}}, \bibinfo
  {author} {\bibfnamefont {S.~F.}\ \bibnamefont {Huelga}}, \ and\ \bibinfo
  {author} {\bibfnamefont {M.~B.}\ \bibnamefont {Plenio}},\ }\href {\doibase
  10.1063/1.3223548} {\bibfield  {journal} {\bibinfo  {journal} {The Journal of
  Chemical Physics}\ }\textbf {\bibinfo {volume} {131}},\ \bibinfo {pages}
  {105106} (\bibinfo {year} {2009})}\BibitemShut {NoStop}%
\bibitem [{\citenamefont {Caruso}\ \emph {et~al.}(2016)\citenamefont {Caruso},
  \citenamefont {Crespi}, \citenamefont {Ciriolo}, \citenamefont {Sciarrino},\
  and\ \citenamefont {Osellame}}]{caruso_fast_2016}%
  \BibitemOpen
  \bibfield  {author} {\bibinfo {author} {\bibfnamefont {F.}~\bibnamefont
  {Caruso}}, \bibinfo {author} {\bibfnamefont {A.}~\bibnamefont {Crespi}},
  \bibinfo {author} {\bibfnamefont {A.~G.}\ \bibnamefont {Ciriolo}}, \bibinfo
  {author} {\bibfnamefont {F.}~\bibnamefont {Sciarrino}}, \ and\ \bibinfo
  {author} {\bibfnamefont {R.}~\bibnamefont {Osellame}},\ }\href {\doibase
  10.1038/ncomms11682} {\bibfield  {journal} {\bibinfo  {journal} {Nature
  Communications}\ }\textbf {\bibinfo {volume} {7}},\ \bibinfo {pages} {11682}
  (\bibinfo {year} {2016})}\BibitemShut {NoStop}%
\bibitem [{\citenamefont {Chin}\ \emph {et~al.}(2012)\citenamefont {Chin},
  \citenamefont {Huelga},\ and\ \citenamefont {Plenio}}]{chin_coherence_2012}%
  \BibitemOpen
  \bibfield  {author} {\bibinfo {author} {\bibfnamefont {A.~W.}\ \bibnamefont
  {Chin}}, \bibinfo {author} {\bibfnamefont {S.~F.}\ \bibnamefont {Huelga}}, \
  and\ \bibinfo {author} {\bibfnamefont {M.~B.}\ \bibnamefont {Plenio}},\
  }\href {\doibase 10.1098/rsta.2011.0224} {\bibfield  {journal} {\bibinfo
  {journal} {Philosophical Transactions of the Royal Society A: Mathematical,
  Physical and Engineering Sciences}\ }\textbf {\bibinfo {volume} {370}},\
  \bibinfo {pages} {3638} (\bibinfo {year} {2012})}\BibitemShut {NoStop}%
\bibitem [{\citenamefont {Collini}(2013)}]{collini_spectroscopic_2013}%
  \BibitemOpen
  \bibfield  {author} {\bibinfo {author} {\bibfnamefont {E.}~\bibnamefont
  {Collini}},\ }\href {\doibase 10.1039/C3CS35444J} {\bibfield  {journal}
  {\bibinfo  {journal} {Chemical Society Reviews}\ }\textbf {\bibinfo {volume}
  {42}},\ \bibinfo {pages} {4932} (\bibinfo {year} {2013})}\BibitemShut
  {NoStop}%
\bibitem [{\citenamefont {Biggerstaff}\ \emph {et~al.}(2016)\citenamefont
  {Biggerstaff}, \citenamefont {Heilmann}, \citenamefont {Zecevik},
  \citenamefont {Gräfe}, \citenamefont {Broome}, \citenamefont {Fedrizzi},
  \citenamefont {Nolte}, \citenamefont {Szameit}, \citenamefont {White},\ and\
  \citenamefont {Kassal}}]{biggerstaff_enhancing_2016}%
  \BibitemOpen
  \bibfield  {author} {\bibinfo {author} {\bibfnamefont {D.~N.}\ \bibnamefont
  {Biggerstaff}}, \bibinfo {author} {\bibfnamefont {R.}~\bibnamefont
  {Heilmann}}, \bibinfo {author} {\bibfnamefont {A.~A.}\ \bibnamefont
  {Zecevik}}, \bibinfo {author} {\bibfnamefont {M.}~\bibnamefont {Gräfe}},
  \bibinfo {author} {\bibfnamefont {M.~A.}\ \bibnamefont {Broome}}, \bibinfo
  {author} {\bibfnamefont {A.}~\bibnamefont {Fedrizzi}}, \bibinfo {author}
  {\bibfnamefont {S.}~\bibnamefont {Nolte}}, \bibinfo {author} {\bibfnamefont
  {A.}~\bibnamefont {Szameit}}, \bibinfo {author} {\bibfnamefont {A.~G.}\
  \bibnamefont {White}}, \ and\ \bibinfo {author} {\bibfnamefont
  {I.}~\bibnamefont {Kassal}},\ }\href {\doibase 10.1038/ncomms11282}
  {\bibfield  {journal} {\bibinfo  {journal} {Nature Communications}\ }\textbf
  {\bibinfo {volume} {7}},\ \bibinfo {pages} {11282} (\bibinfo {year}
  {2016})}\BibitemShut {NoStop}%
\bibitem [{\citenamefont {Faisal}\ \emph {et~al.}(2008)\citenamefont {Faisal},
  \citenamefont {Selen},\ and\ \citenamefont {Wolpert}}]{faisal_noise_2008}%
  \BibitemOpen
  \bibfield  {author} {\bibinfo {author} {\bibfnamefont {A.~A.}\ \bibnamefont
  {Faisal}}, \bibinfo {author} {\bibfnamefont {L.~P.~J.}\ \bibnamefont
  {Selen}}, \ and\ \bibinfo {author} {\bibfnamefont {D.~M.}\ \bibnamefont
  {Wolpert}},\ }\href {\doibase 10.1038/nrn2258} {\bibfield  {journal}
  {\bibinfo  {journal} {Nature Reviews Neuroscience}\ }\textbf {\bibinfo
  {volume} {9}},\ \bibinfo {pages} {292} (\bibinfo {year} {2008})}\BibitemShut
  {NoStop}%
\bibitem [{\citenamefont {Adolphs}\ and\ \citenamefont
  {Renger}(2006)}]{adolphs_how_2006}%
  \BibitemOpen
  \bibfield  {author} {\bibinfo {author} {\bibfnamefont {J.}~\bibnamefont
  {Adolphs}}\ and\ \bibinfo {author} {\bibfnamefont {T.}~\bibnamefont
  {Renger}},\ }\href {\doibase 10.1529/biophysj.105.079483} {\bibfield
  {journal} {\bibinfo  {journal} {Biophysical Journal}\ }\textbf {\bibinfo
  {volume} {91}},\ \bibinfo {pages} {2778} (\bibinfo {year}
  {2006})}\BibitemShut {NoStop}%
\bibitem [{\citenamefont {Gaab}\ and\ \citenamefont
  {Bardeen}(2004)}]{gaab_effects_2004}%
  \BibitemOpen
  \bibfield  {author} {\bibinfo {author} {\bibfnamefont {K.~M.}\ \bibnamefont
  {Gaab}}\ and\ \bibinfo {author} {\bibfnamefont {C.~J.}\ \bibnamefont
  {Bardeen}},\ }\href {\doibase 10.1063/1.1786922} {\bibfield  {journal}
  {\bibinfo  {journal} {The Journal of Chemical Physics}\ }\textbf {\bibinfo
  {volume} {121}},\ \bibinfo {pages} {7813} (\bibinfo {year}
  {2004})}\BibitemShut {NoStop}%
\bibitem [{\citenamefont {Mohseni}\ \emph {et~al.}(2008)\citenamefont
  {Mohseni}, \citenamefont {Rebentrost}, \citenamefont {Lloyd},\ and\
  \citenamefont {Aspuru-Guzik}}]{mohseni_environment-assisted_2008}%
  \BibitemOpen
  \bibfield  {author} {\bibinfo {author} {\bibfnamefont {M.}~\bibnamefont
  {Mohseni}}, \bibinfo {author} {\bibfnamefont {P.}~\bibnamefont {Rebentrost}},
  \bibinfo {author} {\bibfnamefont {S.}~\bibnamefont {Lloyd}}, \ and\ \bibinfo
  {author} {\bibfnamefont {A.}~\bibnamefont {Aspuru-Guzik}},\ }\href {\doibase
  10.1063/1.3002335} {\bibfield  {journal} {\bibinfo  {journal} {The Journal of
  Chemical Physics}\ }\textbf {\bibinfo {volume} {129}},\ \bibinfo {pages}
  {174106} (\bibinfo {year} {2008})}\BibitemShut {NoStop}%
\bibitem [{\citenamefont {Rebentrost}\ \emph
  {et~al.}(2009{\natexlab{b}})\citenamefont {Rebentrost}, \citenamefont
  {Mohseni},\ and\ \citenamefont {Aspuru-Guzik}}]{rebentrost_role_2009}%
  \BibitemOpen
  \bibfield  {author} {\bibinfo {author} {\bibfnamefont {P.}~\bibnamefont
  {Rebentrost}}, \bibinfo {author} {\bibfnamefont {M.}~\bibnamefont {Mohseni}},
  \ and\ \bibinfo {author} {\bibfnamefont {A.}~\bibnamefont {Aspuru-Guzik}},\
  }\href {\doibase 10.1021/jp901724d} {\bibfield  {journal} {\bibinfo
  {journal} {The Journal of Physical Chemistry B}\ }\textbf {\bibinfo {volume}
  {113}},\ \bibinfo {pages} {9942} (\bibinfo {year}
  {2009}{\natexlab{b}})}\BibitemShut {NoStop}%
\bibitem [{\citenamefont {Lindblad}(1976)}]{lindblad_generators_1976}%
  \BibitemOpen
  \bibfield  {author} {\bibinfo {author} {\bibfnamefont {G.}~\bibnamefont
  {Lindblad}},\ }\href {\doibase 10.1007/BF01608499} {\bibfield  {journal}
  {\bibinfo  {journal} {Communications in Mathematical Physics}\ }\textbf
  {\bibinfo {volume} {48}},\ \bibinfo {pages} {119} (\bibinfo {year}
  {1976})}\BibitemShut {NoStop}%
\bibitem [{\citenamefont {Gorini}\ \emph {et~al.}(1976)\citenamefont {Gorini},
  \citenamefont {Kossakowski},\ and\ \citenamefont
  {Sudarshan}}]{gorini_completely_1976}%
  \BibitemOpen
  \bibfield  {author} {\bibinfo {author} {\bibfnamefont {V.}~\bibnamefont
  {Gorini}}, \bibinfo {author} {\bibfnamefont {A.}~\bibnamefont {Kossakowski}},
  \ and\ \bibinfo {author} {\bibfnamefont {E.~C.~G.}\ \bibnamefont
  {Sudarshan}},\ }\href {\doibase 10.1063/1.522979} {\bibfield  {journal}
  {\bibinfo  {journal} {Journal of Mathematical Physics}\ }\textbf {\bibinfo
  {volume} {17}},\ \bibinfo {pages} {821} (\bibinfo {year} {1976})}\BibitemShut
  {NoStop}%
\bibitem [{\citenamefont {Cattaneo}\ \emph {et~al.}(2019)\citenamefont
  {Cattaneo}, \citenamefont {Giorgi}, \citenamefont {Maniscalco},\ and\
  \citenamefont {Zambrini}}]{cattaneo_local_2019}%
  \BibitemOpen
  \bibfield  {author} {\bibinfo {author} {\bibfnamefont {M.}~\bibnamefont
  {Cattaneo}}, \bibinfo {author} {\bibfnamefont {G.~L.}\ \bibnamefont
  {Giorgi}}, \bibinfo {author} {\bibfnamefont {S.}~\bibnamefont {Maniscalco}},
  \ and\ \bibinfo {author} {\bibfnamefont {R.}~\bibnamefont {Zambrini}},\
  }\href {\doibase 10.1088/1367-2630/ab54ac} {\bibfield  {journal} {\bibinfo
  {journal} {New Journal of Physics}\ }\textbf {\bibinfo {volume} {21}},\
  \bibinfo {pages} {113045} (\bibinfo {year} {2019})}\BibitemShut {NoStop}%
\bibitem [{\citenamefont {González}\ \emph {et~al.}(2017)\citenamefont
  {González}, \citenamefont {Correa}, \citenamefont {Nocerino}, \citenamefont
  {Palao}, \citenamefont {Alonso},\ and\ \citenamefont
  {Adesso}}]{gonzalez_testing_2017}%
  \BibitemOpen
  \bibfield  {author} {\bibinfo {author} {\bibfnamefont {J.~O.}\ \bibnamefont
  {González}}, \bibinfo {author} {\bibfnamefont {L.~A.}\ \bibnamefont
  {Correa}}, \bibinfo {author} {\bibfnamefont {G.}~\bibnamefont {Nocerino}},
  \bibinfo {author} {\bibfnamefont {J.~P.}\ \bibnamefont {Palao}}, \bibinfo
  {author} {\bibfnamefont {D.}~\bibnamefont {Alonso}}, \ and\ \bibinfo {author}
  {\bibfnamefont {G.}~\bibnamefont {Adesso}},\ }\href {\doibase
  10.1142/S1230161217400108} {\bibfield  {journal} {\bibinfo  {journal} {Open
  Systems \& Information Dynamics}\ }\textbf {\bibinfo {volume} {24}},\
  \bibinfo {pages} {1740010} (\bibinfo {year} {2017})}\BibitemShut {NoStop}%
\bibitem [{\citenamefont {Hofer}\ \emph {et~al.}(2017)\citenamefont {Hofer},
  \citenamefont {Perarnau-Llobet}, \citenamefont {Miranda}, \citenamefont
  {Haack}, \citenamefont {Silva}, \citenamefont {Brask},\ and\ \citenamefont
  {Brunner}}]{hofer_markovian_2017}%
  \BibitemOpen
  \bibfield  {author} {\bibinfo {author} {\bibfnamefont {P.~P.}\ \bibnamefont
  {Hofer}}, \bibinfo {author} {\bibfnamefont {M.}~\bibnamefont
  {Perarnau-Llobet}}, \bibinfo {author} {\bibfnamefont {L.~D.~M.}\ \bibnamefont
  {Miranda}}, \bibinfo {author} {\bibfnamefont {G.}~\bibnamefont {Haack}},
  \bibinfo {author} {\bibfnamefont {R.}~\bibnamefont {Silva}}, \bibinfo
  {author} {\bibfnamefont {J.~B.}\ \bibnamefont {Brask}}, \ and\ \bibinfo
  {author} {\bibfnamefont {N.}~\bibnamefont {Brunner}},\ }\href {\doibase
  10.1088/1367-2630/aa964f} {\bibfield  {journal} {\bibinfo  {journal} {New
  Journal of Physics}\ }\textbf {\bibinfo {volume} {19}},\ \bibinfo {pages}
  {123037} (\bibinfo {year} {2017})}\BibitemShut {NoStop}%
\bibitem [{\citenamefont {Levy}\ and\ \citenamefont
  {Kosloff}(2014)}]{levy_local_2014}%
  \BibitemOpen
  \bibfield  {author} {\bibinfo {author} {\bibfnamefont {A.}~\bibnamefont
  {Levy}}\ and\ \bibinfo {author} {\bibfnamefont {R.}~\bibnamefont {Kosloff}},\
  }\href {\doibase 10.1209/0295-5075/107/20004} {\bibfield  {journal} {\bibinfo
   {journal} {EPL (Europhysics Letters)}\ }\textbf {\bibinfo {volume} {107}},\
  \bibinfo {pages} {20004} (\bibinfo {year} {2014})}\BibitemShut {NoStop}%
\bibitem [{\citenamefont {Chiara}\ \emph {et~al.}(2018)\citenamefont {Chiara},
  \citenamefont {Landi}, \citenamefont {Hewgill}, \citenamefont {Reid},
  \citenamefont {Ferraro}, \citenamefont {Roncaglia},\ and\ \citenamefont
  {Antezza}}]{chiara_reconciliation_2018}%
  \BibitemOpen
  \bibfield  {author} {\bibinfo {author} {\bibfnamefont {G.~D.}\ \bibnamefont
  {Chiara}}, \bibinfo {author} {\bibfnamefont {G.}~\bibnamefont {Landi}},
  \bibinfo {author} {\bibfnamefont {A.}~\bibnamefont {Hewgill}}, \bibinfo
  {author} {\bibfnamefont {B.}~\bibnamefont {Reid}}, \bibinfo {author}
  {\bibfnamefont {A.}~\bibnamefont {Ferraro}}, \bibinfo {author} {\bibfnamefont
  {A.~J.}\ \bibnamefont {Roncaglia}}, \ and\ \bibinfo {author} {\bibfnamefont
  {M.}~\bibnamefont {Antezza}},\ }\href {\doibase 10.1088/1367-2630/aaecee}
  {\bibfield  {journal} {\bibinfo  {journal} {New Journal of Physics}\ }\textbf
  {\bibinfo {volume} {20}},\ \bibinfo {pages} {113024} (\bibinfo {year}
  {2018})}\BibitemShut {NoStop}%
\bibitem [{\citenamefont {Barra}(2015)}]{barra_thermodynamic_2015}%
  \BibitemOpen
  \bibfield  {author} {\bibinfo {author} {\bibfnamefont {F.}~\bibnamefont
  {Barra}},\ }\href {\doibase 10.1038/srep14873} {\bibfield  {journal}
  {\bibinfo  {journal} {Scientific Reports}\ }\textbf {\bibinfo {volume} {5}},\
  \bibinfo {pages} {14873} (\bibinfo {year} {2015})}\BibitemShut {NoStop}%
\bibitem [{\citenamefont {Esposito}\ \emph {et~al.}(2010)\citenamefont
  {Esposito}, \citenamefont {Lindenberg},\ and\ \citenamefont
  {Broeck}}]{esposito_entropy_2010}%
  \BibitemOpen
  \bibfield  {author} {\bibinfo {author} {\bibfnamefont {M.}~\bibnamefont
  {Esposito}}, \bibinfo {author} {\bibfnamefont {K.}~\bibnamefont
  {Lindenberg}}, \ and\ \bibinfo {author} {\bibfnamefont {C.~V.~d.}\
  \bibnamefont {Broeck}},\ }\href {\doibase 10.1088/1367-2630/12/1/013013}
  {\bibfield  {journal} {\bibinfo  {journal} {New Journal of Physics}\ }\textbf
  {\bibinfo {volume} {12}},\ \bibinfo {pages} {013013} (\bibinfo {year}
  {2010})}\BibitemShut {NoStop}%
\bibitem [{\citenamefont {Reeb}\ and\ \citenamefont
  {Wolf}(2014)}]{reeb_improved_2014}%
  \BibitemOpen
  \bibfield  {author} {\bibinfo {author} {\bibfnamefont {D.}~\bibnamefont
  {Reeb}}\ and\ \bibinfo {author} {\bibfnamefont {M.~M.}\ \bibnamefont
  {Wolf}},\ }\href {\doibase 10.1088/1367-2630/16/10/103011} {\bibfield
  {journal} {\bibinfo  {journal} {New Journal of Physics}\ }\textbf {\bibinfo
  {volume} {16}},\ \bibinfo {pages} {103011} (\bibinfo {year}
  {2014})}\BibitemShut {NoStop}%
\bibitem [{\citenamefont {Attal}\ and\ \citenamefont
  {Pautrat}(2006)}]{attal_repeated_2006}%
  \BibitemOpen
  \bibfield  {author} {\bibinfo {author} {\bibfnamefont {S.}~\bibnamefont
  {Attal}}\ and\ \bibinfo {author} {\bibfnamefont {Y.}~\bibnamefont
  {Pautrat}},\ }\href {\doibase 10.1007/s00023-005-0242-8} {\bibfield
  {journal} {\bibinfo  {journal} {Annales Henri Poincaré}\ }\textbf {\bibinfo
  {volume} {7}},\ \bibinfo {pages} {59} (\bibinfo {year} {2006})}\BibitemShut
  {NoStop}%
\bibitem [{\citenamefont {Karevski}\ and\ \citenamefont
  {Platini}(2009)}]{karevski_quantum_2009}%
  \BibitemOpen
  \bibfield  {author} {\bibinfo {author} {\bibfnamefont {D.}~\bibnamefont
  {Karevski}}\ and\ \bibinfo {author} {\bibfnamefont {T.}~\bibnamefont
  {Platini}},\ }\href {\doibase 10.1103/PhysRevLett.102.207207} {\bibfield
  {journal} {\bibinfo  {journal} {Physical Review Letters}\ }\textbf {\bibinfo
  {volume} {102}},\ \bibinfo {pages} {207207} (\bibinfo {year}
  {2009})}\BibitemShut {NoStop}%
\bibitem [{\citenamefont {Ciccarello}\ \emph {et~al.}(2022)\citenamefont
  {Ciccarello}, \citenamefont {Lorenzo}, \citenamefont {Giovannetti},\ and\
  \citenamefont {Palma}}]{ciccarello_quantum_2022}%
  \BibitemOpen
  \bibfield  {author} {\bibinfo {author} {\bibfnamefont {F.}~\bibnamefont
  {Ciccarello}}, \bibinfo {author} {\bibfnamefont {S.}~\bibnamefont {Lorenzo}},
  \bibinfo {author} {\bibfnamefont {V.}~\bibnamefont {Giovannetti}}, \ and\
  \bibinfo {author} {\bibfnamefont {G.~M.}\ \bibnamefont {Palma}},\ }\href
  {\doibase 10.1016/j.physrep.2022.01.001} {\bibfield  {journal} {\bibinfo
  {journal} {Physics Reports}\ }\bibinfo {series} {Quantum collision models:
  {Open} system dynamics from repeated interactions},\ \textbf {\bibinfo
  {volume} {954}},\ \bibinfo {pages} {1} (\bibinfo {year} {2022})}\BibitemShut
  {NoStop}%
\bibitem [{\citenamefont {Campbell}\ and\ \citenamefont
  {Vacchini}(2021)}]{campbell_collision_2021}%
  \BibitemOpen
  \bibfield  {author} {\bibinfo {author} {\bibfnamefont {S.}~\bibnamefont
  {Campbell}}\ and\ \bibinfo {author} {\bibfnamefont {B.}~\bibnamefont
  {Vacchini}},\ }\href {\doibase 10.1209/0295-5075/133/60001} {\bibfield
  {journal} {\bibinfo  {journal} {Europhysics Letters}\ }\textbf {\bibinfo
  {volume} {133}},\ \bibinfo {pages} {60001} (\bibinfo {year}
  {2021})}\BibitemShut {NoStop}%
\bibitem [{\citenamefont {Scarani}\ \emph {et~al.}(2002)\citenamefont
  {Scarani}, \citenamefont {Ziman}, \citenamefont {Štelmachovič},
  \citenamefont {Gisin},\ and\ \citenamefont
  {Bužek}}]{scarani_thermalizing_2002}%
  \BibitemOpen
  \bibfield  {author} {\bibinfo {author} {\bibfnamefont {V.}~\bibnamefont
  {Scarani}}, \bibinfo {author} {\bibfnamefont {M.}~\bibnamefont {Ziman}},
  \bibinfo {author} {\bibfnamefont {P.}~\bibnamefont {Štelmachovič}},
  \bibinfo {author} {\bibfnamefont {N.}~\bibnamefont {Gisin}}, \ and\ \bibinfo
  {author} {\bibfnamefont {V.}~\bibnamefont {Bužek}},\ }\href {\doibase
  10.1103/PhysRevLett.88.097905} {\bibfield  {journal} {\bibinfo  {journal}
  {Physical Review Letters}\ }\textbf {\bibinfo {volume} {88}},\ \bibinfo
  {pages} {097905} (\bibinfo {year} {2002})}\BibitemShut {NoStop}%
\bibitem [{\citenamefont {Ziman}\ \emph {et~al.}(2002)\citenamefont {Ziman},
  \citenamefont {Štelmachovič}, \citenamefont {Bužek}, \citenamefont
  {Hillery}, \citenamefont {Scarani},\ and\ \citenamefont
  {Gisin}}]{ziman_diluting_2002}%
  \BibitemOpen
  \bibfield  {author} {\bibinfo {author} {\bibfnamefont {M.}~\bibnamefont
  {Ziman}}, \bibinfo {author} {\bibfnamefont {P.}~\bibnamefont
  {Štelmachovič}}, \bibinfo {author} {\bibfnamefont {V.}~\bibnamefont
  {Bužek}}, \bibinfo {author} {\bibfnamefont {M.}~\bibnamefont {Hillery}},
  \bibinfo {author} {\bibfnamefont {V.}~\bibnamefont {Scarani}}, \ and\
  \bibinfo {author} {\bibfnamefont {N.}~\bibnamefont {Gisin}},\ }\href
  {\doibase 10.1103/PhysRevA.65.042105} {\bibfield  {journal} {\bibinfo
  {journal} {Physical Review A}\ }\textbf {\bibinfo {volume} {65}},\ \bibinfo
  {pages} {042105} (\bibinfo {year} {2002})}\BibitemShut {NoStop}%
\bibitem [{\citenamefont {Trushechkin}\ and\ \citenamefont
  {Volovich}(2016)}]{trushechkin_perturbative_2016}%
  \BibitemOpen
  \bibfield  {author} {\bibinfo {author} {\bibfnamefont {A.~S.}\ \bibnamefont
  {Trushechkin}}\ and\ \bibinfo {author} {\bibfnamefont {I.~V.}\ \bibnamefont
  {Volovich}},\ }\href {\doibase 10.1209/0295-5075/113/30005} {\bibfield
  {journal} {\bibinfo  {journal} {Europhysics Letters}\ }\textbf {\bibinfo
  {volume} {113}},\ \bibinfo {pages} {30005} (\bibinfo {year}
  {2016})}\BibitemShut {NoStop}%
\bibitem [{\citenamefont {Soret}\ \emph {et~al.}(2022)\citenamefont {Soret},
  \citenamefont {Cavina},\ and\ \citenamefont
  {Esposito}}]{soret_thermodynamic_2022}%
  \BibitemOpen
  \bibfield  {author} {\bibinfo {author} {\bibfnamefont {A.}~\bibnamefont
  {Soret}}, \bibinfo {author} {\bibfnamefont {V.}~\bibnamefont {Cavina}}, \
  and\ \bibinfo {author} {\bibfnamefont {M.}~\bibnamefont {Esposito}},\ }\href
  {http://arxiv.org/abs/2207.05719} {\enquote {\bibinfo {title} {Thermodynamic
  consistency of quantum master equations},}\ } (\bibinfo {year}
  {2022})\BibitemShut {NoStop}%
\bibitem [{\citenamefont {Hewgill}\ \emph {et~al.}(2021)\citenamefont
  {Hewgill}, \citenamefont {De~Chiara},\ and\ \citenamefont
  {Imparato}}]{hewgill_quantum_2021}%
  \BibitemOpen
  \bibfield  {author} {\bibinfo {author} {\bibfnamefont {A.}~\bibnamefont
  {Hewgill}}, \bibinfo {author} {\bibfnamefont {G.}~\bibnamefont {De~Chiara}},
  \ and\ \bibinfo {author} {\bibfnamefont {A.}~\bibnamefont {Imparato}},\
  }\href {\doibase 10.1103/PhysRevResearch.3.013165} {\bibfield  {journal}
  {\bibinfo  {journal} {Physical Review Research}\ }\textbf {\bibinfo {volume}
  {3}},\ \bibinfo {pages} {013165} (\bibinfo {year} {2021})}\BibitemShut
  {NoStop}%
\bibitem [{\citenamefont {Potts}\ \emph {et~al.}(2021)\citenamefont {Potts},
  \citenamefont {Kalaee},\ and\ \citenamefont
  {Wacker}}]{potts_thermodynamically_2021}%
  \BibitemOpen
  \bibfield  {author} {\bibinfo {author} {\bibfnamefont {P.~P.}\ \bibnamefont
  {Potts}}, \bibinfo {author} {\bibfnamefont {A.~A.~S.}\ \bibnamefont
  {Kalaee}}, \ and\ \bibinfo {author} {\bibfnamefont {A.}~\bibnamefont
  {Wacker}},\ }\href {\doibase 10.1088/1367-2630/ac3b2f} {\bibfield  {journal}
  {\bibinfo  {journal} {New Journal of Physics}\ }\textbf {\bibinfo {volume}
  {23}},\ \bibinfo {pages} {123013} (\bibinfo {year} {2021})}\BibitemShut
  {NoStop}%
\bibitem [{\citenamefont {van~der Wiel}\ \emph {et~al.}(2002)\citenamefont
  {van~der Wiel}, \citenamefont {De~Franceschi}, \citenamefont {Elzerman},
  \citenamefont {Fujisawa}, \citenamefont {Tarucha},\ and\ \citenamefont
  {Kouwenhoven}}]{van_der_wiel_electron_2002}%
  \BibitemOpen
  \bibfield  {author} {\bibinfo {author} {\bibfnamefont {W.~G.}\ \bibnamefont
  {van~der Wiel}}, \bibinfo {author} {\bibfnamefont {S.}~\bibnamefont
  {De~Franceschi}}, \bibinfo {author} {\bibfnamefont {J.~M.}\ \bibnamefont
  {Elzerman}}, \bibinfo {author} {\bibfnamefont {T.}~\bibnamefont {Fujisawa}},
  \bibinfo {author} {\bibfnamefont {S.}~\bibnamefont {Tarucha}}, \ and\
  \bibinfo {author} {\bibfnamefont {L.~P.}\ \bibnamefont {Kouwenhoven}},\
  }\href {\doibase 10.1103/RevModPhys.75.1} {\bibfield  {journal} {\bibinfo
  {journal} {Reviews of Modern Physics}\ }\textbf {\bibinfo {volume} {75}},\
  \bibinfo {pages} {1} (\bibinfo {year} {2002})}\BibitemShut {NoStop}%
\bibitem [{\citenamefont {Cuetara}\ and\ \citenamefont
  {Esposito}(2015)}]{cuetara_double_2015}%
  \BibitemOpen
  \bibfield  {author} {\bibinfo {author} {\bibfnamefont {G.~B.}\ \bibnamefont
  {Cuetara}}\ and\ \bibinfo {author} {\bibfnamefont {M.}~\bibnamefont
  {Esposito}},\ }\href {\doibase 10.1088/1367-2630/17/9/095005} {\bibfield
  {journal} {\bibinfo  {journal} {New Journal of Physics}\ }\textbf {\bibinfo
  {volume} {17}},\ \bibinfo {pages} {095005} (\bibinfo {year}
  {2015})}\BibitemShut {NoStop}%
\bibitem [{\citenamefont {Barato}\ and\ \citenamefont
  {Seifert}(2015)}]{barato_thermodynamic_2015}%
  \BibitemOpen
  \bibfield  {author} {\bibinfo {author} {\bibfnamefont {A.~C.}\ \bibnamefont
  {Barato}}\ and\ \bibinfo {author} {\bibfnamefont {U.}~\bibnamefont
  {Seifert}},\ }\href {\doibase 10.1103/PhysRevLett.114.158101} {\bibfield
  {journal} {\bibinfo  {journal} {Physical Review Letters}\ }\textbf {\bibinfo
  {volume} {114}},\ \bibinfo {pages} {158101} (\bibinfo {year}
  {2015})}\BibitemShut {NoStop}%
\bibitem [{\citenamefont {Pietzonka}\ and\ \citenamefont
  {Seifert}(2018)}]{pietzonka_universal_2018}%
  \BibitemOpen
  \bibfield  {author} {\bibinfo {author} {\bibfnamefont {P.}~\bibnamefont
  {Pietzonka}}\ and\ \bibinfo {author} {\bibfnamefont {U.}~\bibnamefont
  {Seifert}},\ }\href {\doibase 10.1103/PhysRevLett.120.190602} {\bibfield
  {journal} {\bibinfo  {journal} {Physical Review Letters}\ }\textbf {\bibinfo
  {volume} {120}},\ \bibinfo {pages} {190602} (\bibinfo {year}
  {2018})}\BibitemShut {NoStop}%
\bibitem [{\citenamefont {Pietzonka}\ \emph {et~al.}(2017)\citenamefont
  {Pietzonka}, \citenamefont {Ritort},\ and\ \citenamefont
  {Seifert}}]{pietzonka_finite-time_2017}%
  \BibitemOpen
  \bibfield  {author} {\bibinfo {author} {\bibfnamefont {P.}~\bibnamefont
  {Pietzonka}}, \bibinfo {author} {\bibfnamefont {F.}~\bibnamefont {Ritort}}, \
  and\ \bibinfo {author} {\bibfnamefont {U.}~\bibnamefont {Seifert}},\ }\href
  {\doibase 10.1103/PhysRevE.96.012101} {\bibfield  {journal} {\bibinfo
  {journal} {Physical Review E}\ }\textbf {\bibinfo {volume} {96}},\ \bibinfo
  {pages} {012101} (\bibinfo {year} {2017})}\BibitemShut {NoStop}%
\bibitem [{\citenamefont {Pietzonka}\ \emph {et~al.}(2016)\citenamefont
  {Pietzonka}, \citenamefont {Barato},\ and\ \citenamefont
  {Seifert}}]{pietzonka_universal_2016}%
  \BibitemOpen
  \bibfield  {author} {\bibinfo {author} {\bibfnamefont {P.}~\bibnamefont
  {Pietzonka}}, \bibinfo {author} {\bibfnamefont {A.~C.}\ \bibnamefont
  {Barato}}, \ and\ \bibinfo {author} {\bibfnamefont {U.}~\bibnamefont
  {Seifert}},\ }\href {\doibase 10.1103/PhysRevE.93.052145} {\bibfield
  {journal} {\bibinfo  {journal} {Physical Review E}\ }\textbf {\bibinfo
  {volume} {93}},\ \bibinfo {pages} {052145} (\bibinfo {year}
  {2016})}\BibitemShut {NoStop}%
\bibitem [{\citenamefont {Horowitz}\ and\ \citenamefont
  {Gingrich}(2017)}]{horowitz_proof_2017}%
  \BibitemOpen
  \bibfield  {author} {\bibinfo {author} {\bibfnamefont {J.~M.}\ \bibnamefont
  {Horowitz}}\ and\ \bibinfo {author} {\bibfnamefont {T.~R.}\ \bibnamefont
  {Gingrich}},\ }\href {\doibase 10.1103/PhysRevE.96.020103} {\bibfield
  {journal} {\bibinfo  {journal} {Physical Review E}\ }\textbf {\bibinfo
  {volume} {96}},\ \bibinfo {pages} {020103} (\bibinfo {year}
  {2017})}\BibitemShut {NoStop}%
\bibitem [{\citenamefont {Timpanaro}\ \emph {et~al.}(2019)\citenamefont
  {Timpanaro}, \citenamefont {Guarnieri}, \citenamefont {Goold},\ and\
  \citenamefont {Landi}}]{timpanaro_thermodynamic_2019}%
  \BibitemOpen
  \bibfield  {author} {\bibinfo {author} {\bibfnamefont {A.~M.}\ \bibnamefont
  {Timpanaro}}, \bibinfo {author} {\bibfnamefont {G.}~\bibnamefont
  {Guarnieri}}, \bibinfo {author} {\bibfnamefont {J.}~\bibnamefont {Goold}}, \
  and\ \bibinfo {author} {\bibfnamefont {G.~T.}\ \bibnamefont {Landi}},\ }\href
  {\doibase 10.1103/PhysRevLett.123.090604} {\bibfield  {journal} {\bibinfo
  {journal} {Physical Review Letters}\ }\textbf {\bibinfo {volume} {123}},\
  \bibinfo {pages} {090604} (\bibinfo {year} {2019})}\BibitemShut {NoStop}%
\bibitem [{\citenamefont {Carollo}\ \emph {et~al.}(2019)\citenamefont
  {Carollo}, \citenamefont {Jack},\ and\ \citenamefont
  {Garrahan}}]{carollo_unraveling_2019}%
  \BibitemOpen
  \bibfield  {author} {\bibinfo {author} {\bibfnamefont {F.}~\bibnamefont
  {Carollo}}, \bibinfo {author} {\bibfnamefont {R.~L.}\ \bibnamefont {Jack}}, \
  and\ \bibinfo {author} {\bibfnamefont {J.~P.}\ \bibnamefont {Garrahan}},\
  }\href {\doibase 10.1103/PhysRevLett.122.130605} {\bibfield  {journal}
  {\bibinfo  {journal} {Phys. Rev. Lett.}\ }\textbf {\bibinfo {volume} {122}},\
  \bibinfo {pages} {130605} (\bibinfo {year} {2019})}\BibitemShut {NoStop}%
\bibitem [{\citenamefont {Hasegawa}(2020)}]{hasegawa_quantum_2020}%
  \BibitemOpen
  \bibfield  {author} {\bibinfo {author} {\bibfnamefont {Y.}~\bibnamefont
  {Hasegawa}},\ }\href {\doibase 10.1103/PhysRevLett.125.050601} {\bibfield
  {journal} {\bibinfo  {journal} {Physical Review Letters}\ }\textbf {\bibinfo
  {volume} {125}},\ \bibinfo {pages} {050601} (\bibinfo {year}
  {2020})}\BibitemShut {NoStop}%
\bibitem [{\citenamefont {Hasegawa}(2021)}]{hasegawa_thermodynamic_2021}%
  \BibitemOpen
  \bibfield  {author} {\bibinfo {author} {\bibfnamefont {Y.}~\bibnamefont
  {Hasegawa}},\ }\href {\doibase 10.1103/PhysRevLett.126.010602} {\bibfield
  {journal} {\bibinfo  {journal} {Physical Review Letters}\ }\textbf {\bibinfo
  {volume} {126}},\ \bibinfo {pages} {010602} (\bibinfo {year}
  {2021})}\BibitemShut {NoStop}%
\bibitem [{\citenamefont {Van~Vu}\ and\ \citenamefont
  {Saito}(2022)}]{van_vu_thermodynamics_2022}%
  \BibitemOpen
  \bibfield  {author} {\bibinfo {author} {\bibfnamefont {T.}~\bibnamefont
  {Van~Vu}}\ and\ \bibinfo {author} {\bibfnamefont {K.}~\bibnamefont {Saito}},\
  }\href {\doibase 10.1103/PhysRevLett.128.140602} {\bibfield  {journal}
  {\bibinfo  {journal} {Physical Review Letters}\ }\textbf {\bibinfo {volume}
  {128}},\ \bibinfo {pages} {140602} (\bibinfo {year} {2022})}\BibitemShut
  {NoStop}%
\bibitem [{\citenamefont {Agarwalla}\ and\ \citenamefont
  {Segal}(2018)}]{agarwalla_assessing_2018}%
  \BibitemOpen
  \bibfield  {author} {\bibinfo {author} {\bibfnamefont {B.~K.}\ \bibnamefont
  {Agarwalla}}\ and\ \bibinfo {author} {\bibfnamefont {D.}~\bibnamefont
  {Segal}},\ }\href {\doibase 10.1103/PhysRevB.98.155438} {\bibfield  {journal}
  {\bibinfo  {journal} {Physical Review B}\ }\textbf {\bibinfo {volume} {98}},\
  \bibinfo {pages} {155438} (\bibinfo {year} {2018})}\BibitemShut {NoStop}%
\bibitem [{\citenamefont {Liu}\ and\ \citenamefont
  {Segal}(2019)}]{liu_thermodynamic_2019}%
  \BibitemOpen
  \bibfield  {author} {\bibinfo {author} {\bibfnamefont {J.}~\bibnamefont
  {Liu}}\ and\ \bibinfo {author} {\bibfnamefont {D.}~\bibnamefont {Segal}},\
  }\href {\doibase 10.1103/PhysRevE.99.062141} {\bibfield  {journal} {\bibinfo
  {journal} {Physical Review E}\ }\textbf {\bibinfo {volume} {99}},\ \bibinfo
  {pages} {062141} (\bibinfo {year} {2019})}\BibitemShut {NoStop}%
\bibitem [{\citenamefont {Prech}\ \emph {et~al.}(2022)\citenamefont {Prech},
  \citenamefont {Johansson}, \citenamefont {Nyholm}, \citenamefont {Landi},
  \citenamefont {Verdozzi}, \citenamefont {Samuelsson},\ and\ \citenamefont
  {Potts}}]{Prech_2022}%
  \BibitemOpen
  \bibfield  {author} {\bibinfo {author} {\bibfnamefont {K.}~\bibnamefont
  {Prech}}, \bibinfo {author} {\bibfnamefont {P.}~\bibnamefont {Johansson}},
  \bibinfo {author} {\bibfnamefont {E.}~\bibnamefont {Nyholm}}, \bibinfo
  {author} {\bibfnamefont {G.~T.}\ \bibnamefont {Landi}}, \bibinfo {author}
  {\bibfnamefont {C.}~\bibnamefont {Verdozzi}}, \bibinfo {author}
  {\bibfnamefont {P.}~\bibnamefont {Samuelsson}}, \ and\ \bibinfo {author}
  {\bibfnamefont {P.~P.}\ \bibnamefont {Potts}},\ }\href@noop {} {\  (\bibinfo
  {year} {2022})},\ \Eprint {http://arxiv.org/abs/2212.03835v1}
  {arXiv:2212.03835v1 [cond-mat.mes-hall]} \BibitemShut {NoStop}%
\bibitem [{\citenamefont {Breuer}(2003)}]{breuer_quantum_2003}%
  \BibitemOpen
  \bibfield  {author} {\bibinfo {author} {\bibfnamefont {H.-P.}\ \bibnamefont
  {Breuer}},\ }\href {\doibase 10.1103/PhysRevA.68.032105} {\bibfield
  {journal} {\bibinfo  {journal} {Physical Review A}\ }\textbf {\bibinfo
  {volume} {68}},\ \bibinfo {pages} {032105} (\bibinfo {year}
  {2003})}\BibitemShut {NoStop}%
\bibitem [{\citenamefont {Mitchison}\ and\ \citenamefont
  {Plenio}(2018)}]{mitchison_non-additive_2018}%
  \BibitemOpen
  \bibfield  {author} {\bibinfo {author} {\bibfnamefont {M.~T.}\ \bibnamefont
  {Mitchison}}\ and\ \bibinfo {author} {\bibfnamefont {M.~B.}\ \bibnamefont
  {Plenio}},\ }\href {\doibase 10.1088/1367-2630/aa9f70} {\bibfield  {journal}
  {\bibinfo  {journal} {New Journal of Physics}\ }\textbf {\bibinfo {volume}
  {20}},\ \bibinfo {pages} {033005} (\bibinfo {year} {2018})}\BibitemShut
  {NoStop}%
\bibitem [{\citenamefont {Humphrey}\ \emph {et~al.}(2002)\citenamefont
  {Humphrey}, \citenamefont {Newbury}, \citenamefont {Taylor},\ and\
  \citenamefont {Linke}}]{humphrey_reversible_2002}%
  \BibitemOpen
  \bibfield  {author} {\bibinfo {author} {\bibfnamefont {T.~E.}\ \bibnamefont
  {Humphrey}}, \bibinfo {author} {\bibfnamefont {R.}~\bibnamefont {Newbury}},
  \bibinfo {author} {\bibfnamefont {R.~P.}\ \bibnamefont {Taylor}}, \ and\
  \bibinfo {author} {\bibfnamefont {H.}~\bibnamefont {Linke}},\ }\href
  {\doibase 10.1103/PhysRevLett.89.116801} {\bibfield  {journal} {\bibinfo
  {journal} {Physical Review Letters}\ }\textbf {\bibinfo {volume} {89}},\
  \bibinfo {pages} {116801} (\bibinfo {year} {2002})}\BibitemShut {NoStop}%
\bibitem [{\citenamefont {Goan}\ \emph {et~al.}(2001)\citenamefont {Goan},
  \citenamefont {Milburn}, \citenamefont {Wiseman},\ and\ \citenamefont
  {Bi~Sun}}]{goan_continuous_2001}%
  \BibitemOpen
  \bibfield  {author} {\bibinfo {author} {\bibfnamefont {H.-S.}\ \bibnamefont
  {Goan}}, \bibinfo {author} {\bibfnamefont {G.~J.}\ \bibnamefont {Milburn}},
  \bibinfo {author} {\bibfnamefont {H.~M.}\ \bibnamefont {Wiseman}}, \ and\
  \bibinfo {author} {\bibfnamefont {H.}~\bibnamefont {Bi~Sun}},\ }\href
  {\doibase 10.1103/PhysRevB.63.125326} {\bibfield  {journal} {\bibinfo
  {journal} {Physical Review B}\ }\textbf {\bibinfo {volume} {63}},\ \bibinfo
  {pages} {125326} (\bibinfo {year} {2001})}\BibitemShut {NoStop}%
\bibitem [{\citenamefont {Itano}\ \emph {et~al.}(1990)\citenamefont {Itano},
  \citenamefont {Heinzen}, \citenamefont {Bollinger},\ and\ \citenamefont
  {Wineland}}]{Itano_quantum_1990}%
  \BibitemOpen
  \bibfield  {author} {\bibinfo {author} {\bibfnamefont {W.~M.}\ \bibnamefont
  {Itano}}, \bibinfo {author} {\bibfnamefont {D.~J.}\ \bibnamefont {Heinzen}},
  \bibinfo {author} {\bibfnamefont {J.~J.}\ \bibnamefont {Bollinger}}, \ and\
  \bibinfo {author} {\bibfnamefont {D.~J.}\ \bibnamefont {Wineland}},\ }\href
  {\doibase 10.1103/PhysRevA.41.2295} {\bibfield  {journal} {\bibinfo
  {journal} {Physical Review A}\ }\textbf {\bibinfo {volume} {41}},\ \bibinfo
  {pages} {2295} (\bibinfo {year} {1990})}\BibitemShut {NoStop}%
\bibitem [{\citenamefont {Kalaee}\ \emph {et~al.}(2021)\citenamefont {Kalaee},
  \citenamefont {Wacker},\ and\ \citenamefont {Potts}}]{kalaee_violating_2021}%
  \BibitemOpen
  \bibfield  {author} {\bibinfo {author} {\bibfnamefont {A.~A.~S.}\
  \bibnamefont {Kalaee}}, \bibinfo {author} {\bibfnamefont {A.}~\bibnamefont
  {Wacker}}, \ and\ \bibinfo {author} {\bibfnamefont {P.~P.}\ \bibnamefont
  {Potts}},\ }\href {\doibase 10.1103/PhysRevE.104.L012103} {\bibfield
  {journal} {\bibinfo  {journal} {Physical Review E}\ }\textbf {\bibinfo
  {volume} {104}},\ \bibinfo {pages} {L012103} (\bibinfo {year}
  {2021})}\BibitemShut {NoStop}%
\bibitem [{\citenamefont {Cangemi}\ \emph {et~al.}(2020)\citenamefont
  {Cangemi}, \citenamefont {Cataudella}, \citenamefont {Benenti}, \citenamefont
  {Sassetti},\ and\ \citenamefont {De~Filippis}}]{cangemi_violation_2020}%
  \BibitemOpen
  \bibfield  {author} {\bibinfo {author} {\bibfnamefont {L.~M.}\ \bibnamefont
  {Cangemi}}, \bibinfo {author} {\bibfnamefont {V.}~\bibnamefont {Cataudella}},
  \bibinfo {author} {\bibfnamefont {G.}~\bibnamefont {Benenti}}, \bibinfo
  {author} {\bibfnamefont {M.}~\bibnamefont {Sassetti}}, \ and\ \bibinfo
  {author} {\bibfnamefont {G.}~\bibnamefont {De~Filippis}},\ }\href {\doibase
  10.1103/PhysRevB.102.165418} {\bibfield  {journal} {\bibinfo  {journal}
  {Physical Review B}\ }\textbf {\bibinfo {volume} {102}},\ \bibinfo {pages}
  {165418} (\bibinfo {year} {2020})}\BibitemShut {NoStop}%
\bibitem [{\citenamefont
  {Ptaszyński}(2018)}]{ptaszynski_coherence-enhanced_2018}%
  \BibitemOpen
  \bibfield  {author} {\bibinfo {author} {\bibfnamefont {K.}~\bibnamefont
  {Ptaszyński}},\ }\href {\doibase 10.1103/PhysRevB.98.085425} {\bibfield
  {journal} {\bibinfo  {journal} {Physical Review B}\ }\textbf {\bibinfo
  {volume} {98}},\ \bibinfo {pages} {085425} (\bibinfo {year}
  {2018})}\BibitemShut {NoStop}%
\bibitem [{\citenamefont {Rignon-Bret}\ \emph {et~al.}(2021)\citenamefont
  {Rignon-Bret}, \citenamefont {Guarnieri}, \citenamefont {Goold},\ and\
  \citenamefont {Mitchison}}]{rignon-bret_thermodynamics_2021}%
  \BibitemOpen
  \bibfield  {author} {\bibinfo {author} {\bibfnamefont {A.}~\bibnamefont
  {Rignon-Bret}}, \bibinfo {author} {\bibfnamefont {G.}~\bibnamefont
  {Guarnieri}}, \bibinfo {author} {\bibfnamefont {J.}~\bibnamefont {Goold}}, \
  and\ \bibinfo {author} {\bibfnamefont {M.~T.}\ \bibnamefont {Mitchison}},\
  }\href {\doibase 10.1103/PhysRevE.103.012133} {\bibfield  {journal} {\bibinfo
   {journal} {Physical Review E}\ }\textbf {\bibinfo {volume} {103}},\ \bibinfo
  {pages} {012133} (\bibinfo {year} {2021})}\BibitemShut {NoStop}%
\bibitem [{\citenamefont {Kewming}\ \emph {et~al.}(2022)\citenamefont
  {Kewming}, \citenamefont {Mitchison},\ and\ \citenamefont
  {Landi}}]{kewming_diverging_2022}%
  \BibitemOpen
  \bibfield  {author} {\bibinfo {author} {\bibfnamefont {M.~J.}\ \bibnamefont
  {Kewming}}, \bibinfo {author} {\bibfnamefont {M.~T.}\ \bibnamefont
  {Mitchison}}, \ and\ \bibinfo {author} {\bibfnamefont {G.~T.}\ \bibnamefont
  {Landi}},\ }\href {http://arxiv.org/abs/2205.02622} {\bibfield  {journal}
  {\bibinfo  {journal} {arXiv:2205.02622 [quant-ph]}\ } (\bibinfo {year}
  {2022})}\BibitemShut {NoStop}%
\bibitem [{\citenamefont {Schaller}(2014)}]{schaller_open_2014-1}%
  \BibitemOpen
  \bibfield  {author} {\bibinfo {author} {\bibfnamefont {G.}~\bibnamefont
  {Schaller}},\ }\href {\doibase 10.1007/978-3-319-03877-3} {\emph {\bibinfo
  {title} {Open {Quantum} {Systems} {Far} from {Equilibrium}}}},\ \bibinfo
  {series} {Lecture {Notes} in {Physics}}, Vol.\ \bibinfo {volume} {881}\
  (\bibinfo  {publisher} {Springer International Publishing},\ \bibinfo
  {address} {Cham},\ \bibinfo {year} {2014})\BibitemShut {NoStop}%
\bibitem [{\citenamefont {Wiseman}\ and\ \citenamefont
  {Milburn}(2009)}]{wiseman_quantum_2009}%
  \BibitemOpen
  \bibfield  {author} {\bibinfo {author} {\bibfnamefont {H.~M.}\ \bibnamefont
  {Wiseman}}\ and\ \bibinfo {author} {\bibfnamefont {G.~J.}\ \bibnamefont
  {Milburn}},\ }\href {\doibase 10.1017/CBO9780511813948} {\emph {\bibinfo
  {title} {Quantum {Measurement} and {Control}}}}\ (\bibinfo  {publisher}
  {Cambridge University Press},\ \bibinfo {address} {Cambridge},\ \bibinfo
  {year} {2009})\BibitemShut {NoStop}%
\bibitem [{\citenamefont {Lacerda}\ \emph {et~al.}(2022)\citenamefont
  {Lacerda}, \citenamefont {Purkayastha}, \citenamefont {Kewming},
  \citenamefont {Landi},\ and\ \citenamefont {Goold}}]{lacerda_quantum_2022}%
  \BibitemOpen
  \bibfield  {author} {\bibinfo {author} {\bibfnamefont {A.~M.}\ \bibnamefont
  {Lacerda}}, \bibinfo {author} {\bibfnamefont {A.}~\bibnamefont
  {Purkayastha}}, \bibinfo {author} {\bibfnamefont {M.}~\bibnamefont
  {Kewming}}, \bibinfo {author} {\bibfnamefont {G.~T.}\ \bibnamefont {Landi}},
  \ and\ \bibinfo {author} {\bibfnamefont {J.}~\bibnamefont {Goold}},\ }\href
  {http://arxiv.org/abs/2206.01090} {\emph {\bibinfo {title} {Quantum
  thermodynamics with fast driving and strong coupling via the mesoscopic leads
  approach}}},\ \bibinfo {type} {Tech. Rep.}\ \bibinfo {number}
  {arXiv:2206.01090}\ (\bibinfo  {institution} {arXiv},\ \bibinfo {year}
  {2022})\BibitemShut {NoStop}%
\bibitem [{\citenamefont {Brenes}\ \emph {et~al.}(2020)\citenamefont {Brenes},
  \citenamefont {Mendoza-Arenas}, \citenamefont {Purkayastha}, \citenamefont
  {Mitchison}, \citenamefont {Clark},\ and\ \citenamefont
  {Goold}}]{brenes_tensor-network_2020}%
  \BibitemOpen
  \bibfield  {author} {\bibinfo {author} {\bibfnamefont {M.}~\bibnamefont
  {Brenes}}, \bibinfo {author} {\bibfnamefont {J.~J.}\ \bibnamefont
  {Mendoza-Arenas}}, \bibinfo {author} {\bibfnamefont {A.}~\bibnamefont
  {Purkayastha}}, \bibinfo {author} {\bibfnamefont {M.~T.}\ \bibnamefont
  {Mitchison}}, \bibinfo {author} {\bibfnamefont {S.~R.}\ \bibnamefont
  {Clark}}, \ and\ \bibinfo {author} {\bibfnamefont {J.}~\bibnamefont
  {Goold}},\ }\href {\doibase 10.1103/PhysRevX.10.031040} {\bibfield  {journal}
  {\bibinfo  {journal} {Physical Review X}\ }\textbf {\bibinfo {volume} {10}},\
  \bibinfo {pages} {031040} (\bibinfo {year} {2020})}\BibitemShut {NoStop}%
\bibitem [{\citenamefont {Brenes}\ \emph {et~al.}(2022)\citenamefont {Brenes},
  \citenamefont {Guarnieri}, \citenamefont {Purkayastha}, \citenamefont
  {Eisert}, \citenamefont {Segal},\ and\ \citenamefont {Landi}}]{Brenes_2022}%
  \BibitemOpen
  \bibfield  {author} {\bibinfo {author} {\bibfnamefont {M.}~\bibnamefont
  {Brenes}}, \bibinfo {author} {\bibfnamefont {G.}~\bibnamefont {Guarnieri}},
  \bibinfo {author} {\bibfnamefont {A.}~\bibnamefont {Purkayastha}}, \bibinfo
  {author} {\bibfnamefont {J.}~\bibnamefont {Eisert}}, \bibinfo {author}
  {\bibfnamefont {D.}~\bibnamefont {Segal}}, \ and\ \bibinfo {author}
  {\bibfnamefont {G.}~\bibnamefont {Landi}},\ }\href@noop {} {\  (\bibinfo
  {year} {2022})},\ \Eprint {http://arxiv.org/abs/2211.13832v1}
  {arXiv:2211.13832v1 [quant-ph]} \BibitemShut {NoStop}%
\bibitem [{\citenamefont {Purkayastha}\ \emph {et~al.}(2021)\citenamefont
  {Purkayastha}, \citenamefont {Guarnieri}, \citenamefont {Campbell},
  \citenamefont {Prior},\ and\ \citenamefont
  {Goold}}]{purkayastha_periodically_2021}%
  \BibitemOpen
  \bibfield  {author} {\bibinfo {author} {\bibfnamefont {A.}~\bibnamefont
  {Purkayastha}}, \bibinfo {author} {\bibfnamefont {G.}~\bibnamefont
  {Guarnieri}}, \bibinfo {author} {\bibfnamefont {S.}~\bibnamefont {Campbell}},
  \bibinfo {author} {\bibfnamefont {J.}~\bibnamefont {Prior}}, \ and\ \bibinfo
  {author} {\bibfnamefont {J.}~\bibnamefont {Goold}},\ }\href {\doibase
  10.1103/PhysRevB.104.045417} {\bibfield  {journal} {\bibinfo  {journal}
  {Physical Review B}\ }\textbf {\bibinfo {volume} {104}},\ \bibinfo {pages}
  {045417} (\bibinfo {year} {2021})}\BibitemShut {NoStop}%
\bibitem [{\citenamefont {Purkayastha}\ \emph {et~al.}(2022)\citenamefont
  {Purkayastha}, \citenamefont {Guarnieri}, \citenamefont {Campbell},
  \citenamefont {Prior},\ and\ \citenamefont
  {Goold}}]{purkayastha_periodically_2022}%
  \BibitemOpen
  \bibfield  {author} {\bibinfo {author} {\bibfnamefont {A.}~\bibnamefont
  {Purkayastha}}, \bibinfo {author} {\bibfnamefont {G.}~\bibnamefont
  {Guarnieri}}, \bibinfo {author} {\bibfnamefont {S.}~\bibnamefont {Campbell}},
  \bibinfo {author} {\bibfnamefont {J.}~\bibnamefont {Prior}}, \ and\ \bibinfo
  {author} {\bibfnamefont {J.}~\bibnamefont {Goold}},\ }\href {\doibase
  10.22331/q-2022-09-08-801} {\bibfield  {journal} {\bibinfo  {journal}
  {Quantum}\ }\textbf {\bibinfo {volume} {6}},\ \bibinfo {pages} {801}
  (\bibinfo {year} {2022})}\BibitemShut {NoStop}%
\bibitem [{\citenamefont {Wojtowicz}\ \emph {et~al.}(2022)\citenamefont
  {Wojtowicz}, \citenamefont {Purkayastha}, \citenamefont {Zwolak},\ and\
  \citenamefont {Rams}}]{wojtowicz_accumulative_2022}%
  \BibitemOpen
  \bibfield  {author} {\bibinfo {author} {\bibfnamefont {G.}~\bibnamefont
  {Wojtowicz}}, \bibinfo {author} {\bibfnamefont {A.}~\bibnamefont
  {Purkayastha}}, \bibinfo {author} {\bibfnamefont {M.}~\bibnamefont {Zwolak}},
  \ and\ \bibinfo {author} {\bibfnamefont {M.~M.}\ \bibnamefont {Rams}},\
  }\href {\doibase 10.48550/arXiv.2210.04890} {\enquote {\bibinfo {title}
  {Accumulative reservoir construction: {Bridging} continuously relaxed and
  periodically refreshed extended reservoirs},}\ } (\bibinfo {year}
  {2022})\BibitemShut {NoStop}%
\end{thebibliography}%
\bibliographystyle{apsrev4-1}
\end{document}